\documentclass[11pt,a4paper]{article}
\pdfoutput=1
\usepackage{jheppub}
\usepackage{amsmath}
\usepackage{epsfig}
\usepackage{amssymb}
\usepackage{graphics}
\usepackage[active]{srcltx}
\usepackage{epstopdf}
\usepackage{pdfsync}
\usepackage{shuffle}
\usepackage{slashed}
\usepackage{soul}

\usepackage{tikz}
\usetikzlibrary{decorations.pathmorphing}
\usetikzlibrary{arrows.meta}

\def\spaa #1{\langle #1\rangle}

\def\I{\tiny \mbox{I}}
\def\II{\tiny \mbox{II}}

\def\del #1{\mbox{\st{$#1$}}}
\def\delSub #1{\mbox{\st{{\scriptsize $#1$}}}}
\def\delSubsub #1{\mbox{\st{{\tiny $#1$}}}}

\def\cf{\mathbb{C}_{\footnotesize\mbox{f}}}
\def\uv{\mathbb{U}_{\footnotesize\mbox{v}}}
\def\sym{\mathbb{S}_{\footnotesize\mbox{ym}}}

\title{Five-loop Anomalous Dimensions of Cubic Scalar Theory from Operator Product Expansion}
\author[a]{Rijun Huang}
\author[b]{, Qingjun Jin}
\author[b]{and Yi Li}

\affiliation[a]{School of Physics and Technology, Nanjing Normal University, No.1 Wenyuan Road, Nanjing 210046, P.R.China}
\affiliation[b]{Graduate School of China Academy of Engineering Physics, No. 10 Xibeiwang East Road, Haidian District, Beijing, 100193, P.R.China}

\emailAdd{huang@njnu.edu.cn}
\emailAdd{qjin@gscaep.ac.cn}
\emailAdd{yili@gscaep.ac.cn}

\abstract{In this work, we compute the anomalous dimensions of the $\phi^Q$ operator in six-dimensional cubic scalar theory. The renormalization analysis is carried out within the framework of the Operator Product Expansion method, while the ultraviolet divergences of Feynman integrals are evaluated using the graphical function method. Inspired by the intrinsic connection between Wilson coefficients and anomalous dimensions, an algorithm was proposed recently, which provides a practical and systematic framework for calculating the anomalous dimensions of masses, fields, and composite operators, with broad potential applicability to generic quantum field theories. Notably, the HyperlogProcedures package, developed based on the graphical function method, enables the computation of two-point propagator-type integrals, derived herein for capturing ultraviolet divergences, to very high loop orders. With these advanced techniques, we have successfully computed the anomalous dimensions of the $\phi^Q$ operator up to five loops. Furthermore, we present a large $N$ expansion of the scaling dimensions at the Wilson-Fisher fixed point, extended to the $1/N^5$ order. This computation sets a new loop-order record for the anomalous dimension of the $\phi^Q$ operator in cubic scalar theory, while further validating the efficiency and versatility of the proposed algorithm in renormalization analyses.}


\begin{document}

\maketitle \flushbottom

\section{Introduction}
\label{sec:introduction}

In the development of the renormalization group (RG) method, scalar field theories have emerged as an ideal testing ground for validating multi-loop computation techniques while offering valuable insights into more intricate field theories, such as non-Abelian gauge theories. Nevertheless, the renormalization of scalar field theories is inherently significant in its own right. As elucidated in Wilson's theory of critical phenomena \cite{Wilson:1971dc}, critical exponents near second-order phase transitions in certain statistical systems are intimately linked to the anomalous dimensions of fields or composite operators \cite{PELISSETTO2002549,Moshe:2003xn}. Two prominent examples of renormalizable scalar field theories are the six-dimensional cubic scalar field theory and the four-dimensional quartic scalar field theory. Both theories belong to the same $O(N)$ universality class, and a key conjecture posits that the $O(N)$ quartic scalar theory at its ultraviolet (UV) fixed point is equivalent to the $O(N)$ cubic scalar theory at its infrared (IR) fixed point \cite{Fei:2014yja,Fei:2014xta}. Namely, they approach to the same critical theory at their respective fixed points. Verifying this conjecture with high-loop precision for the scaling dimensions of fields or composite operators represents an intriguing problem, yet it requires perturbative computations of  renormalization functions to higher-loop orders. Thanks to recent technical advancements, there has been a growing surge of interest in pushing renormalization computations to the next loop order.

The renormalization computations of quartic and cubic scalar theories possess a long-standing history, tracing back to early-era studies of critical exponents up to $\epsilon^2$ orders for generalized Ising and Baxter models \cite{Wilson:1971dc}, as well as the generalized Heisenberg model \cite{Wilson:1971vs}. The 3-loop computations of the beta functions, field renormalization and mass renormalization for the quartic scalar theory were completed just one year later \cite{BREZIN1973227,Brezin:1973igb}, with 4-loop results following five years thereafter \cite{Kazakov:1979ik}. The full 5-loop renormalization of the quartic scalar theory, however, demanded over a decade of collaborative efforts from numerous researchers \cite{Chetyrkin:1981jq,Gorishnii:1983gp,Kazakov:1983dyk,Gorishnii:1983jr,Kleinert:1991rg,Adzhemyan:2013jra}. Throughout this development, techniques such as the IR-rearrangement \cite{Vladimirov:1979zm,Chetyrkin:1980pr,Caswell:1981ek}, the $R$-operation ($R^\ast$-operation) \cite{Chetyrkin:1982nn,CHETYRKIN1984419,Larin:2002sc}, and modern Feynman diagram generating and Feynman integral reduction algorithms were refined and played major roles. Early attempts were also made to push renormalization computations to higher-loop orders, for instance the primitive divergences of the 7-loop beta function \cite{Broadhurst:1995km}. Or extend them to composite operators, for instance the 4-loop anomalous dimension of operators with derivatives \cite{Derkachov:1997pf}. Nevertheless, the complete 6-loop computation of the quartic scalar theory only became feasible approximately 15 years after the 5-loop result. Firstly with the 6-loop field renormalization result \cite{Batkovich:2016jus}, followed by the beta function and mass renormalization for the $O(1)$ model \cite{Kompaniets:2016hct}, and subsequently for the $O(N)$ model \cite{Kompaniets:2017yct}. A major breakthrough emerged with the development of the graphical function method \cite{Schnetz:2013hqa,Golz:2015rea,Borinsky:2021gkd,Schnetz:2024qqt,HP}, which enabled the successful computation of the 7-loop beta function, mass renormalization, and field renormalization \cite{Schnetz:2016fhy}, as well as the 8-loop field renormalization function \cite{Schnetz:2025opm}. The history of renormalization computations for the cubic scalar theory has closely paralleled that of the quartic scalar theory, which is not surprising since both have benefited from the same technical advancements. However, due to its greater complexity, progress in the cubic scalar theory has consistently lagged by 1 or 2 loops compared to its quartic counterpart at similar stages. Early 1-loop and 2-loop analysis of the cubic scalar theory revealed its asymptotic freedom \cite{Macfarlane:1974vp,Ma:1975vn}. Over the subsequent 40 years, 3-loop \cite{deAlcantaraBonfim:1980pe,deAlcantaraBonfim:1981sy,Fei:2014xta} and 4-loop \cite{Gracey:2015tta} renormalization computations were gradually completed, with the 5-loop result finished only very recently \cite{Kompaniets:2021hwg,Borinsky:2021jdb}. Once again, the frontier record stems from the graphical function method, through which the 6-loop beta function, field renormalization and mass renormalization functions of the cubic scalar theory have been successfully acquired this year \cite{Schnetz:2025wtu}.

Besides the beta functions and the anomalous dimensions of fields and mass, the renormalization of composite operators constitutes an equally important yet more challenging problem. Phenomenologically, this is relevant to specific critical exponents in phase transition phenomena. One class of composite operators of interest is the fixed-charge symmetric traceless operator $\phi^Q$. In statistical physics, the anomalous dimensions of $\phi^Q$ for certain values of $Q$ are relevant to phase transitions in diverse systems \cite{PELISSETTO2002549}, such as liquid crystals, graphite-intercalation compounds and polymers \cite{DePrato:2003yd,Calabrese:2004ca}. Interest in computing the scaling dimensions of the $\phi^Q$ operator in cubic and quartic scalar theories has grown since 2019, when the semi-classical method is introduced to compute the scaling dimension of $\phi^Q$ at the large $Q$ limit. This advancement leads to all-loop results at leading and sub-leading orders in the $Q$ expansion for the quartic scalar theory \cite{Arias-Tamargo:2019xld,Badel:2019oxl,Antipin:2020abu,Giombi:2020enj} and the cubic scalar theory \cite{Arias-Tamargo:2020fow,Antipin:2021jiw}. Progress in the semi-classical approach has stimulated
increasing attention to the perturbative method approach towards renormalization computation of the $\phi^Q$ operator, aiming to validate the semi-classical results and also to obtain results for complete $Q$ dependence. For the quartic scalar theory, the scaling dimension of the $\phi^Q$ operator has been computed perturbatively at 4-loop \cite{Jack:2021ypd} and 5-loop orders \cite{Jin:2022nqq}. While the 6-loop result is also available \cite{Bednyakov:2022guj}, where the data of $Q=1,\ldots,5$ was computed via perturbative Feynman diagram method and the data of $Q=6,7$ was adopted from the semi-classical method. Concurrently, for the cubic scalar theory, perturbative computations have been carried out at 3-loop \cite{Jack:2021ziq} and 4-loop orders \cite{Huang:2024hsn}.

High-loop order renormalization computations would never be feasible without the continuous evolution of both traditional and modern techniques. One notable approach is the graphical function method \cite{Schnetz:2013hqa,Golz:2015rea,Borinsky:2021gkd,Schnetz:2024qqt,HP}, which takes the advantage of computing scalar integrals in coordinate space. To date, it has advanced renormalization computations to 8-loop order in the four-dimensional quartic scalar theory and 6-loop order in the six-dimensional cubic scalar theory. Furthermore,  with some tricks, it has for the first time pushed renormalization computations to 7-loop order in the Gross-Neveu model \cite{Huang:2025ree}. However, the graphical function method is inherently limited to scalar integrals with no more than four external points. When investigating the anomalous dimensions of higher-length composite operators, for instance the $\phi^Q$ operator, the number of external legs can be very large, rendering the graphical function method inapplicable directly. To address this challenge, a novel renormalization algorithm based on the Operator Product Expansion (OPE) has been proposed \cite{Huang:2024hsn}. As established in the literature, the renormalization group equation (RGE) flow of Wilson coefficients is governed by the anomalous dimensions of operators, {\sl e.g.}, as described in \cite{Collins_1984}. This principle enables the extraction of anomalous dimensions from Wilson coefficients and has been occasionally employed to compute the anomalous dimensions of specific operators. For instance, the OPE of two scalar fields was used to determine the anomalous dimension of the $\phi^2$ operator \cite{Collins_1984}, and similarly, the OPE of two stress-energy tensors was applied to compute the anomalous dimension of the Konishi operator in $\mathcal{N}=4$ super-Yang-Mills theory \cite{Eden:2012fe}. In practice, however, standard renormalization computations typically rely on traditional techniques such as the $R^\ast$-operation method and the massive vacuum bubble method. By considering the OPE of a composite operator with a fundamental field, the proposed algorithm \cite{Huang:2024hsn} extends the applicability of the OPE framework to general composite operators in a systematic manner, making it suitable for implementation in modern computational tools. In this algorithm, the UV finiteness of the corresponding OPE coefficients links the anomalous dimensions of higher-dimensional composite operators to those of lower-dimensional operators, thereby establishing recursive relations for anomalous dimensions. Utilizing these recursion relations, the anomalous dimensions of complicated operators can be non-trivially derived from those of simple operators. Additionally, the UV divergences required for renormalization computation are treated globally without the need to subtract sub-divergences, and the latter is often computationally cumbersome in other renormalization techniques such as the $R^\ast$-operation method. The algorithm also provides a systematic approach to deform integrals of correlation functions involving fields and/or operators into two-point propagator-type integrals, which is particularly advantageous when dealing with systems with numerous external legs. We have explicitly validated the OPE based algorithm in the cubic scalar theory and Gross-Neveu-Yukawa theory \cite{Huang:2024hsn}, confirming its highly versatile and efficient. The utility of determining anomalous dimensions via OPE has also been tested in condensate calculations \cite{Marino:2024uco,Liu:2025bqq}. Notably, the generality of OPE across quantum field theories allows the proposed algorithm to be applied to diverse scenarios, ranging from theories involving scalars or particles with spins to situations that include operator mixing, as well as operators carrying derivatives and Lorentz indices.

In this study, we apply the OPE based algorithm proposed in \cite{Huang:2024hsn}, combined with the graphical function method \cite{Schnetz:2013hqa,Golz:2015rea,Borinsky:2021gkd,Schnetz:2024qqt,HP}, to calculate the 5-loop scaling dimension of the $\phi^Q$ operators in the six-dimensional $O(N)$ cubic scalar theory, thereby extending beyond the 4-loop result. This paper is structured as follows. In \S\ref{sec:review}, we provide a concise review of the cubic scalar theory, its renormalization functions, and the aforementioned renormalization algorithm. Additionally, we present approaches for organizing and simplifying Feynman graphs of correlation functions involving the $\phi^Q$ operator. \S\ref{sec:L2demonstration} is dedicated to a detailed demonstration of the OPE based algorithm via 2-loop examples, illustrating how the field renormalization function, beta function, and anomalous dimensions of the $\phi^Q$ operator can be computed efficiently. In \S\ref{sec:L5result}, we present a general discussion on the computation of the 5-loop anomalous dimension of the $\phi^Q$ operator, with various 5-loop results provided in this section and the Appendix. In \S\ref{sec:conclusion}, conclusion and discussion are presented.

\section{Review of the theory and the method}
\label{sec:review}

In this section, we review the six-dimensional $O(N)$ cubic scalar theory, with a focus on its renormalization computations and the techniques for organizing and simplifying Feynman diagrams. To make this paper self-contained, we also present a brief overview of the OPE based renormalization algorithm \cite{Huang:2024hsn}, particular in the context of its application to the cubic scalar theory.

\subsection{The cubic scalar theory and its renormalization}
\label{subsec:theory}

The theory under consideration is the six-dimensional $O(N)$ cubic scalar theory, defined by the Lagrangian,
\begin{equation}
\mathcal{L}=\frac{1}{2}(\partial\phi_i)^2+\frac{1}{2}(\partial\sigma)^2+\frac{g}{2!}\sigma\phi_i^2+\frac{h}{3!}\sigma^3~,
\end{equation}
where $\sigma$ and $\phi_i,i=1,\ldots,N$ denote two types of real scalar fields, with the $\phi_i$ transforming as an $O(N)$ vector. This theory is perturbatively renormalizable in six-dimensional spacetime. The primary focus of this paper is the renormalization analysis of the symmetric traceless $\phi^Q$ operator, defined as,
\begin{equation}
\phi^Q:=T_{i_1\cdots i_Q}\phi^{i_1}\cdots \phi^{i_Q}~,
\end{equation}
where $T_{i_1\cdots i_Q}$ is a symmetric traceless tensor and $Q$ is a fixed charge number. We work in the $\overline{\mbox{MS}}$ regularization scheme \cite{tHooft:1973mfk} in $d=(6-\epsilon)$-dimension. The classical scaling dimension of $\phi^Q$ is $Q(d/2-1)$, while the total scaling dimension reads
\begin{equation}
\Delta_Q=Q\left(\frac{d}{2}-1\right)+\gamma_{\varphi^Q}~.
\end{equation}
The difference between the total and classical scaling dimensions defines the anomalous dimension $\gamma_{\phi^Q}$ of the $\phi^Q$ operator. It can be computed perturbatively via Feynman diagrams involving the insertion of the $\phi^Q$ operator into 1PI Green functions with $Q$ external legs. However, explicit evaluation of these correlation functions is complicated by the cumbersome $O(N)$ tensor structures inherent to the operator, which cause the number of terms in the Feynman rules to grow exponentially with increasing $Q$. To address this challenge, it is convenient to introduce an auxiliary $O(N)$ zero-norm complex vector $q$ satisfying $q^2=0$, $q\cdot \bar{q}=1$, and define the tensor $T$ as \cite{Antipin:2020abu,Jin:2022nqq}
\begin{equation}
T_{i_1\ldots i_Q}:=\frac{1}{Q!}q_{i_1}q_{i_2}\cdots q_{i_Q}~.
\end{equation}
This allows us to reexpress the $\phi^Q$ operator in terms of a $\varphi^Q$ operator,
\begin{equation}
\phi^Q:=\frac{1}{Q!}\varphi^Q~,
\end{equation}
where the complex field $\varphi$ is defined as $\varphi:=q\cdot \phi$, with its complex conjugate $\bar{\varphi}:=\bar{q}\cdot \phi$. This redefinition effectively eliminates $O(N)$ indices from the correlation functions, and the anomalous dimension of $\phi^Q$ operator is equivalent to that of $\varphi^Q$ operator. The latter can now be computed by $\varphi^Q$-$Q\varphi$ correlation functions, which are free of $O(N)$ index structures. In the $\varphi$-field representation, the $\sigma\phi^2$ vertices along the operator and external legs are reformulated as $\sigma\varphi\bar{\varphi}$ vertices, while other $\sigma\phi^2$ vertices appearing in loop structures of $\varphi$ fields contribute factors of $N$ after summing over $O(N)$ indices. Owing to charge conservation, scalar charge flows along the $\varphi$ fields in the corresponding Feynman diagrams.

For the purpose of renormalization computation, we follow standard textbook procedures and introduce renormalized fields as
\begin{equation}
\varphi_{0}=Z^{\frac{1}{2}}_{\varphi}\varphi~~~,~~~\sigma_{0}=Z^{\frac{1}{2}}_{\sigma}\sigma~,\label{eqn:field-relation}
\end{equation}
along with renormalized coupling constants as
\begin{equation}
g_0=\mu^{\frac{\epsilon}{2}}\widetilde{Z}_gg~~~,~~~
h_0=\mu^{\frac{\epsilon}{2}}\widetilde{Z}_{h}h~~~
\mbox{with}~~~\widetilde{Z}_g=Z_gZ^{-\frac{1}{2}}_{\sigma}Z^{-1}_{\varphi}~~~,~~~
\widetilde{Z}_h=Z_hZ^{-\frac{3}{2}}_{\sigma}~.\label{eqn:couplingRelation}
\end{equation}
Here, $g_0,h_0$ denote bare coupling constants and $\varphi_{0}, \sigma_0$ denote bare fields. We also define the renormalized operator $\varphi^Q_{R}=Z_{\varphi^Q}\varphi_0^Q$, where the anomalous dimension $\gamma_{\varphi^Q}$ is derived from $Z_{\varphi^Q}$. Since $\varphi_0^Q=Z_{\varphi}^{\frac{Q}{2}}\varphi^Q$, we obtain
\begin{equation} \varphi_{R}^{Q}=\widetilde{Z}_{\varphi^Q}\varphi^Q~~~
\mbox{with}~~~\widetilde{Z}_{\varphi^Q}=Z_{\varphi^Q}Z_{\varphi}^{\frac{Q}{2}}~.\label{eqn:zPhi-relation}
\end{equation}
The factor $\widetilde{Z}_{\varphi^Q}$ can then be assigned as the Feynman rule for the $\varphi_R^{Q}$-$Q\varphi$ vertex.

Once $Z_{\varphi}$, $Z_{\sigma}$, $Z_{g}$ and $Z_h$ are determined, beta functions can be computed using standard field theory methods. These beta functions follow from $\frac{\partial g}{\partial\ln \mu}$ and $\frac{\partial h}{\partial\ln \mu}$, which are derived by solving matrix equation,
\begin{equation}\label{eqn:compute-beta}
\begin{pmatrix}
 \frac{1}{g}+\frac{\partial \ln \widetilde{Z}_g}{\partial g} & \frac{\partial \ln \widetilde{Z}_g}{\partial h} \\
  \frac{\partial \ln \widetilde{Z}_h}{\partial g} & \frac{1}{h}+\frac{\partial \ln \widetilde{Z}_h}{\partial h} \\
\end{pmatrix}
\begin{pmatrix}
\frac{\partial g}{\partial\ln \mu}\\
\frac{\partial h}{\partial \ln \mu}\\
\end{pmatrix}=\begin{pmatrix}
-\frac{\epsilon}{2}\\
-\frac{\epsilon}{2}\\
\end{pmatrix}~.
\end{equation}
The field anomalous dimensions of $\varphi$ and $\sigma$ are then computed by definition as
\begin{equation}
\gamma_{\varphi}=\frac{1}{2}\left(\frac{\partial \ln Z_{\varphi}}{\partial g}\beta(g)+\frac{\partial \ln Z_{\varphi}}{\partial h}\beta(h)\right)~~~,~~~\gamma_{\sigma}=\frac{1}{2}\left(\frac{\partial \ln Z_{\sigma}}{\partial g}\beta(g)+\frac{\partial \ln Z_{\sigma}}{\partial h}\beta(h)\right)~.\label{eqn:compute-field-anomalous}
\end{equation}
Furthermore, once $Z_{\varphi^Q}$ is determined, the anomalous dimension of the $\varphi^Q$ operator is given by
\begin{equation}
\gamma_{\varphi^Q}:=-\frac{\partial \ln Z_{\varphi^Q}}{\partial \ln \mu}=-\frac{\partial \ln \widetilde{Z}_{\varphi^Q}Z_{\varphi}^{-\frac{Q}{2}}}{\partial \ln \mu}=-\frac{\partial \ln \widetilde{Z}_{\varphi^Q}}{\partial \ln \mu}+Q\left(\frac{1}{2}\frac{\partial \ln Z_{\varphi}}{\partial \ln \mu}\right)~.\label{eqn:compute-operator-anomalous}
\end{equation}
We denote the first term as $\gamma_Q$, and the second term evaluates to $Q\gamma_\varphi$, such that $\gamma_{\varphi^Q}=\gamma_Q+Q\gamma_\varphi$.

To study the large $N$ expansion of the anomalous dimensions around the Wilson-Fisher fixed point, we first solve for the fixed points of the beta functions,
\begin{equation}
\beta(g^{\ast})=0~~~,~~~\beta(h^{\ast})=0~.
\end{equation}
Substituting the solutions $g^\ast,h^\ast$ into the anomalous dimensions and expanding the results in a series of $1/N$ yields the large $N$ expansion of the cubic scalar theory at the IR fixed point.

Renormalization computations require determining the $Z$-factors. In general, a $Z$-factor including up to $L$-loop contributions takes the schematic form
\begin{equation}
Z_{{\tiny \texttt{X}}}=1+\sum_{n_\ell=1}^{L}\sum_{n_\epsilon=1}^{n_\ell}\frac{z_{\tiny\texttt{X}}^{n_\ell n_{\epsilon}}}{\epsilon^{n_\epsilon}}~,\label{eqn:Z-ansatz}
\end{equation}
where \texttt{X} represents the fields $\varphi,\sigma$, the couplings $h,g$ or the operator $\varphi^Q$. The $Z_{{\tiny \texttt{X}}}$-factor is obtained by determining the coefficients $z_{\tiny\texttt{X}}^{n_\ell n_{\epsilon}}$ up to the relevant loop orders. All such $Z$-factors can be conveniently computed using two-point propagator-type integrals, following the spirit of the OPE based renormalization algorithm.

\subsection{The OPE based algorithm for renormalization}
\label{subsec:OPE}

The OPE is a classical technique for investigating physical quantities in the asymptotic expansion limit, and it is textbook knowledge that the anomalous dimension of an operator is linked to the Wilson coefficients of the OPE \cite{Collins_1984}. In principle, this connection enables the computation of anomalous dimensions directly from Wilson coefficients. In \cite{Huang:2024hsn}, we proposed an algorithm that explicitly realize this connection by relating renormalization $Z$-factors to the UV finiteness of Wilson coefficients. This algorithm was validated in practical renormalization calculations for the cubic scalar theory and Gross-Neveu-Yukawa theory, confirming its versatility and efficiency. Unlike other renormalization computational techniques, such as the $R^\ast$-operation method, this algorithm treats UV divergences globally in renormalization procedures, eliminating the need to subtract sub-divergences arising from certain Feynman diagrams. It also provides a systematic approach to extracting the essential UV divergences from two-point integrals, applicable not only to beta functions and the anomalous dimensions of fundamental fields, but also to the anomalous dimensions of composite operators, including those involving correlation functions with a large number of external legs. 

To describe this algorithm, we begin with a general OPE. The product of two time-ordered operators is asymptotically expanded into a basis of composite operators, where the expansion basis is ordered by the operators' dimensions. In momentum space, the expansion is performed in the large momentum limit $p\to \infty$, schematically expressed as
\begin{equation}
\widehat{O}_{\II}^{\tiny\mbox{ren}}(k-p)\widehat{O}_{\I}^{\tiny\mbox{ren}}(p)~\sim~ C_i^{\tiny\mbox{ren}}(g_i;p)\widehat{O}_{i}^{\tiny\mbox{ren}}(k)+\cdots\label{eqn:OPE-ren}
\end{equation}
where $k$ is a finite momentum, and the momenta of the two operators approach infinity in opposite directions\footnote{Pictorially, we can regard the two large momentum operators as hard fields propagating in a background of soft fields. This idea has also been implemented in the study of boundary operators in BCFW recursion relations \cite{Huang:2016bmv,Huang:2022qnx}. By shifting two external momenta $p_1,p_2$ with a complex auxiliary momentum $zq$ as $p_1\to p_1+z q$, $p_2\to p_2-zq$, at the large $z$ limit we get two hard fields propagating in the background of soft fields.}. The superscript $^{\tiny\mbox{ren}}$ denotes a renormalized operator within the dimensional regularization scheme. Alternatively, we may consider the OPE of bare operators, schematically written as
\begin{equation}
\widehat{O}_{\II}^{\tiny\mbox{bare}}(k-p)\widehat{O}_{\I}^{\tiny\mbox{bare}}(p)~\sim~ C_i^{\tiny\mbox{bare}}(g_{i0};p)\widehat{O}_{i}^{\tiny\mbox{bare}}(k)+\cdots\label{eqn:OPE-bare}
\end{equation}
where the Wilson coefficients are now functions of $p^2$ and the bare coupling constants. Since renormalized and bare operators are related by their corresponding $Z$-factor,
\begin{equation}
\widehat{O}^{{\tiny\mbox{ren}}}=Z_{\widehat{O}} \widehat{O}^{{\tiny\mbox{bare}}}~,\label{eqn:def-Zo}
\end{equation}
combing eqn.(\ref{eqn:OPE-ren}) and eqn.(\ref{eqn:OPE-bare}) yields the relation
\begin{equation}
C_i^{\tiny\mbox{ren}}(g_i;p)=
\frac{Z_{\widehat{O}_{\I}}Z_{\widehat{O}_{\II}}}{Z_{\widehat{O}_i}}C_i^{\tiny\mbox{bare}}(g_{i0};p)~.\label{eqn:C-relation}
\end{equation}
This relation explicitly realize the connection between anomalous dimensions and Wilson coefficients. By carefully selecting the basis operator $\widehat{O}_i$, we can compute the desired $Z$-factors using simpler correlation functions. 

Special attention must be paid to the OPE of fundamental fields. For a fundamental field $\phi$, its associated $Z$-factor does not directly correspond to $Z_{\phi}$. Instead, the field renormalization is defined as $\phi=Z_{\phi}^{-\frac{1}{2}}\phi_0$. Furthermore, amputated correlation functions are typically computed, and an additional factor of $Z_{\phi}$ must be included for each fundamental field in the OPE of amputated correlation functions. Incorporating these two considerations, a fundamental field $\phi$ contributes a factor of $Z^{\frac{1}{2}}_{\phi}$ to the numerator of the Wilson coefficient relation (\ref{eqn:C-relation}) in the context of amputated OPEs. 

Since the Wilson coefficients of renormalized operators, $C_i^{\tiny\mbox{ren}}(g_i;p)$, are UV finite, once the bare Wilson coefficients $C_i^{\tiny\mbox{bare}}(g_{i0};p)$ are computed, the ratios of $Z$-factors can be determined by enforcing UV finiteness conditions, specifically, the cancellation of $\epsilon$ poles. It is possible for basis operators to have identical dimensions, leading to operator mixing. In such cases, the relation (\ref{eqn:def-Zo}) generalizes to vectors of mixed operators related by matrices of $Z$-factors. These $Z$-matrices are determined by imposing UV finiteness on all OPE coefficients corresponding to the mixed operators. Notably, the above discussion applies to generic scalar and tensor operators within the framework of general quantum field theories.

The bare Wilson coefficient $C_i^{\tiny\mbox{bare}}(g_{i0};p)$ for the bare operator $\widehat{O}_i^{\tiny\mbox{bare}}$ is computed via the large momentum expansion of bare correlation functions. For an operator $\widehat{O}_i^{\tiny\mbox{bare}}$ composed of $n$ fields, the corresponding Wilson coefficient is given by
\begin{equation}
C_i^{\tiny\mbox{bare}}(g_{i0};p)=\Big\langle \widehat{O}_{\II}^{\tiny\mbox{bare}}(k-p)\widehat{O}_{\I}^{\tiny\mbox{bare}}(p)\varphi_0(k_1)\varphi_0(k_2)\cdots \varphi_0(k_n)\Big\rangle \Big|_{k_i\to 0~,~i=1,\ldots,n}~,\label{eqn:two-point-Wilson}
\end{equation}
where $k=-\sum_{i=1}^n k_i$. Pictorially, this corresponds to splitting the Feynman graph of the original correlation function into hard and soft sub-graphs by cutting $n$ internal propagators and/or external legs. The Wilson coefficient receives contributions from all possible $n$-cut hard sub-graphs. For operators involving derivative fields, for instance $\partial_\mu \varphi_0(k)$, the coefficient is computed by acting with $i\frac{\partial}{\partial k^\mu}$ on the bare correlation function. For hard sub-graphs, since all $k_i, i=1,\ldots,n$ are set to zero, leaving only two operators with hard momenta, we effectively reduce the problem to two-point graphs. Thus, the bare Wilson coefficients $C_i^{\tiny\mbox{bare}}(g_{i0};p)$ are evaluated using two-point propagator-type integrals. This significantly simplifies computations, particularly when dealing with correlation functions with a large number of external legs. 

The above discussion outlines the core idea of the OPE based algorithm for computing renormalization functions. To determine beta functions and anomalous dimensions, we first extract $Z$-factors from UV finiteness conditions implied by eqn.(\ref{eqn:C-relation}),
\begin{equation}
\frac{Z_{\widehat{O}_{\I}}Z_{\widehat{O}_{\II}}}{Z_{\widehat{O}_i}}C_i^{\tiny\mbox{bare}}(g_{i0};p)=\mbox{UV~finite}~.\label{eqn:UV-condition}
\end{equation}
To establish these UV finiteness conditions, firstly we compute the bare coefficients $C_i^{\tiny\mbox{bare}}(g_{i0};p)$ from the corresponding two-point integrals. Then using the general {\sl ansatz} (\ref{eqn:Z-ansatz}) for $Z$-factors, we expand the full expression as a series in $\epsilon$, and generate equations by setting the coefficients of all $\epsilon$ poles to zero. Solving these equations yields the coefficients
$z_{\tiny\texttt{X}}^{n_\ell n_{\epsilon}}$ of the $Z_{\tiny\texttt{X}}$-factor, thereby determining $Z_{\tiny\texttt{X}}$-factor to the desired loop order, provided the two-point integrals can be computed to that order.

We now apply this algorithm to the cubic scalar theory of interest, working in coordinate space. As standard, the $Z_\varphi$ and $Z_\sigma$-factors are computed from two-point correlation functions. This computation naturally fits within this algorithmic framework. Consider the OPEs,
\begin{equation}
\varphi\bar{\varphi}~\sim~ c_{\mathbf{1}}~\mathbf{1}+\cdots~~~,~~~\sigma\sigma~\sim~c'_{\mathbf{1}}~\mathbf{1}+\cdots~,
\end{equation}
where $\mathbf{1}$ denotes the zero-dimensional unit operator, and $c,c'$ are distinct Wilson coefficients for the two expansions. For the OPE of two fundamental fields, the large momentum expansion yields
\begin{equation}
\spaa{\varphi(x_1)\bar{\varphi}(x_2)}~\cong~ Z_{\varphi}G(g_0,h_0;\varphi(x_1),\bar{\varphi}(x_2))~~~,~~~
\spaa{\sigma(x_1)\sigma(x_2)}~\cong~ Z_{\sigma}G(g_0,h_0;\sigma(x_1),\sigma(x_2))~,\label{eqn:P2fieldOPE}
\end{equation}
where we use the notation $G(g_{i0};\varphi(x_1),\varphi(x_2),\varphi(\del{x}_3),\ldots)$ to represent the bare Wilson coefficient $C_i^{\tiny\mbox{bare}}(g_{i0};p)$. As defined in eqn.(\ref{eqn:two-point-Wilson}), this coefficient is computed via two-point integrals derived from bare correlation functions, with $\del{x}$ indicating the removal of the corresponding external leg. Since they are already two-point functions, no additional legs need ro be removal. Thus $G(g_0,h_0;\varphi(x_1),\bar{\varphi}(x_2))$ and $G(g_0,h_0;\sigma(x_1),\sigma(x_2))$ are evaluated directly from the Feynman graphs of the two-point correlation functions. The $Z$-factors $Z_\varphi$ and $Z_\sigma$ are then determined by imposing UV finiteness conditions of (\ref{eqn:P2fieldOPE}).

The $Z$-factors for the coupling constants $g$ and $h$ are computed from the OPEs,
\begin{equation}
\varphi\bar{\varphi}~\sim~ c_{\sigma}~\sigma+\cdots~~~,~~~\sigma\sigma~\sim~c'_{\sigma}~\sigma+\cdots~.
\end{equation}
In the large momentum limit, these expansions yield $Z$-factor ratios  $Z^{\frac{1}{2}}_{\varphi}Z^{\frac{1}{2}}_{\varphi}/Z^{-\frac{1}{2}}_{\sigma}$ and $Z^{\frac{1}{2}}_{\sigma}Z^{\frac{1}{2}}_{\sigma}/Z^{-\frac{1}{2}}_{\sigma}$, respectively. Here, the soft $\sigma$ field is treated as an operator and dose not require amputation. This leads to
\begin{eqnarray}
&&\spaa{\varphi(x_1)\bar{\varphi}(x_2)\sigma(x_3)}~\cong~ Z_{\sigma}^{\frac{1}{2}}Z_{\varphi}G(g_0,h_0;\varphi(x_1),\bar{\varphi}(x_2),\sigma(\del{x}_3))~,\\
&&\spaa{\sigma(x_1)\sigma(x_2)\sigma(x_3)}~\cong~Z^{\frac{3}{2}}_{\sigma}G(g_0,h_0;\sigma(x_1),\sigma(x_2),\sigma(\del{x}_3))~.
\end{eqnarray}
Using the coupling renormalization relations (\ref{eqn:couplingRelation}), we further obtain\footnote{In the previous paper \cite{Huang:2024hsn}, we adopted the convention that $g_0$ or $h_0$ in the denominator is absorbed into the bare correlation function for three-point functions. This leads to slightly different expressions in that paper, where $G$ represents the bare correlation function divided by $g_0$ or $h_0$, respectively.}
\begin{eqnarray}
&&\spaa{\varphi(x_1)\bar{\varphi}(x_2)\sigma(x_3)}~\cong~\frac{Z_{g}g }{g_0}\mu^{\frac{\epsilon}{2}} G(g_0,h_0;\varphi(x_1),\bar{\varphi}(x_2),\sigma(\del{x}_3))~,\label{eqn:P3couplingOPE-1}\\
&&\spaa{\sigma(x_1)\sigma(x_2)\sigma(x_3)}~\cong~\frac{Z_{h}h}{h_0} \mu^{\frac{\epsilon}{2}} G(g_0,h_0;\sigma(x_1),\sigma(x_2),\sigma(\del{x}_3))~.\label{eqn:P3couplingOPE-2}
\end{eqnarray}
The coefficients $G(g_0,h_0;\varphi(x_1),\bar{\varphi}(x_2),\sigma(\del{x}_3))$ and $G(g_0,h_0;\sigma(x_1),\sigma(x_2),\sigma(\del{x}_3))$ are computed directly from two-point integrals of the corresponding three-point functions, with one $\sigma$ field leg removed. The factors $Z_g$ and $Z_h$ are then determined by enforcing UV finiteness on (\ref{eqn:P3couplingOPE-1}) and (\ref{eqn:P3couplingOPE-2}). With $Z_\varphi,Z_{\sigma}$ and $Z_g,Z_h$ in hand, the beta functions and field anomalous dimensions are readily computed via definitions (\ref{eqn:compute-beta}) and (\ref{eqn:compute-field-anomalous}).

Finally, the $Z$-factor of the $\varphi^Q$ operator is computed from the OPE,
\begin{equation}
\varphi^Q\bar{\varphi}~\sim~c_{Q}~\varphi^{Q-1}~.
\end{equation}
After large momentum expansion of this OPE yields the $Z$-factors ratio
\begin{equation}
\frac{Z_{\varphi^Q}Z^{\frac{1}{2}}_\varphi}{Z_{\varphi^{Q-1}}}
=\frac{Z_{\varphi^Q}Z^{\frac{Q}{2}}_\varphi}{Z_{\varphi^{Q-1}}Z^{\frac{Q-1}{2}}_\varphi}
=\frac{\widetilde{Z}_{\varphi^Q}}{\widetilde{Z}_{\varphi^{Q-1}}}~.
\end{equation}
Thus, $Z_{\varphi^Q}$ can be computed recursively by imposing UV finiteness condition,
\begin{equation}\label{eqn:phiQ-UV-finiteness}
\frac{\widetilde{Z}_{\varphi^Q}}{\widetilde{Z}_{\varphi^{Q-1}}}
G_{Q}(g_0,h_0;x_0,\del{x}_{\sigma_1},\ldots,\del{x}_{\sigma_{Q-1}},x_Q)=\mbox{UV~finite}~,
\end{equation}
where $G_{Q}$ is computed from Feynman graphs of the $\varphi^Q$-$Q\varphi$ correlation function with $(Q-1)$ external legs removed. These integrals are two-point propagator-type integrals, with large momentum localized at the $\varphi^Q$ operator and one remaining external leg. Using the resulting $Z_{\varphi^Q}$ and the previously computed beta functions, the anomalous dimension of the $\varphi^Q$ operator os obtained via eqn.(\ref{eqn:compute-operator-anomalous}).

As a side remark, the transformation of general integrals into two-point integrals is a common technique for evaluating UV divergences across various scenarios. Examples include the graphical function method \cite{Schnetz:2013hqa,Golz:2015rea,Borinsky:2021gkd,Schnetz:2024qqt,HP}, the IR-rearrangement technique \cite{Vladimirov:1979zm,Chetyrkin:1980pr,Caswell:1981ek}, and works that do not rely on the OPE approach \cite{Fei:2014xta,Jack:2021ziq}.  In all these cases, two-point integrals serve as a convenient computational tool for extracting UV divergences from Feynman integrals. However, with the aid of the OPE, two-point integrals are encoded as Wilson coefficients, from which anomalous dimensions can be systematically derived.

\subsection{From Feynman diagrams to two-point integrals}
\label{subsec:diagram}

The UV finiteness conditions discussed in the preceding subsection require computing two-point integrals derived from two-point, three-point, and $\varphi^Q$-$Q\varphi$ correlation functions. In this subsection, we review key techniques that facilitate simplifying the generation of Feynman diagrams and transforming them into two-point graphs for the subsequent evaluation of two-point integrals.

\subsubsection*{The $\Phi$-field representation}

The cubic scalar theory under consideration involves two types of fields $\varphi,\sigma$ and two classes of cubic interactions. The complex field $\varphi$ carries a scalar charge that flows along $\varphi$ and $\bar{\varphi}$ lines. This charge flow can be represented by an arrow indicating its direction. Generating Feynman diagrams requires accounting for all possible internal field configurations, a task that becomes increasing cumbersome as the number of internal lines grows. Notably, for a fixed external field configuration, different internal field configurations only affect the combinations of couplings, while the underlying scalar integrals remain identical. Thus, there is no need to generate all Feynman diagrams explicitly or redundantly compute identical integrals. To decouple the coupling structure from the integral calculations, we adopt the $\Phi$-field representation. In this framework, we unify the $\varphi_i,i=1,\ldots,N$ fields and the $\sigma$ field into a single $(N+1)$ component scalar field
\begin{equation}
\Phi_I=(\Phi_0,\Phi_1,\ldots,\Phi_N):=(\sigma,\varphi_1,\ldots,\varphi_N)~,
\end{equation}
with the cubic interactions universally encoded as
\begin{equation}
\frac{1}{6} \lambda_{IJK}\Phi_I\Phi_J\Phi_K~~~\mbox{where}~~~\lambda_{000}=h~~\mbox{and}~~\lambda_{0ij}=g\delta_{ij}~~\forall i,j\neq 0~,
\end{equation}
and all other components of the symmetric coupling tensor $\lambda_{IJK}$ vanish. In the $\Phi$-field representation, we generate Feynman diagram topologies without specifying the explicit component fields of $\Phi_I$. This effectively separates the scalar integral information from the coupling-dependent factors. The Feynman scalar integral for a given topology is derived by assigning propagators according to standard Feynman rules, while the coupling factors for all diagrams sharing that topology are encapsulated in combinations of the $\lambda_{IJK}$ couplings. We refer to this combined term as the {\sl coupling factor}.

\subsubsection*{The symmetry factor and the coupling factor}

When constructing the Feynman integrals for correlation functions, we first generate all corresponding topologies using custom Mathematica code. For each topology, the scalar integral is derived from the product of propagators, multiplied by the standard symmetry factor $\sym$, accounting for the total symmetry of the graph and external legs. Each topology effectively represents a set of numerous Feynman diagrams that share the same external state configuration but differ in their internal state configurations. By summing over the indices of the $\lambda_{IJK}$, we obtain the coupling factor $\cf$ for the contributing Feynman graphs.

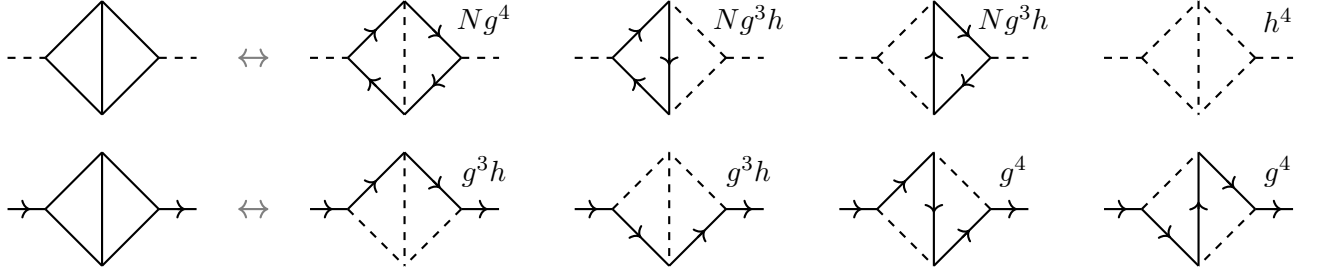
\begin{figure}
  \centering
    \begin{tikzpicture}
      \node [] at (1.25,3.5) {
      \begin{tikzpicture}
        \draw [thick] (0,0.75)--(0,-0.75);
        \draw [thick] (-0.75,0)--(0,0.75);
        \draw [thick] (-0.75,0)--(0,-0.75);
        \draw [thick] (0.75,0)--(0,0.75);
        \draw [thick] (0.75,0)--(0,-0.75);
        \draw [dashed, thick] (-1.25,0)--(-0.75,0) (0.75,0)--(1.25,0);
      \end{tikzpicture}
      };
      \node [] at (5.25,3.5) {
      \begin{tikzpicture}
        \draw [thick,dashed] (0,0.75)--(0,-0.75);
        \draw [thick] (-0.75,0)--(0,0.75);
        \draw [thick,->] (-0.375,0.375)--(-0.374,0.376);
        \draw [thick] (-0.75,0)--(0,-0.75);
        \draw [thick,->] (-0.375,-0.375)--(-0.475,-0.275);
        \draw [thick] (0.75,0)--(0,0.75);
        \draw [thick,->] (0.375,0.375)--(0.475,0.275);
        \draw [thick] (0.75,0)--(0,-0.75);
        \draw [thick,->] (0.375,-0.375)--(0.374,-0.376);
        \draw [dashed, thick] (-1.25,0)--(-0.75,0) (0.75,0)--(1.25,0);
      \end{tikzpicture}
      };
      \node [] at (8.75,3.5) {
      \begin{tikzpicture}
        \draw [thick] (0,0.75)--(0,-0.75);
        \draw [thick,->] (0,0)--(0,-0.1);
        \draw [thick] (-0.75,0)--(0,0.75);
        \draw [thick,->] (-0.375,0.375)--(-0.374,0.376);
        \draw [thick] (-0.75,0)--(0,-0.75);
        \draw [thick,->] (-0.375,-0.375)--(-0.475,-0.275);
        \draw [thick,dashed] (0.75,0)--(0,0.75);
        \draw [thick,dashed] (0.75,0)--(0,-0.75);
        \draw [dashed, thick] (-1.25,0)--(-0.75,0) (0.75,0)--(1.25,0);
      \end{tikzpicture}
      };
      \node [] at (12.25,3.5) {
      \begin{tikzpicture}
        \draw [thick] (0,0.75)--(0,-0.75);
        \draw [thick,->] (0,0)--(0,0.1);
        \draw [thick,dashed] (-0.75,0)--(0,0.75);
        \draw [thick,dashed] (-0.75,0)--(0,-0.75);
        \draw [thick] (0.75,0)--(0,0.75);
        \draw [thick,->] (0.375,0.375)--(0.475,0.275);
        \draw [thick] (0.75,0)--(0,-0.75);
        \draw [thick,->] (0.375,-0.375)--(0.374,-0.376);
        \draw [dashed, thick] (-1.25,0)--(-0.75,0) (0.75,0)--(1.25,0);
      \end{tikzpicture}
      };
      \node [] at (15.75,3.5) {
      \begin{tikzpicture}
        \draw [thick,dashed] (0,0.75)--(0,-0.75);
        \draw [thick,dashed] (-0.75,0)--(0,0.75);
        \draw [thick,dashed] (-0.75,0)--(0,-0.75);
        \draw [thick,dashed] (0.75,0)--(0,0.75);
        \draw [thick,dashed] (0.75,0)--(0,-0.75);
        \draw [dashed, thick] (-1.25,0)--(-0.75,0) (0.75,0)--(1.25,0);
      \end{tikzpicture}
      };
      \draw [gray, thick, <->] (3.05,3.5)--(3.45,3.5);
      \node [] at (6.3,4) {$Ng^4$};
      \node [] at (9.8,4) {$Ng^3h$};
      \node [] at (13.3,4) {$Ng^3h$};
      \node [] at (16.8,4) {$h^4$};
      \node [] at (1.25,1.5) {
      \begin{tikzpicture}
        \draw [thick] (0,0.75)--(0,-0.75);
        \draw [thick] (-0.75,0)--(0,0.75);
        \draw [thick] (-0.75,0)--(0,-0.75);
        \draw [thick] (0.75,0)--(0,0.75);
        \draw [thick] (0.75,0)--(0,-0.75);
        \draw [thick] (-1.25,0)--(-0.75,0) (0.75,0)--(1.25,0);
        \draw [thick,->] (-1,0)--(-0.95,0);
        \draw [thick,->] (1,0)--(1.1,0);
      \end{tikzpicture}
      };
      \node [] at (5.25,1.5) {
      \begin{tikzpicture}
        \draw [thick,dashed] (0,0.75)--(0,-0.75);
        \draw [thick] (-0.75,0)--(0,0.75);
        \draw [thick,->] (-0.375,0.375)--(-0.374,0.376);
        \draw [thick,dashed] (-0.75,0)--(0,-0.75);
        \draw [thick] (0.75,0)--(0,0.75);
        \draw [thick,->] (0.375,0.375)--(0.475,0.275);
        \draw [thick,dashed] (0.75,0)--(0,-0.75);
        \draw [thick] (-1.25,0)--(-0.75,0) (0.75,0)--(1.25,0);
        \draw [thick,->] (-1,0)--(-0.95,0);
        \draw [thick,->] (1,0)--(1.1,0);
      \end{tikzpicture}
      };
      \node [] at (8.75,1.5) {
      \begin{tikzpicture}
        \draw [thick,dashed] (0,0.75)--(0,-0.75);
        \draw [thick,dashed] (-0.75,0)--(0,0.75);
        \draw [thick] (-0.75,0)--(0,-0.75);
        \draw [thick,->] (-0.375,-0.375)--(-0.374,-0.376);
        \draw [thick,dashed] (0.75,0)--(0,0.75);
        \draw [thick] (0.75,0)--(0,-0.75);
        \draw [thick,->] (0.375,-0.375)--(0.475,-0.275);
        \draw [thick] (-1.25,0)--(-0.75,0) (0.75,0)--(1.25,0);
        \draw [thick,->] (-1,0)--(-0.95,0);
        \draw [thick,->] (1,0)--(1.1,0);
      \end{tikzpicture}
      };
      \node [] at (12.25,1.5) {
      \begin{tikzpicture}
        \draw [thick] (0,0.75)--(0,-0.75);
        \draw [thick,->] (0,0)--(0,-0.05);
        \draw [thick] (-0.75,0)--(0,0.75);
        \draw [thick,->] (-0.375,0.375)--(-0.374,0.376);
        \draw [thick,dashed] (-0.75,0)--(0,-0.75);
        \draw [thick,dashed] (0.75,0)--(0,0.75);
        \draw [thick] (0.75,0)--(0,-0.75);
        \draw [thick,->] (0.375,-0.375)--(0.475,-0.275);
        \draw [thick] (-1.25,0)--(-0.75,0) (0.75,0)--(1.25,0);
        \draw [thick,->] (-1,0)--(-0.95,0);
        \draw [thick,->] (1,0)--(1.1,0);
      \end{tikzpicture}
      };
      \node [] at (15.75,1.5) {
      \begin{tikzpicture}
        \draw [thick] (0,0.75)--(0,-0.75);
        \draw [thick,->] (0,0)--(0,0.1);
        \draw [thick,dashed] (-0.75,0)--(0,0.75);
        \draw [thick] (-0.75,0)--(0,-0.75);
        \draw [thick,->] (-0.375,-0.375)--(-0.374,-0.376);
        \draw [thick] (0.75,0)--(0,0.75);
        \draw [thick,->] (0.375,0.375)--(0.475,0.275);
        \draw [thick,dashed] (0.75,0)--(0,-0.75);
        \draw [thick] (-1.25,0)--(-0.75,0) (0.75,0)--(1.25,0);
        \draw [thick,->] (-1,0)--(-0.95,0);
        \draw [thick,->] (1,0)--(1.1,0);
      \end{tikzpicture}
      };
      \draw [thick, gray, <->] (3.05,1.5)--(3.45,1.5);
      \node [] at (6.3,2) {$g^3h$};
      \node [] at (9.8,2) {$g^3h$};
      \node [] at (13.3,2) {$g^4$};
      \node [] at (16.8,2) {$g^4$};
     \end{tikzpicture}
  \caption{Examples of coupling factors in the $\Phi$-field representations and their corresponding component field representations. The internal $\Phi$ fields represent all possible internal state configurations in component fields.}\label{fig:coupling-factor}
\end{figure}

To illustrate the concept of the coupling factor, we consider the two-point two-loop topology shown below, 
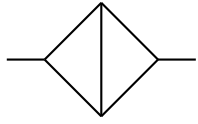
\begin{center}
\begin{tikzpicture}
\draw [thick] (0,0.75)--(0,-0.75);
\draw [thick] (-0.75,0)--(0,0.75);
\draw [thick] (-0.75,0)--(0,-0.75);
\draw [thick] (0.75,0)--(0,0.75);
\draw [thick] (0.75,0)--(0,-0.75);
\draw [thick] (-1.25,0)--(-0.75,0) (0.75,0)--(1.25,0);
\end{tikzpicture}
\end{center}
For this topology, we analyze two external field configurations: (1) two external $\sigma$ field represented by dashed lines, (2) two external $\varphi_i$ and $\bar{\varphi}_j$ fields represented by arrowed lines, as depicted in the LHS of Fig.(\ref{fig:coupling-factor}). Each vertex is assigned a $\lambda_{IJK}$ tensor, and summing over the internal indices yields the respective coupling factors
\begin{eqnarray}
&&\cf(\Gamma^{\tiny \mbox{L2-II}}_{\sigma\sigma})=\sum_{I_1,I_2,I_3,I_4,I_5}\lambda_{0I_1I_2}\lambda_{I_1I_4I_3}\lambda_{I_2I_3I_5}
\lambda_{I_5I_4 0}=Ng^4+2Ng^3h+h^4~,\\
&&\cf(\Gamma^{\tiny \mbox{L2-II}}_{\varphi_{i}\bar{\varphi}_{j}})=\sum_{I_1,I_2,I_3,I_4,I_5}\lambda_{iI_1I_2}\lambda_{I_1I_4I_3}\lambda_{I_2I_3I_5}
\lambda_{I_5I_4 j}=(2g^3h+2g^4)\delta_{ij}~.
\end{eqnarray}
Alternatively, we can  draw all contributing Feynman diagrams with their internal field configurations, as shown in the RHS of Fig.(\ref{fig:coupling-factor}). It can be verified that summing over the contributions of all four Feynman diagrams for each external field configuration reproduces the coupling factors derived above. Since generating topologies is significantly more efficient than generating full Feynman diagrams, and evaluating coupling factors involves only algebraic operations, the $\Phi$-field representation substantially enhances computational efficiency.

\subsubsection*{The Feynman graphs of $\varphi^Q$-$Q\varphi$ correlation functions}
In the UV finiteness conditions, the function $G$ is schematically computed as
\begin{equation}
  G=\sum \sym(\Gamma^{{\tiny\mbox{topo}}})\cf(\Gamma^{{\tiny\mbox{topo}}})\uv(\Gamma^{{\tiny\mbox{topo}}})~,
\end{equation}
where the summation runs over all possible topologies, and the UV divergence $\uv(\Gamma^{{\tiny\mbox{topo}}})$ is evaluated from the two-point integrals generated by these topologies. It is worth noting that the coupling factor $\cf$ of a topology depends on the explicit external field configuration, whereas the symmetry factor $\sym$ and UV divergence $\uv$ are topology-dependent quantities that are independent of external fields. For two-point correlation functions, $G$ is directly evaluated using all 1PI two-point topologies. For three-point correlation functions, we start from all 1PI three-point topologies and construct independent two-point graphs by removing one external leg from these topologies in every possible manner. For the $\varphi^Q$-$Q\varphi$ correlation function, we begin with all 1PI $\varphi^Q\to Q\varphi$ topologies. $G_Q$ is then computed from all independent two-point graphs generated by removing $(Q-1)$ external legs from these topologies in all possible ways.

By analyzing the structure of 1PI $\varphi^Q\to Q\varphi$ topologies, it has been proven that only a subset of these topologies contributes to the two-point integrals \cite{Huang:2024hsn}. These contributing topologies are termed {\sl irreducible graphs}, defined by the property that deleting the operator vertex leaves all remaining external legs connected within a single graph (with the exception of configurations that reduce to a single line). The remaining topologies are classified as {\sl reducible graphs}, which exhibit the property that deleting the operator vertex partitions the external legs into two or more disjoint connected components. Thus, we only need to generate irreducible graphs from the set of 1PI $\varphi^Q\to Q\varphi$ topologies, and compute $G_Q$ from their corresponding two-point graphs after all possible removals of $(Q-1)$ legs. The summation of these graphs yields
\begin{equation}
G_{Q}(g_0,h_0;x_0,\del{x}_{\sigma_1},\ldots,\del{x}_{\sigma_{Q-1}},x_Q)
=\uv\left( \frac{1}{x_Q^{2\lambda}}\right)+\sum_{n=2}^{Q}C_{Q-1}^{n-1}\mathtt{F}_n(\del{x}_{\sigma_1},\cdots, \del{x}_{\sigma_{n-1}}, x_Q)~.\label{eqn:phiQ-G}
\end{equation}
The first term corresponds to the tree-level contribution and evaluates to 1, while
\begin{equation}
\mathtt{F}_{n}(\del{x}_{\sigma_1},\cdots, \del{x}_{\sigma_{n-1}}, x_Q)=(n-1)!\sum_{n_\ell=n-1}^{L}
G_{\scriptsize\mbox{irre}}^{n_\ell{\tiny\mbox{-loop}}}(g_0,h_0;x_0,\del{x}_{\sigma_1},\ldots,\del{x}_{\sigma_{Q-1}},x_Q)~.
\end{equation}
$G_{\scriptsize\mbox{irre}}^{n_\ell{\tiny\mbox{-loop}}}$ receives contributions from all $n_\ell$-loop irreducible topologies. Following standard procedures, it can be evaluated using the symmetry factor, coupling factor and UV divergence as
\begin{equation}
G_{\scriptsize\mbox{irre}}^{n_\ell{\tiny\mbox{-loop}}}(g_0,h_0;x_0,\del{x}_{\sigma_1},\ldots,\del{x}_{\sigma_{Q-1}},x_Q)=
\sum_{{\tiny\mbox{irreducible}}\atop {\tiny\mbox{topologies}}}\sym(\Gamma^{n_\ell}_{\otimes\varphi\delSub{\varphi}\cdots\delSub{\varphi}})
\cf(\Gamma^{n_\ell}_{\otimes\varphi\delSub{\varphi}\cdots\delSub{\varphi}})
\uv(\Gamma^{n_\ell}_{\otimes\varphi\delSub{\varphi}\cdots\delSub{\varphi}})~,
\end{equation}
where the summation is performed over all non-isomorphic graphs generated by removing $(Q-1)$ external legs from the irreducible 1PI topologies of the $\varphi^Q$-$Q\varphi$ correlation function.

Let us summarize this section as follows. After generating the independent topologies of $\varphi^Q$-$Q\varphi$ correlation functions, we compute the symmetry factor and coupling factor of each topology, along with the UV divergences of two-point integrals in the dimensional regularization scheme up to the desired $\epsilon$ order. The quantities $G$ and $G_Q$ can then be readily obtained by summing over all two-point graphs. The $Z$-factors are determined from the UV finiteness conditions, and the anomalous dimensions are ultimately computed from these $Z$-factors.

\section{Two-loop demonstration}
\label{sec:L2demonstration}

For illustrative purposes, this section presents detailed computations of the two-loop anomalous dimensions of the $\varphi^Q$ operator. Beta functions and anomalous dimensions are derived from $Z$-factors following standard quantum field theory definitions, where the $Z$-factors themselves are determined via the OPE based algorithm. This algorithm provides a non-trivial implementation of the connection between OPE Wilson coefficients and anomalous dimensions, enabling the systematic determination of $Z$-factors through summation over two-point integrals. The HyperlogProcedures package \cite{HP} developed from graphical function method assures the computation of two-point integral to very high loop order. The computational work-flow is formalized as a sequential algorithm, proceeding through four key steps: (1) computing the symmetry factors of topologies, (2) evaluating the coupling factors of Feynman diagrams, (3) determining the UV divergences of two-point integrals, and (4) extracting $Z$-factors from UV finiteness conditions.

\subsection{Two-point correlation functions and the field renormalization}
\label{subsec:L2field}

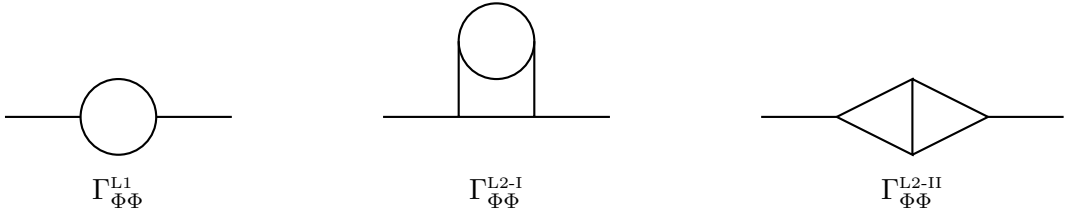
\begin{figure}
  \centering
    \begin{tikzpicture}
    \node [] at (1.5,0) {
      \begin{tikzpicture}
        \draw [thick] (0,0)--(1,0) (2,0)--(3,0);
        \draw [thick] (1.5,0) circle [radius=0.5];
      \end{tikzpicture}
    };
    \node [] at (6.5,0.75) {
      \begin{tikzpicture}
        \draw [thick] (0,0)--(3,0) (1,0)--(1,1) (2,0)--(2,1);
        \draw [thick] (1.5,1) circle [radius=0.5];
      \end{tikzpicture}
    };
    \node [] at (12,0) {
      \begin{tikzpicture}
        \draw [thick] (0,0)--(1,0)--(2,-0.5)--(2,0.5)--(1,0) (2,0.5)--(3,0) (2,-0.5)--(3,0)--(4,0);
      \end{tikzpicture}
    };
    \node [] at (1.5,-1) {$\Gamma^{\tiny \mbox{L1}}_{\Phi\Phi}$};
    \node [] at (6.5,-1) {$\Gamma^{\tiny \mbox{L2-I}}_{\Phi\Phi}$};
    \node [] at (12,-1) {$\Gamma^{\tiny \mbox{L2-II}}_{\Phi\Phi}$};
    \end{tikzpicture}
  \caption{The 1-loop and 2-loop 1PI topologies of two-point functions in $\Phi$-field representation.}\label{fig:P2topo}
\end{figure}

We begin with the computation of field renormalization. As specified in eqn.(\ref{eqn:P2fieldOPE}), the $Z$-factors $Z_\varphi$ and $Z_\sigma$ are derived from the two-point correlation functions $\spaa{\varphi\bar{\varphi}}$ and $\spaa{\sigma\sigma}$, respectively. Notably, while the derivations coincide with those of traditional renormalization methods, we frame it explicitly within the OPE-based algorithm for consistency. The one-loop and two-loop topologies in the $\Phi$-field representation are illustrate in Fig.(\ref{fig:P2topo}).
Each topology represents a set of diagrams with distinct valid internal configurations of the $\sigma$ and $\varphi_i$ fields, where these sets are characterized by their respective coupling factors. The symmetry factors and coupling factors for these topologies are listed in the table below,
\begin{center}
\begin{tabular}{|c||c|c|c|}
  \hline
  topology & $\sym(\Gamma_{\Phi\Phi})$ & $\cf(\Gamma_{\sigma\sigma})$ & $\cf(\Gamma_{\varphi\bar{\varphi}})$ \\ \hline \hline
  $\Gamma^{\tiny \mbox{L1}}_{\Phi\Phi}$ & 1/2 & $h^2+Ng^2$ & $2g^2$ \\ \hline
  $\Gamma^{\tiny \mbox{L2-I}}_{\Phi\Phi}$ & 1/2 & $N(2 g^4 +g^2 h^2)+h^4$ & $Ng^4+2 g^4+g^2 h^2$ \\ \hline
  $\Gamma^{\tiny \mbox{L2-II}}_{\Phi\Phi}$ & 1/2 & $N(g^4 +2 g^3 h )+h^4$ & $2 g^4+2 g^3 h$ \\
  \hline
\end{tabular}
\end{center}
No external legs need to be cut from these topologies, and their UV divergences are computed directly using the \verb!TwoPoint! command in the HyperlogProcedures package. Feynman integrals in the $\overline{\mbox{MS}}$ dimensional regularization scheme are evaluated as power series in $\epsilon$, with results presented below (an overall factor $(1/p^2)^{\frac{\epsilon}{2}L}$ for $L$-loop integral is implicitly included and not explicitly shown),
\begin{eqnarray}
  &&\uv(\Gamma^{\tiny \mbox{L1}}_{\Phi\Phi})=-\frac{1}{3 \epsilon}-\frac{4}{9}+\left(\frac{\pi^2}{72}-\frac{13}{27}\right)\epsilon +\mathcal{O}(\epsilon^2)~, \\
  &&\uv(\Gamma^{\tiny \mbox{L2-I}}_{\Phi\Phi})=\frac{1}{18 \epsilon^2}+\frac{43}{216 \epsilon}+\left(\frac{1171}{2592}-\frac{\pi^2}{216}\right)+\left(-\frac{\zeta_3}{108}-\frac{43 \pi^2}{2592}+\frac{25195}{31104}\right)\epsilon +\mathcal{O}(\epsilon^2)~, \\
  &&\uv(\Gamma^{\tiny \mbox{L2-II}}_{\Phi\Phi})=-\frac{1}{3 \epsilon^2}-\frac{1}{\epsilon}+\left(\frac{\pi^2}{36}-\frac{15}{8}\right)+ \left(-\frac{7\zeta_3}{36} +\frac{\pi^2}{12}-\frac{769}{288}\right)\epsilon+\mathcal{O}(\epsilon^2)~.
\end{eqnarray}
The bare correlation functions are then obtained as
\begin{equation}\label{eqn:compute-2point}
G(g_0,h_0;\Phi,\Phi)=1-\sum_{L={\tiny\mbox{L1}},\atop {\tiny\mbox{L2-I}},{\tiny\mbox{L2-II}}}\sym(\Gamma^{L}_{\Phi\Phi})\cf(\Gamma^{L}_{\Phi\Phi})\uv(\Gamma^{L}_{\Phi\Phi})~,
\end{equation}
where $\Phi$ corresponds to either $\sigma$ or $\varphi$ in the coupling factor $\cf$. Care must be taken with the sign of each term in the summation. Bare correlation functions include contributions from all tree-level and loop Feynman diagrams, but the \verb!TwoPoint! command only computes the scalar integrals, omitting $i$-factors from vertices, propagators, or other possible sources. These $i$-factors must be recovered using Feynman rules derived for the Euclidean-space Lagrangian. Graphical analysis shows the overall factor is $(-1)^{n_p}i$, where $n_p$ is the number of external legs. For two-point integrals ($n_p=2$), each loop order carries a factor $i$, while the tree-level contrition has a factor $-i$. Factoring out an overall $-i$, the two-point correlation function is expressed as the tree-level contribution minus the sum of loop contributions.

The renormalized two-point correlation functions eqn.(\ref{eqn:P2fieldOPE}) must be UV finite, leading to the UV finiteness conditions
\begin{equation}
Z_{\varphi}G(g_0,h_0;\varphi(x_0),\bar{\varphi}(x_1))=\mbox{UV~finite}~~~,~~~Z_{\sigma}G(g_0,h_0;\sigma(x_0),\sigma(x_1))=\mbox{UV~finite}~.
\end{equation}
To derive these conditions explicitly, firstly we rewrite the bare couplings $g_0,h_0$ in the bare correlation functions in terms of renormalized couplings $g,h$ using eqn.(\ref{eqn:couplingRelation}), then expand the full expression in the double limit $g,h\to 0$, followed by $\epsilon\to 0$. Note that eqn.(\ref{eqn:couplingRelation}) implies each $g$ or $h$ is accompanied by a factor $\mu^{\frac{\epsilon}{2}}$, so $L$-loop contributions carry a $\mu^{L\epsilon}$ factor. Combined with the implicit $(1/p^2)^{\frac{\epsilon}{2}L}$ factor, $L$-loop terms include an overall $(\mu^2/p^2)^{\frac{\epsilon}{2}L}$ factor. Expanding this factor around $\epsilon\to 0$ generates terms proportional to powers of $\ln\frac{\mu^2}{p^2}$. Since each power of $\ln\frac{\mu^2}{p^2}$ is independent, UV finiteness requires the coefficient of every $\ln\frac{\mu^2}{p^2}$ term to be free of $\epsilon$ poles, introducing additional equations beyond those from the leading terms. Inspecting the UV divergence structures of $L$-loop integrals, the leading divergence is ${1/\epsilon^L}$, so the $\ln^0\frac{\mu^2}{p^2}$ order generally yields $L$ equations for canceling $L$ poles. For $\ln^{k}\frac{\mu^2}{p^2}$ terms, the leading divergence is $1/\epsilon^{L-k}$, producing $(L-k)$ equations. We may either use only equations from leading Log terms\footnote{Leading Log terms refer to those with $\ln^0(\mu^2/p^2)$ dependence, sub-Log terms to $\ln(\mu^2/p^2)$ dependence, and so on.}, or include those from higher Log orders. If only leading Log equations are used, $Z_\varphi,Z_\sigma,Z_g,Z_h$ must be solved simultaneously to fully determine the $Z$-factors. In contrast, incorporating sub-Log equations allows each $Z$-factor to be determined independently via it own UV finiteness conditions.

We now analyze the UV finiteness conditions for the $Z_\varphi$-factor. At one-loop order, following the expansions outlined above, we obtain
\begin{equation}
\left(\frac{1}{\epsilon}\left(z_{\phi}^{11}+\frac{g^2}{3}\right)+\left(1+\frac{7 g^2}{9}\right)+\mathcal{O}(\epsilon)\right)
+\left(\frac{g^2}{3}+\frac{7g^2}{9}\epsilon+\mathcal{O}(\epsilon^2)\right)\ln\frac{\mu^2}{p^2}+\cdots
\end{equation}
The coefficients of the sub-Log terms are already UV finite and do not yield additional equations. For the leading Log term to be finite, the $1/\epsilon$ pole must cancel, leading to the condition,
\begin{equation}
z_{\varphi}^{11}=-\frac{g^2}{3}~.
\end{equation}
At two-loop order, we solve for the coefficients $z^{21}_{\varphi},z^{22}_{\varphi}$. Expanding the expression, the cancelation of $1/\epsilon$ and $1/\epsilon^2$ poles in the leading Log term gives two equations,
\begin{eqnarray}
&& 0=z_\varphi^{21}+z_\varphi^{11}+\frac{7}{3}\left(\frac{2}{3} g^2 z_g^{11}-\frac{1}{3} g^2 z_\varphi^{11}-\frac{1}{3} g^2 z_\sigma^{11}\right)+\left(\frac{g^2}{3}-\frac{67 g^4 N}{432}+\frac{293 g^4}{216}+\frac{5 g^3 h}{3}-\frac{67 g^2 h^2}{432}\right)~,\nonumber\\
&& 0=z_\varphi^{22}+\left(\frac{2}{3} g^2 z_g^{11}-\frac{1}{3} g^2 z_\varphi^{11}-\frac{1}{3} g^2 z_\sigma^{11}\right)+\left(\frac{5 g^4}{18}+\frac{g^3 h}{3}-\frac{g^4 N}{36}-\frac{g^2 h^2}{36}\right)~,\nonumber
\end{eqnarray}
while the cancelation of the $1/\epsilon$ pole in the sub-Log term provides a third equation,
\begin{equation}
0=\left(\frac{2}{3} g^2 z_g^{11}-\frac{1}{3} g^2 z_\varphi^{11}-\frac{1}{3} g^2 z_\sigma^{11}\right)+\left(\frac{5 g^4}{9}+\frac{2 g^3 h}{3}-\frac{g^4 N}{18}-\frac{g^2 h^2}{18}\right)~.\nonumber
\end{equation}
Computations can be performed iteratively by loop order, solving for $Z$-factors using only leading Log term equations. For instance, using the two leading Log equations above, determining $z_{\varphi}^{21}$ and $z_{\varphi}^{22}$ requires prior knowledge of one-loop $Z$-factors, including $Z_g,Z_h,Z_\varphi,Z_\sigma$.  Alternatively, incorporating UV finiteness conditions from all Log terms of a correlation function generates a sufficient set of equations to solve for the corresponding $Z$-factor independently, without relying on lower-loop $Z$-factors from other channels. For instance, the sub-Log term equation can be solved to obtain a combination of one-loop $Z$-factors. This combination directly appears in the leading Log equations. Substituting this result back into the two leading Log equations fully determines the two-loop $Z_\varphi$-factor. As a byproduct, sub-Log term equations also provide consistency checks for lower-loop $Z$-factors computed from other correlation functions.

Following this approach, we determine the two-loop $Z_\varphi$-factor coefficients as
\begin{equation}
z_{\varphi}^{21}=\frac{11 g^4 N}{432}-\frac{13 g^4}{216}-\frac{g^3 h}{9}+\frac{11 g^2 h^2}{432}~~~,~~~z_{\varphi}^{22}=-\frac{g^4 N}{36}+\frac{5 g^4}{18}+\frac{g^3 h}{3}-\frac{g^2 h^2}{36}~.
\end{equation}
The $Z_\sigma$-factor is determined via an identical procedure, with the only difference being the coupling factors. After analogous computations, the two-loop $Z_\sigma$-factor coefficients are
\begin{eqnarray}
&&z_{\sigma}^{11}=-\frac{g^2 N}{6}-\frac{h^2}{6}~,\\
&&z_{\sigma}^{21}=\frac{11 g^2 h^2 N}{432}-\frac{g^4 N}{216}-\frac{g^3 h N}{9} -\frac{13 h^4}{432}~~~,~~~z_{\sigma}^{22}=\frac{g^4 N}{9}+\frac{g^3 h N}{3} -\frac{g^2 h^2 N}{36} +\frac{5 h^4}{36}~.
\end{eqnarray}
Higher-loop $Z_\varphi$ and $Z_\sigma$-factor can be determined using the same iterative procedure. However, this becomes increasingly computationally demanding due to the exponential growth in the number of Feynman diagrams and the heightened complexity of evaluating two-point integrals at higher loops.

\subsection{Three-point correlation functions and the beta functions}
\label{subsec:L2beta}

%
\begin{figure}
\centering
\begin{tikzpicture}
  \node [] at (1,0) {
  \begin{tikzpicture}
    \draw [thick] (0.5,0.5)--(1,1.5)--(1.5,0.5)--(0.5,0.5) (0,0)--(0.5,0.5) (1,1.5)--(1,2) (1.5,0.5)--(2,0);
  \end{tikzpicture}
  };
  \node [] at (6,0) {
  \begin{tikzpicture}
    \draw [thick] (0.75,0.5)--(0.5,0.5)--(1,1.5)--(1.5,0.5)--(1.25,0.5) (0,0)--(0.5,0.5) (1,1.5)--(1,2) (1.5,0.5)--(2,0);
    \draw [thick] (1,0.5) circle [radius=0.25];
  \end{tikzpicture}
  };
  \node [] at (10,0) {
  \begin{tikzpicture}
    \draw [thick] (0.5,1)--(0.5,0.5)--(1.5,0.5)--(1.5,1)--(0.5,1)--(1,1.5)--(1.5,1) (0,0)--(0.5,0.5) (1,1.5)--(1,2) (1.5,0.5)--(2,0);
  \end{tikzpicture}
  };
  \node [] at (14,0) {
  \begin{tikzpicture}
    \draw [thick] (0.5,0.5)--(1,1.5)--(1.5,0.5)--(0.75,1) (0,0)--(0.5,0.5) (1,1.5)--(1,2) (1.5,0.5)--(2,0);
    \draw [thick] (0.5,0.5)--(0.88,0.75) (1.13,0.92)--(1.25,1);
    \draw [thick] (0.88,0.75) to [out=315, in=300] (1.13,0.92);
  \end{tikzpicture}
  };
  \node [] at (2,0.5) {$\Gamma^{\tiny \mbox{L1}}_{\Phi\Phi\Phi}$};
  \node [] at (7,0.5) {$\Gamma^{\tiny \mbox{L2-I}}_{\Phi\Phi\Phi}$};
  \node [] at (11,0.5) {$\Gamma^{\tiny \mbox{L2-II}}_{\Phi\Phi\Phi}$};
  \node [] at (15,0.5) {$\Gamma^{\tiny \mbox{L2-III}}_{\Phi\Phi\Phi}$};
  \draw [gray,thick,->] (1,-1.375)--(1,-1.625);
  \draw [gray,thick,->] (6,-1.375)--(6,-1.625);
  \draw [gray,thick,->] (10,-1.375)--(10,-1.625);
  \draw [gray,thick,->] (14,-1.375)--(14,-1.625);
  \node [] at (1,-2.75) {
  \begin{tikzpicture}
    \draw [thick] (0.5,0.5)--(1,1.5)--(1.5,0.5)--(0.5,0.5) (0,0)--(0.5,0.5) (1.5,0.5)--(2,0);
  \end{tikzpicture}
  };
  \node [] at (6,-2.75) {
  \begin{tikzpicture}
    \draw [thick] (0.75,0.5)--(0.5,0.5)--(1,1.5)--(1.5,0.5)--(1.25,0.5) (0,0)--(0.5,0.5) (1.5,0.5)--(2,0);
    \draw [thick] (1,0.5) circle [radius=0.25];
  \end{tikzpicture}
  };
  \node [] at (5.75,-5) {
  \begin{tikzpicture}
    \draw [thick] (0.75,0.5)--(0.5,0.5)--(1,1.5)--(1.5,0.5)--(1.25,0.5) (0,0)--(0.5,0.5) (1,1.5)--(1,2);
    \draw [thick] (1,0.5) circle [radius=0.25];
  \end{tikzpicture}
  };
  \node [] at (10,-2.75) {
  \begin{tikzpicture}
    \draw [thick] (0.5,1)--(0.5,0.5)--(1.5,0.5)--(1.5,1)--(0.5,1)--(1,1.5)--(1.5,1) (0,0)--(0.5,0.5) (1.5,0.5)--(2,0);
  \end{tikzpicture}
  };
  \node [] at (9.75,-5) {
  \begin{tikzpicture}
    \draw [thick] (0.5,1)--(0.5,0.5)--(1.5,0.5)--(1.5,1)--(0.5,1)--(1,1.5)--(1.5,1) (0,0)--(0.5,0.5) (1,1.5)--(1,2);
  \end{tikzpicture}
  };
  \node [] at (14,-2.75) {
  \begin{tikzpicture}
    \draw [thick] (0.5,0.5)--(1,1.5)--(1.5,0.5)--(0.75,1) (0,0)--(0.5,0.5) (1.5,0.5)--(2,0);
    \draw [thick] (0.5,0.5)--(0.88,0.75) (1.13,0.92)--(1.25,1);
    \draw [thick] (0.88,0.75) to [out=315, in=300] (1.13,0.92);
  \end{tikzpicture}
  };
  \node [] at (2,-2) {$\Gamma^{\tiny \mbox{L1}}_{\Phi\Phi\delSub{\Phi}}$};
  \node [] at (7,-2) {$\Gamma^{\tiny \mbox{L2-I}_1}_{\Phi\Phi\delSub{\Phi}}$};
  \node [] at (7,-4.5) {$\Gamma^{\tiny \mbox{L2-I}_2}_{\Phi\Phi\delSub{\Phi}}$};
  \node [] at (11,-2) {$\Gamma^{\tiny \mbox{L2-II}_1}_{\Phi\Phi\delSub{\Phi}}$};
  \node [] at (11,-4.5) {$\Gamma^{\tiny \mbox{L2-II}_2}_{\Phi\Phi\delSub{\Phi}}$};
  \node [] at (15,-2) {$\Gamma^{\tiny \mbox{L2-III}}_{\Phi\Phi\delSub{\Phi}}$};
\end{tikzpicture}
  \caption{The 1-loop and 2-loop 1PI topologies of three-point functions in $\Phi$-field representation. A gray arrow points to the cut graphs of each topology. They are generated by the non-isomorphic computation of all possible removals of one external leg. }\label{fig:P3topo}
\end{figure}
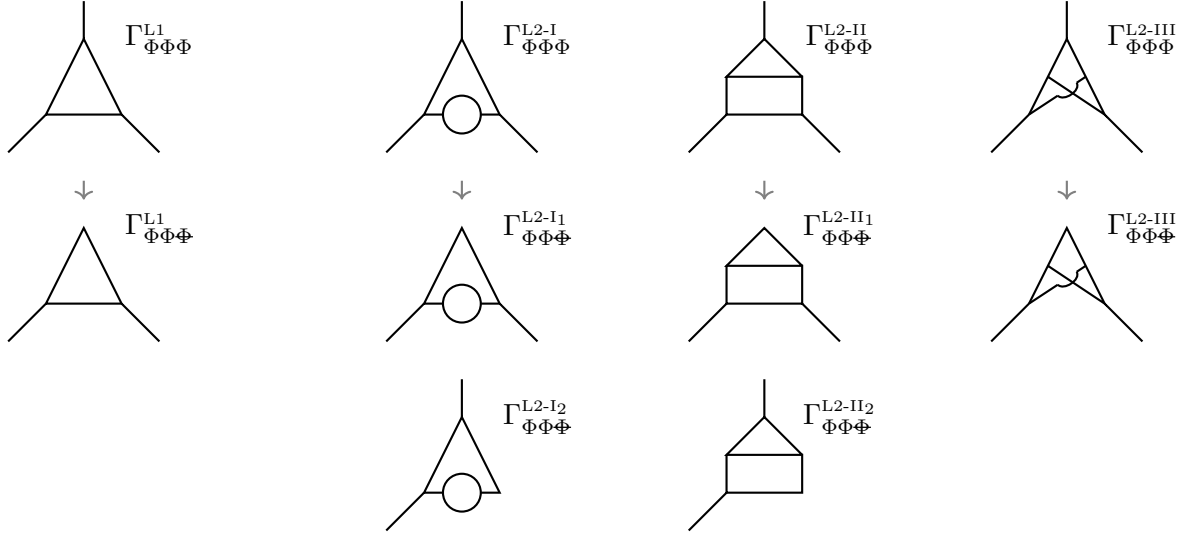
\begin{table}
  \centering
  \begin{tabular}{|c||c|c|c|}
  \hline
  topology & $\sym(\Gamma_{\Phi\Phi\delSub{\Phi}})$ & $\cf(\Gamma_{\sigma\sigma\delSub{\Phi}})$ & $\cf(\Gamma_{\varphi\bar{\varphi}\delSub{\Phi}})$ \\ \hline \hline
 $\Gamma^{\tiny \mbox{L1}}_{\Phi\Phi\delSub{\Phi}}$ & 1 & $g^3 N+h^3$ & $g^3+g^2 h$ \\ \hline
 $\Gamma^{\tiny \mbox{L2-I}_1}_{\Phi\Phi\delSub{\Phi}}$ & 1/2 & $2 g^5 N+g^2 h^3 N+h^5$ & $g^5 N+2 g^4 h+g^3 h^2$\\ \hline
 $\Gamma^{\tiny \mbox{L2-I}_2}_{\Phi\Phi\delSub{\Phi}}$ & 1 & $2 g^5 N+g^2 h^3 N+h^5$ & $2 g^5+g^4 h N+g^2 h^3$ \\ \hline
 $\Gamma^{\tiny \mbox{L2-II}_1}_{\Phi\Phi\delSub{\Phi}}$ & 1 & $g^5 N+g^4 h N+g^3 h^2 N+h^5$ & $g^5 N+g^5+g^4 h+g^2 h^3$ \\ \hline
 $\Gamma^{\tiny \mbox{L2-II}_2}_{\Phi\Phi\delSub{\Phi}}$ & 2 & $g^5 N+g^4 h N+g^3 h^2 N+h^5$ & $g^5+2 g^4 h+g^3 h^2$ \\ \hline
 $\Gamma^{\tiny \mbox{L2-III}}_{\Phi\Phi\delSub{\Phi}}$ & 1/2 & $3 g^4 h N+h^5$ & $2 g^5+2 g^3 h^2$ \\
  \hline
\end{tabular}
  \caption{The symmetry factors and coupling factors of cut graphs for 2-loop three-point topologies.}\label{table:L2P3-sym-cf}
\end{table}

We now apply the proposed algorithm to determine the $Z_h$ and $Z_g$-factors. Starting from the corresponding OPEs, we derive the UV finiteness conditions given in eqn.(\ref{eqn:P3couplingOPE-1}) and (\ref{eqn:P3couplingOPE-2}). These equations non-trivially relate the respective $Z$-factor ratios to the sum of two-point integrals, which are generated by removing one external leg from three-point topologies in all possible ways. Following the algorithmic strategy, the one-loop and two-loop topologies in the $\Phi$-field representation, along with their independent cut graphs, are illustrated in Fig.(\ref{fig:P3topo}). The $Z_g$-factor is determined from $\spaa{\varphi\bar{\varphi}\sigma}$ with one $\sigma$ field removed, while the $Z_h$-factor is determined from $\spaa{\sigma\sigma\sigma}$ with one $\sigma$ field removed. The symmetry factors and coupling factors of the cut graphs are listed in Table.(\ref{table:L2P3-sym-cf}). \footnote{We note that the symmetry factor defined herein is given by $(n-1)!$ divided the internal symmetry of the diagram and the permutation symmetry of external legs, where $n$ is the number of external legs. This differs from the conventional definition, which only accounts for internal symmetry, as used in \cite{Fei:2014xta}. For example, using data from Table.(\ref{table:L2P3-sym-cf}), the conventional symmetry factors for two-loop topologies are $S(\Gamma^{\tiny \mbox{L2-I}_1}_{\Phi\Phi\delSubsub{\Phi}})=\Gamma^{\tiny \mbox{L2-I}_2}_{\Phi\Phi\delSubsub{\Phi}}=\frac{1}{2}$, $S(\Gamma^{\tiny \mbox{L2-II}_1}_{\Phi\Phi\delSubsub{\Phi}})=\Gamma^{\tiny \mbox{L2-II}_2}_{\Phi\Phi\delSubsub{\Phi}}=1$, $S(\Gamma^{\tiny \mbox{L2-III}}_{\Phi\Phi\delSubsub{\Phi}})=\frac{1}{2}$. The ratios of our symmetry factors to the conventional ones are $\sym(\Gamma^{\tiny \mbox{L2-I}_2}_{\Phi\Phi\delSubsub{\Phi}})/S(\Gamma^{\tiny \mbox{L2-I}_2}_{\Phi\Phi\delSubsub{\Phi}})=2$ and $\sym(\Gamma^{\tiny \mbox{L2-II}_2}_{\Phi\Phi\delSubsub{\Phi}})/S(\Gamma^{\tiny \mbox{L2-II}_2}_{\Phi\Phi\delSubsub{\Phi}})=2$. These differences can be absorbed into the coupling factors if the conventional symmetry factor definition is adopted, effectively encoding the summation of external legs into structure factors.}

UV divergences for the two-point integrals of the cut graphs are computed using \verb!TwoPoint! command in HyperlogProcedures package. The one-loop results\footnote{As before, an overall factor $(1/p^2)^{\frac{\epsilon}{2}L}$ for $L$-loop integral is implicit included and not explicitly shown.} are
\begin{eqnarray}
  &&\uv(\Gamma^{\tiny \mbox{L1}}_{\Phi\Phi\delSub{\Phi}})=\frac{1}{\epsilon}+3+\left(\frac{23}{4}-\frac{\pi ^2}{16}\right)\epsilon +\mathcal{O}(\epsilon^2)~,\end{eqnarray}
while the two-loop results are
\begin{eqnarray}
  &&\uv(\Gamma^{\tiny \mbox{L2-I}_1}_{\Phi\Phi\delSub{\Phi}})=-\frac{1}{6 \epsilon^2}-\frac{67}{72 \epsilon}-\left(\frac{2671}{864}-\frac{5 \pi ^2}{288}\right)- \left(\frac{80911}{10368}-\frac{335 \pi ^2}{3456}-\frac{5 \zeta_3}{144}\right)\epsilon+\mathcal{O}(\epsilon^2)~, \\
  &&\uv(\Gamma^{\tiny \mbox{L2-I}_2}_{\Phi\Phi\delSub{\Phi}})=-\frac{1}{6 \epsilon^2}-\frac{67}{72 \epsilon}-\left(\frac{2743}{864}-\frac{5 \pi ^2}{288}\right)- \left(\frac{86167}{10368}-\frac{335 \pi ^2}{3456}-\frac{5 \zeta_3}{144}\right)\epsilon+\mathcal{O}(\epsilon^2)~,
\end{eqnarray}
\begin{eqnarray}
  &&\uv(\Gamma^{\tiny \mbox{L2-II}_1}_{\Phi\Phi\delSub{\Phi}})=\frac{1}{2 \epsilon^2}+\frac{21}{8 \epsilon}+\left(\frac{275}{32}-\frac{5 \pi ^2}{96}\right)+ \left(\frac{2785}{128}-\frac{35 \pi ^2}{128}-\frac{5 \zeta_3}{48} \right)\epsilon+\mathcal{O}(\epsilon^2)~, \\
  &&\uv(\Gamma^{\tiny \mbox{L2-II}_2}_{\Phi\Phi\delSub{\Phi}})=\frac{1}{2 \epsilon^2}+\frac{21}{8 \epsilon}+\left(\frac{259}{32}-\frac{5 \pi ^2}{96}\right)+ \left(\frac{7139}{384}-\frac{35 \pi ^2}{128}+\frac{19 \zeta_3}{48}\right)\epsilon+\mathcal{O}(\epsilon^2)~, \\
  &&\uv(\Gamma^{\tiny \mbox{L2-III}}_{\Phi\Phi\delSub{\Phi}})=\frac{1}{2 \epsilon}+\left(\frac{83}{24}-\zeta_3\right) + \left(\frac{3983}{288}-\frac{5 \pi ^2}{96}-\frac{16 \zeta_3}{3} -\frac{\pi ^4}{120}\right)\epsilon+\mathcal{O}(\epsilon^2)~.
\end{eqnarray}
Special care is required for the tree-level contribution, which is evaluated from the two-point integral of a three-point vertex with one leg removed. In coordinate space, this result is not unity but,
\begin{equation}
\uv(\Gamma^{\tiny \mbox{L0}}_{\Phi\Phi\delSub{\Phi}})=1+\epsilon+\left(\frac{3}{4}-\frac{\pi ^2}{48}\right) \epsilon^2+\left(\frac{1}{2}-\frac{\pi ^2}{48}-\frac{\zeta_3}{24}\right)\epsilon^3 +\left(\frac{5}{16}-\frac{\pi ^2}{64}-\frac{\zeta_3}{24}+\frac{\pi^4}{23040}\right)\epsilon^4+\mathcal{O}(\epsilon^5)~.
\end{equation}
For $L$-loop renormalization computations, the appropriate orders of $\epsilon$ must be included. The bare correlation functions for two-loop computations are then obtained as\footnote{For three-point function, $n_p=3$, so the overall $i$-factors for tree and loop contributions have the same sign.}
\begin{equation}\label{eqn:compute-3point}
G(g_0,h_0;\Phi,\Phi,\del{\Phi})=g_{\Phi}(1+\epsilon)+\sum_{L={\tiny\mbox{L1}},{\tiny\mbox{L2-I}}_1,{\tiny\mbox{L2-I}}_2,\atop {\tiny\mbox{L2-II}}_1,{\tiny\mbox{L2-II}}_2,{\tiny\mbox{L2-III}}}\sym(\Gamma^{L}_{\Phi\Phi\delSub{\Phi}})
\cf(\Gamma^{L}_{\Phi\Phi\delSub{\Phi}})\uv(\Gamma^{L}_{\Phi\Phi\delSub{\Phi}})~,
\end{equation}
where $\Phi$ corresponds to $\sigma$ or $\varphi$ in the coupling factor $\cf$, and $g_\Phi=g$ for $\spaa{\varphi\bar{\varphi}\sigma}$ while $g_\Phi=h$ for $\spaa{\sigma\sigma\sigma}$.

The UV finiteness conditions of eqn.(\ref{eqn:P3couplingOPE-1}) and (\ref{eqn:P3couplingOPE-2}) are explicitly, 
\begin{eqnarray}
\frac{Z_{g}g }{g_0}\mu^{\frac{\epsilon}{2}} G(g_0,h_0;\varphi,\bar{\varphi},\del{\sigma})=\mbox{UV finite}~~,~~\frac{Z_{h}h}{h_0} \mu^{\frac{\epsilon}{2}} G(g_0,h_0;\sigma,\sigma,\del{\sigma})=\mbox{UV~finite}~.
\end{eqnarray}
After rewriting all bare couplings $g_0,h_0$ to $g,h$ using relations eqn.(\ref{eqn:couplingRelation}), and expanding the full expression in the double limit $g,h\to 0$, followed by $\epsilon\to 0$, we obtain equations for $Z$-factors by enforcing the cancelation of $\epsilon$ poles in each $\ln\frac{\mu^2}{p^2}$ term\footnote{The bare coupling in the denominator introduces a $\mu^{\frac{\epsilon}{2}}$, which cancels the $\mu^{\frac{\epsilon}{2}}$ factor in the numerator. In the bare correlation function, $L$-loop contribution include a $(1/p^2)^{\frac{\epsilon}{2}L}$ factor from integration and a $\mu^{\frac{\epsilon}{2}(2L+1)}$ factor from couplings. Factoring our an overall $\mu^{\frac{\epsilon}{2}}$ (which does not affect the $\epsilon$ expansion), each $L$-loop term carries a $(\mu^2/p^2)^{\frac{\epsilon}{2}L}$ factor that contributes to the $\ln\frac{\mu^2}{p^2}$ expansion as $\epsilon\to 0$.}.

For the $Z_g$-factor, the one-loop equation yields,
\begin{equation}
g^3+g^2h+g z_{g}^{11}=0~~~\to~~~z_g^{11}=-g^2-gh~.
\end{equation}
At two-loop order, the $\ln^0\frac{\mu^2}{p^2}$ term provides two equations,
\begin{eqnarray}
&&  0=z_{g}^{22}+ z_{g}^{21}+(9 g^2 +6 g h +1 )z_g^{11}+3g\Big( h z_{h}^{11}-(2 g + h) z_{\phi}^{11}-( g+2 h) z_{\sigma}^{11}\Big)\nonumber\\
&&~~~~~~~~~~~~~~+\left(\frac{311 g^4 N}{144}+\frac{469 g^4}{72}-\frac{67g^3 h N}{72} +\frac{439 g^3 h}{36}+\frac{761 g^2 h^2}{144}+g^2+\frac{61 g h^3}{36}+g h\right)~,\\
&& 0=z_{g}^{22}+(2 g h +3 g^2) z_g^{11}+g\Big(h z_{h}^{11}-(2 g + h) z_{\phi}^{11}-(g +2  h )z_{\sigma}^{11}\Big)\nonumber\\
&&~~~~~~~~~~~~~~~~~~~~~~~~~~~~~~~~~~~~~~~~~~~+\left(\frac{5 g^4 N}{12}+\frac{7 g^4}{6}-\frac{g^3 h N}{6} +\frac{7 g^3 h}{3}+\frac{11 g^2 h^2}{12}+\frac{g h^3}{3}\right)~,
\end{eqnarray}
while the sub-Log term provides a third equation,
\begin{eqnarray}
&&0=(3 g +2 h )z_g^{11}+ \Big(h z_{h}^{11}-(2 g + h )z_{\phi}^{11}-(g +2  h )z_{\sigma}^{11}\Big)\nonumber\\
&&~~~~~~~~~~~~~~~~~~~~~~~~~~~~~~~~~~~~+\left(\frac{5 g^3 N}{6}+\frac{7 g^3}{3}-\frac{g^2 h N}{3} +\frac{14 g^2 h}{3}+\frac{11 g h^2}{6}+\frac{2  h^3}{3}\right)~.
\end{eqnarray}
As before, the sub-Log term equation allows solving for the combination of one-loop $Z$-factors. Substituting this result into the leading Log equations fully determines the two-loop $Z_g$-factor coefficients,
\begin{equation}
z_{g}^{21}=\frac{7 g^3 h N}{72}-\frac{11 g^4 N}{144}-\frac{49 g^4}{72}-\frac{19 g^3 h}{36}-\frac{101 g^2 h^2}{144}-\frac{g h^3}{36}~~,~~z_{g}^{22}=\frac{5 g^4 N}{12}+\frac{7 g^4}{6}-\frac{g^3 h N}{6} +\frac{7 g^3 h}{3}+\frac{11 g^2 h^2}{12}+\frac{g h^3}{3}~.\nonumber
\end{equation}
Following analogous steps for the $Z_h$-factors, we solve its UV finiteness conditions to obtain the two-loop results. The one-loop coefficient is $z_h^{11}=-(g^3 N+h^3)/h$, and the two-loop coefficients are
\begin{equation}
z_h^{21}=\frac{7g^2 h^2 N}{48}-\frac{g^5 N}{12 h}-\frac{9 g^4 N}{8}-\frac{3g^3 h N}{8}  -\frac{23 h^4}{48}~~,~~z_h^{22}=\frac{g^5 N}{h}+\frac{3 g^4 N}{2}+\frac{3 g^3 h N}{2}-\frac{g^2 h^2 N}{4} +\frac{5 h^4}{4}~.\nonumber
\end{equation}
Higher-loop $Z_h$ and $Z_g$-factors can be  determined similarly by including additional topologies and higher-order
$\epsilon$ terms in the two-point integrals.

With $Z_h,Z_g,Z_{\phi}$ and $Z_{\sigma}$ determined, the beta functions are computed by solving  eqn.(\ref{eqn:compute-beta}). Expanding the solutions in the limit $g,h\to 0$, the two-loop results are
\begin{eqnarray}
\beta(g)&=&-\frac{\epsilon }{2}g+\left(
\frac{g^3 N}{12}-\frac{2 g^3}{3}-g^2 h+\frac{g h^2}{12}\right)\nonumber\\
&&+\left(\frac{11g^4 h N}{36}+\frac{13 g h^4}{432}-\frac{43 g^5 N}{216}-\frac{67 g^5}{54} -\frac{5 g^4 h}{6}-\frac{11g^3 h^2 N}{432} -\frac{157 g^3 h^2}{108}-\frac{g^2 h^3}{18}\right)~,~~\label{eqn:beta-g}\\
\beta(h)&=&-\frac{\epsilon}{2} h+\left(\frac{g^2 h N}{4} -g^3N-\frac{3 h^3}{4}\right)+\left(\frac{31g^2 h^3 N}{144} -\frac{g^5 N}{6}-\frac{161g^4 h N}{72} -\frac{5g^3 h^2 N}{12} -\frac{125 h^5}{144}\right)~.~~\label{eqn:beta-h}
\end{eqnarray}
The two-loop anomalous dimension of the $\varphi$ and $\sigma$ fields are then obtained via eqn.(\ref{eqn:compute-field-anomalous}),
\begin{equation}
\gamma_{\varphi}=\frac{g^2}{6}+\frac{13 g^4}{216}+\frac{g^3 h}{9}-\frac{11 g^4 N}{432}-\frac{11 g^2 h^2}{432}~~~,~~~\gamma_{\sigma}=\frac{h^2}{12}+\frac{g^2 N}{12}+\frac{g^4 N}{216}+\frac{g^3 h N}{9} -\frac{11g^2 h^2 N}{432} +\frac{13 h^4}{432}~.\label{eqn:anomalous-field}
\end{equation}
%

\subsection{Anomalous dimensions of the $\varphi^2$ and $\varphi^3$ operators}
\label{subsec:L2anomalous}

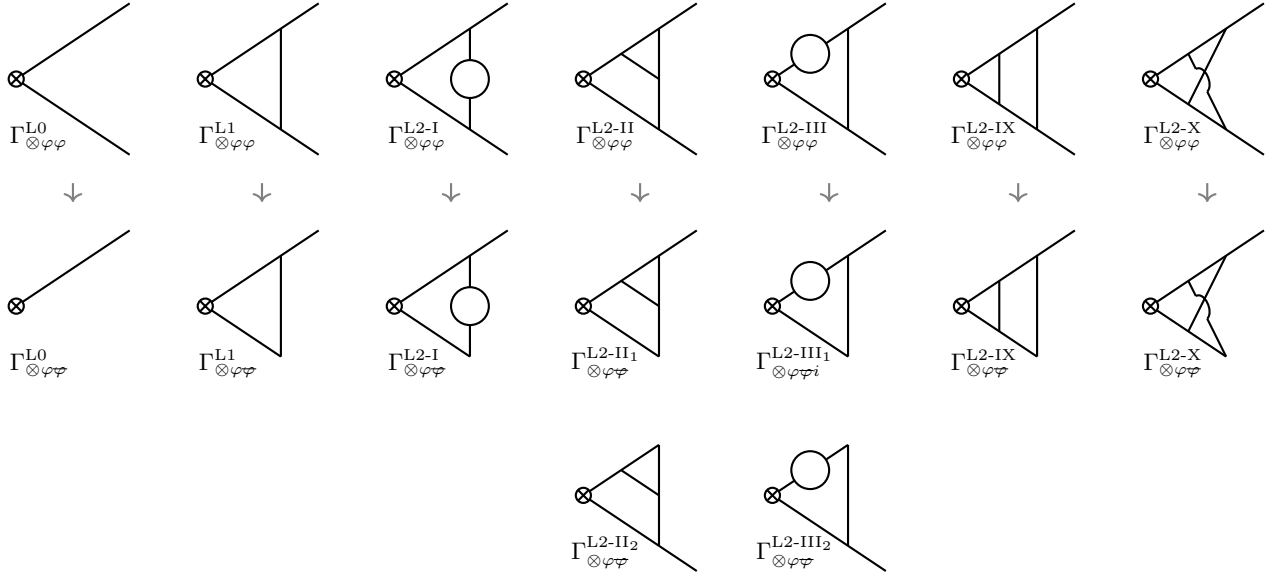
\begin{figure}
\centering
\begin{tikzpicture}
  \node [] at (0.7,-1) {
    \begin{tikzpicture}
      \draw [thick] (0,0)--(1.5,1) (0,0)--(1.5,-1);
      \draw [thick, fill=white] (0,0) circle [radius=0.1];
      \draw [thick] (0.07,0.07)--(-0.07,-0.07)  (0.07,-0.07)--(-0.07,0.07);
    \end{tikzpicture}
  };
  \node [] at (3.2,-1) {
    \begin{tikzpicture}
      \draw [thick] (0,0)--(1.5,1) (0,0)--(1.5,-1);
      \draw [thick, fill=white] (0,0) circle [radius=0.1];
      \draw [thick] (0.07,0.07)--(-0.07,-0.07)  (0.07,-0.07)--(-0.07,0.07);
      \draw [thick] (1,0.667)--(1,-0.667);
    \end{tikzpicture}
  };
  \node [] at (5.7,-1) {
    \begin{tikzpicture}
      \draw [thick] (0,0)--(1.5,1) (0,0)--(1.5,-1);
      \draw [thick, fill=white] (0,0) circle [radius=0.1];
      \draw [thick] (0.07,0.07)--(-0.07,-0.07)  (0.07,-0.07)--(-0.07,0.07);
      \draw [thick] (1,0.667)--(1,-0.667);
      \draw [thick, fill=white] (1,0) circle [radius=0.25];
    \end{tikzpicture}
  };
  \node [] at (8.2,-1) {
    \begin{tikzpicture}
      \draw [thick] (0,0)--(1.5,1) (0,0)--(1.5,-1);
      \draw [thick, fill=white] (0,0) circle [radius=0.1];
      \draw [thick] (0.07,0.07)--(-0.07,-0.07)  (0.07,-0.07)--(-0.07,0.07);
      \draw [thick] (1,0.667)--(1,-0.667);
      \draw [thick] (0.5,0.333)--(1,0);
    \end{tikzpicture}
  };
  \node [] at (10.7,-1) {
    \begin{tikzpicture}
      \draw [thick] (0,0)--(1.5,1) (0,0)--(1.5,-1);
      \draw [thick, fill=white] (0,0) circle [radius=0.1];
      \draw [thick] (0.07,0.07)--(-0.07,-0.07)  (0.07,-0.07)--(-0.07,0.07);
      \draw [thick] (1,0.667)--(1,-0.667);
      \draw [thick, fill=white] (0.5,0.333) circle [radius=0.25];
    \end{tikzpicture}
  };
  \node [] at (13.2,-1) {
    \begin{tikzpicture}
      \draw [thick] (0,0)--(1.5,1) (0,0)--(1.5,-1);
      \draw [thick, fill=white] (0,0) circle [radius=0.1];
      \draw [thick] (0.07,0.07)--(-0.07,-0.07)  (0.07,-0.07)--(-0.07,0.07);
      \draw [thick] (1,0.667)--(1,-0.667);
      \draw [thick] (0.5,0.333)--(0.5,-0.333);
    \end{tikzpicture}
  };
  \node [] at (15.7,-1) {
    \begin{tikzpicture}
      \draw [thick] (0,0)--(1.5,1) (0,0)--(1.5,-1);
      \draw [thick, fill=white] (0,0) circle [radius=0.1];
      \draw [thick] (0.07,0.07)--(-0.07,-0.07)  (0.07,-0.07)--(-0.07,0.07);
      \draw [thick] (1,0.667)--(0.5,-0.333);
      \draw [thick] (0.5,0.333)--(0.6,0.133) (0.75,-0.166)--(1,-0.667);
      \draw [thick] (0.6,0.133) to [out=30,in=45] (0.75,-0.166);
    \end{tikzpicture}
  };
  \node [] at (0.3,-1.75) {{\footnotesize $\Gamma^{\tiny \mbox{L0}}_{\otimes\varphi\varphi}$}};
  \node [] at (2.8,-1.75) {{\footnotesize $\Gamma^{\tiny \mbox{L1}}_{\otimes\varphi\varphi}$}};
  \node [] at (5.3,-1.75) {{\footnotesize $\Gamma^{\tiny \mbox{L2-I}}_{\otimes\varphi\varphi}$}};
  \node [] at (7.8,-1.75) {{\footnotesize $\Gamma^{\tiny \mbox{L2-II}}_{\otimes\varphi\varphi}$}};
  \node [] at (10.3,-1.75) {{\footnotesize $\Gamma^{\tiny \mbox{L2-III}}_{\otimes\varphi\varphi}$}};
  \node [] at (12.8,-1.75) {{\footnotesize $\Gamma^{\tiny \mbox{L2-IX}}_{\otimes\varphi\varphi}$}};
  \node [] at (15.3,-1.75) {{\footnotesize $\Gamma^{\tiny \mbox{L2-X}}_{\otimes\varphi\varphi}$}};
  \node [] at (0.7,-4) {
    \begin{tikzpicture}
      \draw [thick] (0,0)--(1.5,1);
      \draw [thick,white] (0,0)--(1.5,-1);
      \draw [thick, fill=white] (0,0) circle [radius=0.1];
      \draw [thick] (0.07,0.07)--(-0.07,-0.07)  (0.07,-0.07)--(-0.07,0.07);
    \end{tikzpicture}
  };
  \node [] at (3.2,-4) {
    \begin{tikzpicture}
      \draw [thick,white] (1,-0.667)--(1.5,-1);
      \draw [thick] (0,0)--(1.5,1) (0,0)--(1,-0.667);
      \draw [thick, fill=white] (0,0) circle [radius=0.1];
      \draw [thick] (0.07,0.07)--(-0.07,-0.07)  (0.07,-0.07)--(-0.07,0.07);
      \draw [thick] (1,0.667)--(1,-0.667);
    \end{tikzpicture}
  };
  \node [] at (5.7,-4) {
    \begin{tikzpicture}
      \draw [thick,white] (1,-0.667)--(1.5,-1);
      \draw [thick] (0,0)--(1.5,1) (0,0)--(1,-0.667);
      \draw [thick, fill=white] (0,0) circle [radius=0.1];
      \draw [thick] (0.07,0.07)--(-0.07,-0.07)  (0.07,-0.07)--(-0.07,0.07);
      \draw [thick] (1,0.667)--(1,-0.667);
      \draw [thick, fill=white] (1,0) circle [radius=0.25];
    \end{tikzpicture}
  };
  \node [] at (8.2,-4) {
    \begin{tikzpicture}
      \draw [thick,white] (1,-0.667)--(1.5,-1);
      \draw [thick] (0,0)--(1.5,1) (0,0)--(1,-0.667);
      \draw [thick, fill=white] (0,0) circle [radius=0.1];
      \draw [thick] (0.07,0.07)--(-0.07,-0.07)  (0.07,-0.07)--(-0.07,0.07);
      \draw [thick] (1,0.667)--(1,-0.667);
      \draw [thick] (0.5,0.333)--(1,0);
    \end{tikzpicture}
  };
  \node [] at (10.7,-4) {
    \begin{tikzpicture}
      \draw [thick,white] (1,-0.667)--(1.5,-1);
      \draw [thick] (0,0)--(1.5,1) (0,0)--(1,-0.667);
      \draw [thick, fill=white] (0,0) circle [radius=0.1];
      \draw [thick] (0.07,0.07)--(-0.07,-0.07)  (0.07,-0.07)--(-0.07,0.07);
      \draw [thick] (1,0.667)--(1,-0.667);
      \draw [thick, fill=white] (0.5,0.333) circle [radius=0.25];
    \end{tikzpicture}
  };
  \node [] at (13.2,-4) {
    \begin{tikzpicture}
      \draw [thick,white] (1,-0.667)--(1.5,-1);
      \draw [thick] (0,0)--(1.5,1) (0,0)--(1,-0.667);
      \draw [thick, fill=white] (0,0) circle [radius=0.1];
      \draw [thick] (0.07,0.07)--(-0.07,-0.07)  (0.07,-0.07)--(-0.07,0.07);
      \draw [thick] (1,0.667)--(1,-0.667);
      \draw [thick] (0.5,0.333)--(0.5,-0.333);
    \end{tikzpicture}
  };
  \node [] at (15.7,-4) {
    \begin{tikzpicture}
      \draw [thick,white] (1,-0.667)--(1.5,-1);
      \draw [thick] (0,0)--(1.5,1) (0,0)--(1,-0.667);
      \draw [thick, fill=white] (0,0) circle [radius=0.1];
      \draw [thick] (0.07,0.07)--(-0.07,-0.07)  (0.07,-0.07)--(-0.07,0.07);
      \draw [thick] (1,0.667)--(0.5,-0.333);
      \draw [thick] (0.5,0.333)--(0.6,0.133) (0.75,-0.166)--(1,-0.667);
      \draw [thick] (0.6,0.133) to [out=30,in=45] (0.75,-0.166);
    \end{tikzpicture}
  };
  \node [] at (0.3,-4.75) {{\footnotesize $\Gamma^{\tiny \mbox{L0}}_{\otimes\varphi\delSubsub{\varphi}}$}};
  \node [] at (2.8,-4.75) {{\footnotesize $\Gamma^{\tiny \mbox{L1}}_{\otimes\varphi\delSubsub{\varphi}}$}};
  \node [] at (5.3,-4.75) {{\footnotesize $\Gamma^{\tiny \mbox{L2-I}}_{\otimes\varphi\delSubsub{\varphi}}$}};
  \node [] at (7.8,-4.75) {{\footnotesize $\Gamma^{\tiny \mbox{L2-II}_1}_{\otimes\varphi\delSubsub{\varphi}}$}};
  \node [] at (10.3,-4.75) {{\footnotesize $\Gamma^{\tiny \mbox{L2-III}_1}_{\otimes\varphi\delSubsub{\varphi}i}$}};
  \node [] at (12.8,-4.75) {{\footnotesize $\Gamma^{\tiny \mbox{L2-IX}}_{\otimes\varphi\delSubsub{\varphi}}$}};
  \node [] at (15.3,-4.75) {{\footnotesize $\Gamma^{\tiny \mbox{L2-X}}_{\otimes\varphi\delSubsub{\varphi}}$}};
  \node [] at (8.2,-6.5) {
    \begin{tikzpicture}
      \draw [thick,white] (1,0.667)--(1.5,1);
      \draw [thick] (0,0)--(1.5,-1) (0,0)--(1,0.667);
      \draw [thick, fill=white] (0,0) circle [radius=0.1];
      \draw [thick] (0.07,0.07)--(-0.07,-0.07)  (0.07,-0.07)--(-0.07,0.07);
      \draw [thick] (1,0.667)--(1,-0.667);
      \draw [thick] (0.5,0.333)--(1,0);
    \end{tikzpicture}
  };
  \node [] at (10.7,-6.5) {
    \begin{tikzpicture}
      \draw [thick,white] (1,0.667)--(1.5,1);
      \draw [thick] (0,0)--(1.5,-1) (0,0)--(1,0.667);
      \draw [thick, fill=white] (0,0) circle [radius=0.1];
      \draw [thick] (0.07,0.07)--(-0.07,-0.07)  (0.07,-0.07)--(-0.07,0.07);
      \draw [thick] (1,0.667)--(1,-0.667);
      \draw [thick, fill=white] (0.5,0.333) circle [radius=0.25];
    \end{tikzpicture}
  };
  \node [] at (7.8,-7.25) {{\footnotesize $\Gamma^{\tiny \mbox{L2-II}_2}_{\otimes\varphi\delSubsub{\varphi}}$}};
  \node [] at (10.3,-7.25) {{\footnotesize $\Gamma^{\tiny \mbox{L2-III}_2}_{\otimes\varphi\delSubsub{\varphi}}$}};
  \draw [gray,thick,->] (0.75,-2.375)--(0.75,-2.625);
  \draw [gray,thick,->] (3.25,-2.375)--(3.25,-2.625);
  \draw [gray,thick,->] (5.75,-2.375)--(5.75,-2.625);
  \draw [gray,thick,->] (8.25,-2.375)--(8.25,-2.625);
  \draw [gray,thick,->] (10.75,-2.375)--(10.75,-2.625);
  \draw [gray,thick,->] (13.25,-2.375)--(13.25,-2.625);
  \draw [gray,thick,->] (15.75,-2.375)--(15.75,-2.625);
\end{tikzpicture}
\caption{The tree-level, 1-loop and 2-loop 1PI topologies of $\varphi^2$-$2\varphi$ correlation function, where the $_\otimes$ in subscript denotes a $\phi^Q$ operator. A gray arrow points to the cut graphs of each topology. They are generated by the non-isomorphic computation of all possible removals of one external leg. The UV divergences are computed from these cut graphs.}\label{fig:phi2topo}
\end{figure}

The $\widetilde{Z}_{\varphi^Q}$-factors for the $\varphi^Q$ operator are non-trivially related to two-point integrals via UV finiteness conditions in eqn.(\ref{eqn:phiQ-UV-finiteness}), where the two-point integrals are generated by removing $(Q-1)$ external legs from $\varphi^Q\to Q\varphi$ correlation function graphs. These integrals are evaluated using eqn.(\ref{eqn:phiQ-G}). For $Q=2$, all contributing topologies are irreducible. The tree-level, one-loop and two-loop topologies of the $\varphi^2$-$2\varphi$ correlation function, along with their corresponding cut graphs, are illustrated in Fig.(\ref{fig:phi2topo}).
Due to graph symmetry, a single topology may generate multiple non-isomorphic cut graphs (as observed in three-point function computations), and cut graphs derived from the same irreducible topology share identical symmetry and coupling factors. For $Q=2$, eqn.(\ref{eqn:phiQ-G}) yields the explicit expression for $\varphi^2$,
\begin{equation}
G_{Q=2}(g_0,h_0;x_0,\del{x}_{1},x_Q)
=1+\sum_{j=1}^{L}
G_{\scriptsize\mbox{irre}}^{j{\tiny\mbox{-loop}}}(g_0,h_0;x_0,\del{x}_{1},x_{2})~.\label{eqn:phi2-G}
\end{equation}
There is only one 1-loop irreducible topology $\Gamma^{\tiny \mbox{L1}}_{\otimes\varphi{\varphi}}$, which generates the cut graph $\Gamma^{\tiny \mbox{L1}}_{\otimes\varphi\delSub{\varphi}}$ with the following inputs,
\begin{equation}
\sym(\Gamma^{\tiny \mbox{L1}}_{\otimes\varphi\delSub{\varphi}})=1~~~,~~~\cf(\Gamma^{\tiny \mbox{L1}}_{\otimes\varphi\delSub{\varphi}})=g^2_0~~~,~~~\uv(\Gamma^{\tiny \mbox{L1}}_{\otimes\varphi\delSub{\varphi}})=\frac{1}{\epsilon}+2+\epsilon\left(\frac{5}{2}
-\frac{\pi^2}{24}\right)+\mathcal{O}(\epsilon^2)~.\label{eqn:1-loop-Q2-irre-input}
\end{equation}
The one-loop irreducible UV contribution for $\varphi^2\to 2\varphi$ is
\begin{equation}
G_{\scriptsize\mbox{irre}}^{{\tiny\mbox{1-loop}}}(g_0,h_0;x_0,\del{x}_{1},x_{2})=\sym(\Gamma^{\tiny \mbox{L1}}_{\otimes\varphi\delSub{\varphi}})\cf(\Gamma^{\tiny \mbox{L1}}_{\otimes\varphi\delSub{\varphi}})\uv(\Gamma^{\tiny \mbox{L1}}_{\otimes\varphi\delSub{\varphi}})~.\label{eqn:1-loop-Q2-irre}
\end{equation}
Note that the $L=1$ loop order factor $(1/p^2)^{\frac{\epsilon}{2}}$ is implicitly included in the UV divergence result, and $L$-loop coupling factors introduce a $\mu^{\epsilon L}$ factor. Converting bare couplings $g_0,h_0\to g,h$ introduces an overall $\left(\frac{\mu^2}{p^2}\right)^{\frac{\epsilon}{2}}$ factor, which contributes to the $\ln\frac{\mu^2}{p^2}$ expansion. Generally, $G_{\scriptsize\mbox{irre}}^{j{\tiny\mbox{-loop}}}$ carries an overall $\left(\frac{\mu^2}{p^2}\right)^{j\frac{\epsilon}{2}}$ factor, and we always treat the $(1/p^2)^{\frac{\epsilon}{2}L}$ factor implicit in all UV divergence expressions.

The one-loop $\widetilde{Z}_{\varphi^2}$ is determined by enforcing UV finiteness of ${\widetilde{Z}_{\varphi^2}}G_{Q=2}(g_0,h_0;x_0,\del{x}_{1},x_Q)$, where we have used the fact that the 1PI contribution of  $\widetilde{Z}_{\varphi^1}=1$. Rewriting bare couplings via eqn.(\ref{eqn:couplingRelation}) and expanding in the limit $g,h\to 0$ followed by $\epsilon\to 0$, the  $1/\epsilon$ pole cancelation in the leading Log term gives,
\begin{equation}
0=z_{\varphi^2}^{11}+g^2~~~\to~~~z_{\varphi^2}^{11}=-g^2~.\label{eqn:coe-Phi2-1}
\end{equation}
No additional conditions arise from sub-Log terms. For the two-loop $\widetilde{Z}_{\varphi^2}$-factor, $G_{\scriptsize\mbox{irre}}^{{\tiny\mbox{2-loop}}}(g_0,h_0;x_0,\del{x}_{1},x_{2})$ is included. There are 5 independent irreducible topologies generating 7 cut graphs, with symmetry factors 
\begin{equation}
\sym(\Gamma^{\tiny \mbox{L2-I}}_{\otimes\varphi\delSub{\varphi}})=\frac{1}{2}~~,~~\sym(\Gamma^{\tiny \mbox{L2-II}}_{\otimes\varphi\delSub{\varphi}})=1~~,~~\sym(\Gamma^{\tiny \mbox{L2-III}}_{\otimes\varphi\delSub{\varphi}})=\frac{1}{2}~~,~~\sym(\Gamma^{\tiny \mbox{L2-IX}}_{\otimes\varphi\delSub{\varphi}})=1~~,~~\sym(\Gamma^{\tiny \mbox{L2-X}}_{\otimes\varphi\delSub{\varphi}})=\frac{1}{2}~,
\end{equation}
and coupling factors
\begin{equation}
\cf(\Gamma^{\tiny \mbox{L2-I}}_{\otimes\varphi\delSub{\varphi}})=g_0^4 N+g_0^2 h_0^2~~,~~\cf(\Gamma^{\tiny \mbox{L2-II}}_{\otimes\varphi\delSub{\varphi}})=g_0^4+g_0^3 h_0~~,~~\cf(\Gamma^{\tiny \mbox{L2-III}}_{\otimes\varphi\delSub{\varphi}})=\cf(\Gamma^{\tiny \mbox{L2-X}}_{\otimes\varphi\delSub{\varphi}})=2 g_0^4~~,~~\cf(\Gamma^{\tiny \mbox{L2-IX}}_{\otimes\varphi\delSub{\varphi}})=g_0^4~.
\end{equation}
UV divergences are again computed by \verb!TwoPoint! command in HyperlogProcedures, which are
\begin{eqnarray}
  &&\uv(\Gamma^{\tiny \mbox{L2-I}}_{\otimes\varphi\delSub{\varphi}})=-\frac{1}{6 \epsilon^2}-\frac{55}{72 \epsilon}+\left(\frac{\pi ^2}{72}-\frac{1759}{864}\right)+\epsilon \left(\frac{\zeta_3}{36}+\frac{55 \pi ^2}{864}-\frac{42487}{10368}\right)+\mathcal{O}(\epsilon^2)~, \\
  &&\uv(\Gamma^{\tiny \mbox{L2-II}_1}_{\otimes\varphi\delSub{\varphi}})=\frac{1}{2 \epsilon^2}+\frac{17}{8 \epsilon}+\left(\frac{155}{32}-\frac{\pi ^2}{24}\right)+\epsilon \left(\frac{5 \zeta_3}{12}-\frac{17 \pi ^2}{96}+\frac{2915}{384}\right)+\mathcal{O}(\epsilon^2)~, \\
  &&\uv(\Gamma^{\tiny \mbox{L2-II}_2}_{\otimes\varphi\delSub{\varphi}})=\frac{1}{2 \epsilon^2}+\frac{17}{8 \epsilon}+\left(\frac{171}{32}-\frac{\pi ^2}{24}\right)+\epsilon \left(-\frac{\zeta_3}{12}-\frac{17 \pi ^2}{96}+\frac{1313}{128}\right)+\mathcal{O}(\epsilon^2)~,
\end{eqnarray}
and
\begin{eqnarray}
  &&\uv(\Gamma^{\tiny \mbox{L2-III}_1}_{\otimes\varphi\delSub{\varphi}})=-\frac{1}{6 \epsilon^2}-\frac{55}{72 \epsilon}+\left(\frac{\pi ^2}{72}-\frac{1687}{864}\right)+\epsilon \left(\frac{\zeta_3}{36}+\frac{55 \pi ^2}{864}-\frac{38095}{10368}\right)+\mathcal{O}(\epsilon^2)~, \\
  &&\uv(\Gamma^{\tiny \mbox{L2-III}_2}_{\otimes\varphi\delSub{\varphi}})=-\frac{1}{6 \epsilon^2}-\frac{55}{72 \epsilon}+\left(\frac{\pi ^2}{72}-\frac{1759}{864}\right)+\epsilon \left(\frac{\zeta_3}{36}+\frac{55 \pi ^2}{864}-\frac{42487}{10368}\right)+\mathcal{O}(\epsilon^2)~, \\
  &&\uv(\Gamma^{\tiny \mbox{L2-IX}}_{\otimes\varphi\delSub{\varphi}})=\frac{1}{2 \epsilon^2}+\frac{17}{8 \epsilon}+\left(\frac{155}{32}-\frac{\pi ^2}{24}\right)+\epsilon \left(\frac{5 \zeta_3}{12}-\frac{17 \pi ^2}{96}+\frac{2915}{384}\right)+\mathcal{O}(\epsilon^2)~,\\
  &&\uv(\Gamma^{\tiny \mbox{L2-X}}_{\otimes\varphi\delSub{\varphi}})=\frac{1}{2 \epsilon}+\left(\frac{71}{24}-\zeta_3\right)+
\epsilon \left(-\frac{\pi ^4}{120}-\frac{13 \zeta_3}{3} -\frac{\pi ^2}{24}+\frac{2807}{288}\right) +\mathcal{O}(\epsilon^2)~.
\end{eqnarray}
The two-loop irreducible UV contribution for $\varphi^2\to 2\varphi$ is
\begin{equation}
G_{\scriptsize\mbox{irre}}^{{\tiny\mbox{2-loop}}}(g_0,h_0;x_0,\del{x}_{1},x_{2})=
\sum_{L={\tiny\mbox{L2-I}},{\tiny\mbox{L2-II}}_1,{\tiny\mbox{L2-II}}_2,\atop {\tiny\mbox{L2-III}}_1,{\tiny\mbox{L2-III}}_2,{\tiny\mbox{L2-IX}},{\tiny\mbox{L2-X}}}\sym(\Gamma^{\tiny \mbox{L}}_{\otimes\varphi\delSub{\varphi}})\cf(\Gamma^{\tiny \mbox{L}}_{\otimes\varphi\delSub{\varphi}})\uv(\Gamma^{\tiny \mbox{L}}_{\otimes\varphi\delSub{\varphi}})~.\label{eqn:2-loop-Q2-irre}
\end{equation}
Expanding ${\widetilde{Z}_{\varphi^2}}G_{Q=2}(g_0,h_0;x_0,\del{x}_{1},x_Q)$, the UV finiteness conditions, from canceling $1/\epsilon$ and $1/\epsilon^2$ poles in the leading Log term, are
\begin{eqnarray}
&&0=z_{\varphi^2}^{21}+\left(2 g^2+1\right) z_{\varphi^2}^{11}+\left(g^2-\frac{7 g^4 N}{144}+\frac{193 g^4}{72}+\frac{g^3 h}{4}-\frac{7 g^2 h^2}{144}\right)~,\\
&&0=z_{\varphi^2}^{22}+g^2 z_{\varphi^2}^{11}+\left(\frac{g^4 N}{12}-\frac{g^4}{6}-g^3 h+\frac{g^2 h^2}{12}\right)~.
\end{eqnarray}
Using $z_{\varphi^2}^{11}$, solving these equations yields
\begin{equation}
z_{\varphi^2}^{21}=\frac{7 g^4 N}{144}-\frac{49 g^4}{72}-\frac{g^3 h}{4}+\frac{7 g^2 h^2}{144}~~~,~~~z_{\varphi^2}^{22}=-\frac{g^4 N}{12}+\frac{7 g^4}{6}+g^3 h-\frac{g^2 h^2}{12}~.\label{eqn:coe-Phi2-2}
\end{equation}
Notably, the sub-Log term’s $1/\epsilon$ pole cancelation provides an additional equation $0=g^2 z_{\varphi^2}^{11}+g^4$, enabling full determination of $z_{\varphi^2}^{11},z_{\varphi^2}^{21},z_{\varphi^2}^{22}$ even without prior one-loop computations.

\begin{figure}
\centering
\begin{tikzpicture}
  \node [] at (0.7,-1) {
    \begin{tikzpicture}
      \draw [thick] (0,0)--(0.5,1);
      \draw [thick] (0,0)--(1.5,1) (0,0)--(1.5,-1);
      \draw [thick, fill=white] (0,0) circle [radius=0.1];
      \draw [thick] (0.07,0.07)--(-0.07,-0.07)  (0.07,-0.07)--(-0.07,0.07);
    \end{tikzpicture}
  };
  \node [] at (3.2,-1) {
    \begin{tikzpicture}
      \draw [thick] (0,0)--(0.5,1);
      \draw [thick] (0,0)--(1.5,1) (0,0)--(1.5,-1);
      \draw [thick, fill=white] (0,0) circle [radius=0.1];
      \draw [thick] (0.07,0.07)--(-0.07,-0.07)  (0.07,-0.07)--(-0.07,0.07);
      \draw [thick] (1,0.667)--(1,-0.667);
    \end{tikzpicture}
  };
  \node [] at (5.7,-1) {
    \begin{tikzpicture}
      \draw [thick] (0,0)--(0.5,1);
      \draw [thick] (0,0)--(1.5,1) (0,0)--(1.5,-1);
      \draw [thick, fill=white] (0,0) circle [radius=0.1];
      \draw [thick] (0.07,0.07)--(-0.07,-0.07)  (0.07,-0.07)--(-0.07,0.07);
      \draw [thick] (1,0.667)--(1,-0.667);
      \draw [thick, fill=white] (1,0) circle [radius=0.25];
    \end{tikzpicture}
  };
  \node [] at (8.2,-1) {
    \begin{tikzpicture}
      \draw [thick] (0,0)--(0.5,1);
      \draw [thick] (0,0)--(1.5,1) (0,0)--(1.5,-1);
      \draw [thick, fill=white] (0,0) circle [radius=0.1];
      \draw [thick] (0.07,0.07)--(-0.07,-0.07)  (0.07,-0.07)--(-0.07,0.07);
      \draw [thick] (1,0.667)--(1,-0.667);
      \draw [thick] (0.5,0.333)--(1,0);
    \end{tikzpicture}
  };
  \node [] at (10.7,-1) {
    \begin{tikzpicture}
      \draw [thick] (0,0)--(0.5,1);
      \draw [thick] (0,0)--(1.5,1) (0,0)--(1.5,-1);
      \draw [thick, fill=white] (0,0) circle [radius=0.1];
      \draw [thick] (0.07,0.07)--(-0.07,-0.07)  (0.07,-0.07)--(-0.07,0.07);
      \draw [thick] (1,0.667)--(1,-0.667);
      \draw [thick, fill=white] (0.5,0.333) circle [radius=0.25];
    \end{tikzpicture}
  };
  \node [] at (13.2,-1) {
    \begin{tikzpicture}
      \draw [thick] (0,0)--(0.5,1);
      \draw [thick] (0,0)--(1.5,1) (0,0)--(1.5,-1);
      \draw [thick, fill=white] (0,0) circle [radius=0.1];
      \draw [thick] (0.07,0.07)--(-0.07,-0.07)  (0.07,-0.07)--(-0.07,0.07);
      \draw [thick] (1,0.667)--(1,-0.667);
      \draw [thick] (0.5,0.333)--(0.5,-0.333);
    \end{tikzpicture}
  };
  \node [] at (15.7,-1) {
    \begin{tikzpicture}
      \draw [thick] (0,0)--(0.5,1);
      \draw [thick] (0,0)--(1.5,1) (0,0)--(1.5,-1);
      \draw [thick, fill=white] (0,0) circle [radius=0.1];
      \draw [thick] (0.07,0.07)--(-0.07,-0.07)  (0.07,-0.07)--(-0.07,0.07);
      \draw [thick] (1,0.667)--(0.5,-0.333);
      \draw [thick] (0.5,0.333)--(0.6,0.133) (0.75,-0.166)--(1,-0.667);
      \draw [thick] (0.6,0.133) to [out=30,in=45] (0.75,-0.166);
    \end{tikzpicture}
  };
  \node [] at (0.3,-1.75) {{\footnotesize $\Gamma^{\tiny \mbox{L0}}_{\otimes\varphi\varphi\varphi}$}};
  \node [] at (2.8,-1.75) {{\footnotesize $\Gamma^{\tiny \mbox{L1}}_{\otimes\varphi\varphi\varphi}$}};
  \node [] at (5.3,-1.75) {{\footnotesize $\Gamma^{\tiny \mbox{L2-I}}_{\otimes\varphi\varphi\varphi}$}};
  \node [] at (7.8,-1.75) {{\footnotesize $\Gamma^{\tiny \mbox{L2-II}}_{\otimes\varphi\varphi\varphi}$}};
  \node [] at (10.3,-1.75) {{\footnotesize $\Gamma^{\tiny \mbox{L2-III}}_{\otimes\varphi\varphi\varphi}$}};
  \node [] at (12.8,-1.75) {{\footnotesize $\Gamma^{\tiny \mbox{L2-IX}}_{\otimes\varphi\varphi\varphi}$}};
  \node [] at (15.3,-1.75) {{\footnotesize $\Gamma^{\tiny \mbox{L2-X}}_{\otimes\varphi\varphi\varphi}$}};
  \node [] at (0.7,-3.5) {
    \begin{tikzpicture}
      \draw [thick] (0,0)--(1.5,1) (0,0)--(1.5,0) (0,0)--(1.5,-1);
      \draw [thick, fill=white] (0,0) circle [radius=0.1];
      \draw [thick] (0.07,0.07)--(-0.07,-0.07)  (0.07,-0.07)--(-0.07,0.07);
      \draw [thick] (1,0)--(0.75,0.5);
      \draw [thick] (0.875,0.25)--(0.8,0.133);
      \draw [thick] (0.5,-0.333)--(0.625,-0.139);
      \draw [thick] (0.625,-0.139) to [out=115,in=180] (0.8,0.133);
    \end{tikzpicture}
  };
  \node [] at (5.7,-3.5) {
    \begin{tikzpicture}
      \draw [thick] (0,0)--(1.5,1) (0,0)--(1.5,0) (0,0)--(1.5,-1);
      \draw [thick, fill=white] (0,0) circle [radius=0.1];
      \draw [thick] (0.07,0.07)--(-0.07,-0.07)  (0.07,-0.07)--(-0.07,0.07);
      \draw [thick] (1,0)--(1,-0.667) (0.75,0)--(0.75,0.5);
    \end{tikzpicture}
  };
  \node [] at (0.3,-4.25) {{\footnotesize $\Gamma^{\tiny \mbox{L2-XI}}_{\otimes\varphi\varphi\varphi}$}};
  \node [] at (5.3,-4.25) {{\footnotesize $\Gamma^{\tiny \mbox{L2-XII}}_{\otimes\varphi\varphi\varphi}$}};
  \node [] at (0.7,-6.5) {
    \begin{tikzpicture}
      \draw [thick,white] (0.5,-0.333)--(1.5,-1);
      \draw [thick] (0,0)--(1.5,1) (0,0)--(1,0) (0,0)--(0.5,-0.333);
      \draw [thick, fill=white] (0,0) circle [radius=0.1];
      \draw [thick] (0.07,0.07)--(-0.07,-0.07)  (0.07,-0.07)--(-0.07,0.07);
      \draw [thick] (1,0)--(0.75,0.5);
      \draw [thick] (0.875,0.25)--(0.8,0.133);
      \draw [thick] (0.5,-0.333)--(0.625,-0.139);
      \draw [thick] (0.625,-0.139) to [out=115,in=180] (0.8,0.133);
    \end{tikzpicture}
  };
  \node [] at (5.7,-6.5) {
    \begin{tikzpicture}
      \draw [thick,white] (0.75,0.5)--(1.5,1);
      \draw [thick] (0,0)--(0.75,0.5) (0,0)--(1,0) (0,0)--(1.5,-1);
      \draw [thick, fill=white] (0,0) circle [radius=0.1];
      \draw [thick] (0.07,0.07)--(-0.07,-0.07)  (0.07,-0.07)--(-0.07,0.07);
      \draw [thick] (1,0)--(1,-0.667) (0.75,0)--(0.75,0.5);
    \end{tikzpicture}
  };
  \node [] at (8.2,-6.5) {
    \begin{tikzpicture}
      \draw [thick,white] (0.75,0.5)--(1.5,1) (1,-0.667)--(1.5,-1);
      \draw [thick] (0,0)--(0.75,0.5) (0,0)--(1.5,0) (0,0)--(1,-0.667);
      \draw [thick, fill=white] (0,0) circle [radius=0.1];
      \draw [thick] (0.07,0.07)--(-0.07,-0.07)  (0.07,-0.07)--(-0.07,0.07);
      \draw [thick] (1,0)--(1,-0.667) (0.75,0)--(0.75,0.5);
    \end{tikzpicture}
  };
  \node [] at (10.7,-6.5) {
    \begin{tikzpicture}
      \draw [thick,white] (1,-0.667)--(1.5,-1);
      \draw [thick] (0,0)--(1.5,1) (0,0)--(1,0) (0,0)--(1,-0.667);
      \draw [thick, fill=white] (0,0) circle [radius=0.1];
      \draw [thick] (0.07,0.07)--(-0.07,-0.07)  (0.07,-0.07)--(-0.07,0.07);
      \draw [thick] (1,0)--(1,-0.667) (0.75,0)--(0.75,0.5);
    \end{tikzpicture}
  };
  \node [] at (0.3,-7.25) {{\footnotesize $\Gamma^{\tiny \mbox{L2-XI}}_{\otimes\varphi\delSubsub{\varphi}\delSubsub{\varphi}}$}};
  \node [] at (5.3,-7.25) {{\footnotesize $\Gamma^{\tiny \mbox{L2-XII}_1}_{\otimes\varphi\delSubsub{\varphi}\delSubsub{\varphi}}$}};
  \node [] at (7.8,-7.25) {{\footnotesize $\Gamma^{\tiny \mbox{L2-XII}_2}_{\otimes\varphi\delSubsub{\varphi}\delSubsub{\varphi}}$}};
  \node [] at (10.3,-7.25) {{\footnotesize $\Gamma^{\tiny \mbox{L2-XII}_3}_{\otimes\varphi\delSubsub{\varphi}\delSubsub{\varphi}}$}};
  \draw [thick,->] (0.75,-4.875)--(0.75,-5.125);
  \draw [thick,->] (5.75,-4.875)--(5.75,-5.125);
\end{tikzpicture}
\caption{The tree-level, 1-loop and 2-loop irreducible 1PI topologies of $\varphi^3$-$3\varphi$ correlation function. The topologies in the first row are equivalent to those corresponding to the $\varphi^2$-$2\varphi$ correlation function. There are only 2 new irreducible topologies shown in the second row. A gray arrow points to the cut graphs of each new irreducible topology, which are generated by the non-isomorphic computation of all possible removals of two external legs.}\label{fig:phi3topo}
\end{figure}
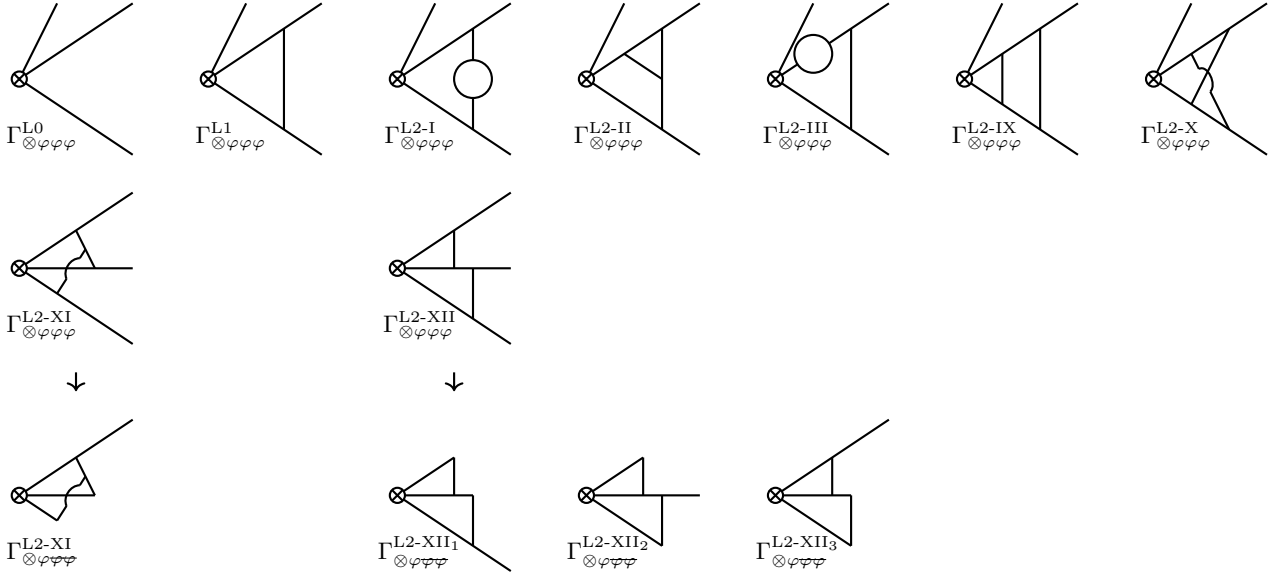

The $\widetilde{Z}_{\varphi^3}$-factor is determined via UV finiteness of $\frac{\widetilde{Z}_{\varphi^3}}{\widetilde{Z}_{\varphi^2}}G_{Q=3}(g_0,h_0;x_0,\del{x}_{1},\del{x}_{2},x_Q)$. The  $\varphi^3$-$3\varphi$ correlation function topologies, as shown in Fig.(\ref{fig:phi3topo}),
include irreducible topologies from $\varphi^2\to 2\varphi$, plus two new two-loop irreducible topologies $\Gamma^{\tiny \mbox{L2-XI}}_{\otimes\varphi{\varphi}\varphi}$ and $\Gamma^{\tiny \mbox{L2-XII}}_{\otimes\varphi{\varphi}\varphi}$. From eqn.(\ref{eqn:phiQ-G}),  we get
\begin{equation}
G_{Q=3}(g_0,h_0;x_0,\del{x}_{1},\del{x}_{2},x_Q)
=1+C_2^1\sum_{j=1}^{L}
G_{\scriptsize\mbox{irre}}^{j{\tiny\mbox{-loop}}}(g_0,h_0;x_0,\del{x}_{1},x_{2})+(2!)C_2^2\sum_{j=2}^{L}
G_{\scriptsize\mbox{irre}}^{j{\tiny\mbox{-loop}}}(g_0,h_0;x_0,\del{x}_{1},\del{x}_{2},x_{3})~.\label{eqn:phi3-G}
\end{equation}
Since $G_{\scriptsize\mbox{irre}}^{{\tiny\mbox{1-loop}}}(g_0,h_0;x_0,\del{x}_{1},x_{2})$ and  $G_{\scriptsize\mbox{irre}}^{{\tiny\mbox{2-loop}}}(g_0,h_0;x_0,\del{x}_{1},x_{2})$ have already been computed in $\varphi^2$ case as (\ref{eqn:1-loop-Q2-irre}) and (\ref{eqn:2-loop-Q2-irre}), only the new two-loop irreducible topologies require computation. These generate 4 cut graphs, with symmetry factors and coupling factors as,
\begin{equation}
\sym(\Gamma^{\tiny \mbox{L2-XI}}_{\otimes\varphi\delSub{\varphi}\delSub{\varphi}})=\frac{1}{2}~~~,~~~\sym(\Gamma^{\tiny \mbox{L2-XII}}_{\otimes\varphi\delSub{\varphi}\delSub{\varphi}})=1~~~,~~~\cf(\Gamma^{\tiny \mbox{L2-XI}}_{\otimes\varphi\delSub{\varphi}\delSub{\varphi}})=g_0^3h_0~~~,~~~\cf(\Gamma^{\tiny \mbox{L2-XII}}_{\otimes\varphi\delSub{\varphi}\delSub{\varphi}})=g_0^4~.
\end{equation}
UV divergences are,
\begin{eqnarray}
  &&\uv(\Gamma^{\tiny \mbox{L2-XI}}_{\otimes\varphi\delSub{\varphi}\delSub{\varphi}})=\frac{1}{2 \epsilon}+\frac{17}{8}+\epsilon\left(\frac{171}{32}-\frac{\pi ^2}{24}\right) +\mathcal{O}(\epsilon^2)~, \\
  &&\uv(\Gamma^{\tiny \mbox{L2-XII}_1}_{\otimes\varphi\delSub{\varphi}\delSub{\varphi}})=\frac{1}{2 \epsilon^2}+\frac{17}{8 \epsilon}+\left(\frac{163}{32}-\frac{\pi ^2}{24}\right)+\epsilon \left(-\frac{\zeta_3}{12}-\frac{17 \pi ^2}{96}+\frac{1161}{128}\right)+\mathcal{O}(\epsilon^2)~, \\
  &&\uv(\Gamma^{\tiny \mbox{L2-XII}_2}_{\otimes\varphi\delSub{\varphi}\delSub{\varphi}})=-\frac{1}{2 \epsilon^2}-\frac{15}{8 \epsilon}+\left(\frac{\pi ^2}{24}-\frac{137}{32}\right)+\epsilon \left(\frac{\zeta_3}{12}+\frac{5 \pi ^2}{32}-\frac{971}{128}\right)+\mathcal{O}(\epsilon^2)~,\\
  &&\uv(\Gamma^{\tiny \mbox{L2-XII}_3}_{\otimes\varphi\delSub{\varphi}\delSub{\varphi}})=\frac{1}{2 \epsilon^2}+\frac{17}{8 \epsilon}+\left(\frac{171}{32}-\frac{\pi ^2}{24}\right)+\epsilon \left(-\frac{\zeta_3}{12}-\frac{17 \pi ^2}{96}+\frac{1313}{128}\right)+\mathcal{O}(\epsilon^2)~.
\end{eqnarray}
Then the required UV contribution of 2-loop irreducible topologies for $\varphi^3\to 3\varphi$ is
\begin{equation}
G_{\scriptsize\mbox{irre}}^{{\tiny\mbox{2-loop}}}(g_0,h_0;x_0,\del{x}_{1},\del{x}_{2},x_{3})=
\sum_{L={\tiny\mbox{L2-XI}},{\tiny\mbox{L2-XII}}_1\atop {\tiny\mbox{L2-XII}}_2,{\tiny\mbox{L2-XII}}_3}\sym(\Gamma^{\tiny \mbox{L}}_{\otimes\varphi\delSub{\varphi}\delSub{\varphi}})\cf(\Gamma^{\tiny \mbox{L}}_{\otimes\varphi\delSub{\varphi}\delSub{\varphi}})\uv(\Gamma^{\tiny \mbox{L}}_{\otimes\varphi\delSub{\varphi}\delSub{\varphi}})~.\label{eqn:2-loop-Q3-irre}
\end{equation}
Expanding $\frac{\widetilde{Z}_{\varphi^3}}{\widetilde{Z}_{\varphi^2}}G_{Q=3}(g_0,h_0;x_0,\del{x}_{1},\del{x}_{2},x_Q)$, the UV finiteness conditions, from canceling $1/\epsilon$ and $1/\epsilon^2$ poles in the leading Log term, and $1/\epsilon$ pole in the sub-Log term, give three equations, 
\begin{eqnarray}
&&0=z_{\varphi^3}^{21}+(4 g^2+1) z_{\varphi^3}^{11}+\left(3 g^2-\frac{7 g^4 N}{48}+\frac{355 g^4}{24}+\frac{5 g^3 h}{4}-\frac{7 g^2 h^2}{48}\right)~,\\
&&0=z_{\varphi^3}^{22}+3 g^2 z_{\varphi^3}^{11}+\left(\frac{g^4 N}{4}+\frac{5 g^4}{2}-3 g^3 h+\frac{g^2 h^2}{4}\right)~~~,~~~0=2 g^2 z_{\varphi^3}^{11}+6 g^4~.
\end{eqnarray}
Solving these yields,
\begin{equation}
z_{\varphi^3}^{11}=-3 g^2~~,~~z_{\varphi^3}^{21}=\frac{7 g^4 N}{48}-\frac{67 g^4}{24}-\frac{5 g^3 h}{4}+\frac{7 g^2 h^2}{48}~~,~~z_{\varphi^3}^{22}=-\frac{g^4 N}{4}+\frac{13 g^4}{2}+3 g^3 h-\frac{g^2 h^2}{4}~.\label{eqn:coe-Phi3}
\end{equation}
With results (\ref{eqn:coe-Phi2-1}), (\ref{eqn:coe-Phi2-2}) and (\ref{eqn:coe-Phi3}), the two-loop $\widetilde{Z}_{\varphi^2}$, $\widetilde{Z}_{\varphi^3}$-factors are finally determined as,
\begin{equation}
\widetilde{Z}_{\varphi^2}=1+\frac{z_{\varphi^2}^{11}}{\epsilon}+\frac{z_{\varphi^2}^{21}}{\epsilon}
+\frac{z_{\varphi^2}^{22}}{\epsilon^2}~~~,~~~\widetilde{Z}_{\varphi^3}=1+\frac{z_{\varphi^3}^{11}}{\epsilon}
+\frac{z_{\varphi^3}^{21}}{\epsilon}
+\frac{z_{\varphi^3}^{22}}{\epsilon^2}~.
\end{equation}

The $\gamma_{Q}:=-\frac{\partial \ln \widetilde{Z}_{\varphi^Q}}{\partial \ln \mu}$ are computed using  $\beta(g)$ and $\beta(h)$ from eqns.(\ref{eqn:beta-g}) and (\ref{eqn:beta-h}),
\begin{eqnarray}
&&\gamma_{2}=-\frac{\partial \ln \widetilde{Z}_{\varphi^2}}{\partial g}\beta(g)-\frac{\partial \ln \widetilde{Z}_{\varphi^2}}{\partial h}\beta(h)=-g^2+\left(
\frac{7 g^4 N}{72}-\frac{49 g^4}{36}-\frac{g^3 h}{2}+\frac{7 g^2 h^2}{72}\right)~,\\
&&\gamma_{3}=-\frac{\partial \ln \widetilde{Z}_{\varphi^3}}{\partial g}\beta(g)-\frac{\partial \ln \widetilde{Z}_{\varphi^3}}{\partial h}\beta(h)=-3 g^2+\left(\frac{7 g^4 N}{24}-\frac{67 g^4}{12}-\frac{5 g^3 h}{2}+\frac{7 g^2 h^2}{24}\right)~.
\end{eqnarray}
Assuming $\gamma_{Q}$ follows the {\sl ansatz} $\gamma_{Q}=c_1Q(Q-1)(Q-2)+c_2Q(Q-1)$, we determine
\begin{equation}
c_1=-\frac{g^4}{4}-\frac{g^3 h}{6}~~~,~~~c_2=-\frac{g^2}{2}+\frac{7 g^4 N}{144}-\frac{49 g^4}{72}-\frac{g^3 h}{4}+\frac{7 g^2 h^2}{144}~.
\end{equation}
Then the scaling dimension of the $\varphi^Q$ operator is 
\begin{equation}
\Delta_Q=\left(\frac{d}{2}-1\right)Q+\gamma_{Q}+Q\gamma_\varphi~,
\end{equation}
where $\gamma_\varphi$ is the field anomalous dimension from eqn.(\ref{eqn:anomalous-field}).

\section{Five-loop anomalous dimensions}
\label{sec:L5result}

The two-loop computation procedures outlined in the preceding section can be straightforwardly generalized to higher-loop calculations following an identical framework. The primary distinction lies in the need to account for a substantially larger number of Feynman diagrams, as topological complexity increases with loop order. The field renormalization factors $Z_{\varphi}$ and $Z_{\sigma}$ are determined from their respective two-point topologies, following the approach analogous to (\ref{eqn:compute-2point}). Similarly, the $Z_h$ and $Z_g$-factors are derived from two-point integrals of cut graphs, generated by removing one external leg from the respective three-point topologies, following the approach outlined in  (\ref{eqn:compute-3point}). For the composite operator renormalization factors $\widetilde{Z}_{\varphi^Q}$, these are determined from two-point integrals of cut graphs produced by removing $(Q-1)$ external legs from irreducible 1PI $\varphi^Q\to Q\varphi$ topologies, in a manner parallel to that described in eqn.(\ref{eqn:phiQ-G}). The anomalous dimension $\gamma_{\varphi^Q}$ of the $\varphi^Q$ operator is then given by $\gamma_{Q}+Q\gamma_\varphi$, where $\gamma_{Q}$ is computed from $\widetilde{Z}_{\varphi^Q}$ and the beta functions. Determining the 5-loop $\gamma_{\varphi^Q}$ for general $Q$ requires results of $L$-loop $\gamma_{{Q_{L}}}$ with $Q_L=2,\ldots,L+1$ up to 5-loops. Notably, since UV divergences are evaluated from scalar Feynman integrals, the specific internal field configurations of the diagrams need not be considered, only their topological structures are relevant.

All topologies for two-point, three-point and $\varphi^Q$-$Q\varphi$ correlation functions, along with their corresponding cut graphs, are generated using custom Mathematica codes. The number of 1PI topologies for two-point,  three-point functions and the irreducible correlation functions is listed in Table.(\ref{table:number-topo}).
\begin{table}
  \centering
  \begin{tabular}{|c||c|c|c|c|c|c||c|c|}
  \hline
   & $Q=2$ & $Q=3$ & $Q=4$ & $Q=5$ & $Q=6$ & Total & 2-point & 3-point \\ \hline \hline
  $L=1$ & 1 & {\color{gray}-} & {\color{gray}-} & {\color{gray}-} & {\color{gray}-} & 1& 1 & 1 \\ \hline
  $L=2$ & 5 & 2 & {\color{gray}-} & {\color{gray}-}  & {\color{gray}-} & 7 & 2 & 3 \\ \hline
  $L=3$ & 35 & 26 & 7 & {\color{gray}-} & {\color{gray}-} & 68 & 9 & 17 \\ \hline
  $L=4$ & 303 & 358 & 168 & 25 & {\color{gray}-} & 854 & 46 & 125 \\ \hline
  $L=5$ & 3136 & 5300 & 3709 & 1103 & 128 & 13376 & 322 & 1170 \\
  \hline
\end{tabular}\vspace{0.15in}
\begin{tabular}{|c||c|c|c|c|c|c||c|c|}
  \hline
   & $Q=2$ & $Q=3$ & $Q=4$ & $Q=5$ & $Q=6$ & Total & 2-point & 3-point \\ \hline \hline
  $L=1$ & 1 & {\color{gray}-} & {\color{gray}-} & {\color{gray}-} & {\color{gray}-} & 1& 1 & 1 \\ \hline
  $L=2$ & 7 & 4 & {\color{gray}-} & {\color{gray}-}  & {\color{gray}-} & 11 & 2 & 5 \\ \hline
  $L=3$ & 56 & 67 & 19 & {\color{gray}-} & {\color{gray}-} & 142 & 9 & 35 \\ \hline
  $L=4$ & 540 & 999 & 586 & 107 & {\color{gray}-} & 2232 & 46 & 303 \\ \hline
  $L=5$ & 5905 & 15238 & 13846 & 5060 & 644 & 40693 & 322 & 3136 \\
  \hline
\end{tabular}
  \caption{(Top Table): the number of 1PI topologies of two-point, three-point correlation functions and the irreducible $\varphi^Q$-$Q\varphi$ correlation functions. (Bottom Table): the number of cut graphs, generated by the removal of one external leg for three-point function and $(Q-1)$ external legs for irreducible $\varphi^Q$-$Q\varphi$ correlation functions. These are the graphs to be computed by TwoPoint command for UV divergences.}\label{table:number-topo}
\end{table}
The number grows rapidly with increasing loop order. Cut graphs are generated via non-isomorphic enumeration of all possible external leg removals, and each cut graph must be individually evaluated to extract its UV divergence. As shown in the Table.(\ref{table:number-topo}), the number of cut graphs to be computed is approximately 3–4 times larger than the number of original topologies.

UV divergences are calculated using \verb!TwoPoint! command in the HyperlogProcedures Maple package. The computational feasibility  and time cost of these evaluations are highly sensitive to both the loop order and the truncation order of the dimensional regulator $\epsilon$. For an $L_{\tiny\mbox{max}}$-loop renormalization computation, each $L$-loop divergence in dimensional regularization must  be computed up to the $\epsilon^{L_{\tiny\mbox{max}}-1-L}$ order. Increasing $L_{\tiny\mbox{max}}$ by one loop requires extending the $\epsilon$ expansion of all loop integrals by an additional order, which substantially increases computational time. For 5-loop computations, each $L$-loop divergence must be evaluated to the $\epsilon^{4-L}$ order, and the current version of HyperlogProcedures is capable of evaluating all required two-point integrals. Time consumption varies depending on the computer's CPU performance and memory capacity. On a standard desktop workstation equipped with a \texttt{3.20GHz}  base frequency CPU and \texttt{64GB} of RAM, the approximate time required to evaluate all necessary loop integrals is summarized in Table.(\ref{table:time-evaluation}).
\begin{table}
  \centering
  \begin{tabular}{|c||c|c|c|c|c|c||c|c|}
  \hline
   & $Q=2$ & $Q=3$ & $Q=4$ & $Q=5$ & $Q=6$ & Total & 2-point & 3-point \\ \hline \hline
  $L=1$ & 0 & {\color{gray}-} & {\color{gray}-} & {\color{gray}-} & {\color{gray}-} & 0 & 0 & 0 \\ \hline
  $L=2$ & 0.02 \texttt{h} & 0 & {\color{gray}-} & {\color{gray}-}  & {\color{gray}-} & 0.02 \texttt{h}& 0  & 0 \\ \hline
  $L=3$ & 0.6 \texttt{h} & 0.2 \texttt{h} & 0 & {\color{gray}-} & {\color{gray}-} & 0.8 \texttt{h} & 0.03 \texttt{h} & 0.15 \texttt{h} \\ \hline
  $L=4$ & 7 \texttt{h} & 5.5 \texttt{h} & 1.5 \texttt{h} & 0 & {\color{gray}-} & 14 \texttt{h} & 2.35 \texttt{h} & 5.5 \texttt{h} \\ \hline
  $L=5$ & 16 \texttt{d} & 24 \texttt{d} & 10 \texttt{d} & 1 \texttt{d} & 0.2 \texttt{h} & 51 \texttt{d} & 5 \texttt{d} & 8 \texttt{d} \\
  \hline
\end{tabular}
  \caption{The approximate time consumption for 5-loop computations in a normal desktop. The symbols \texttt{h} and  \texttt{d} are abbreviation of computer core hours and days. The working time less than 0.01 hour is simply denoted by 0.}\label{table:time-evaluation}
\end{table}
Time consumption also scales rapidly with increasing loop order. However, for 5-loop computations, it remains within a reasonably acceptable range. In contrast, the generation of symmetry factors and coupling factors for these topologies is nearly instantaneous, with negligible computational overhead compared to the time-intensive evaluation of UV divergences.

We have computed the field renormalization factors $Z_{\varphi}$, $Z_{\sigma}$, coupling renormalization factors $Z_h$, $Z_g$, and all composite operator renormalization factors $\widetilde{Z}_{\varphi^{Q_L}}, Q_L=2,\ldots,L+1$ up to 5-loops. From these results, we have derived the beta functions $\beta(g),\beta(h)$, field anomalous dimensions $\gamma_{\varphi}$, $\gamma_{\sigma}$, and scaling dimensions of the ${\varphi^Q}$ operator, all up to 5-loops. These results are provided in ancillary files to the arXiv, and we have cross-checked them against existing literature with excellent agreements. For example, the up to 3-loop pieces of $\gamma_\varphi, \gamma_\sigma, \beta_g,\beta_h$ are identical to those in \cite{Fei:2014xta}, after unifying possible overall factors $(4\pi)$ from different conventions. The up to 4-loop pieces of results are identical to those in the $O(N)$ sector in \cite{Gracey:2015tta}, after unifying possible overall factors $2$ from different conventions, and the fixed point solutions of beta functions up to $\epsilon^4$ order are identical to those in \cite{Kompaniets:2021hwg} after compensating an overall $\sqrt{2}$ factor due to different convention.

The 5-loop correction to the anomalous dimension $\gamma_{Q}+Q\gamma_{\varphi}$ is presented in Appendix \S\ref{appendix:5-loop-result}. The scaling dimension of the $\varphi^Q$ operator at the Wilson-Fisher fixed point is computed as
\begin{equation}
\Delta_{Q\tiny\mbox{FP}}=\left(2-\frac{\epsilon}{2}\right)Q+\left(\gamma_{Q}
+Q\gamma_{\varphi}\right)\Big|_{g=g^{\ast},h=h^{\ast}}~,
\end{equation}
where $g^{\ast},h^{\ast}$ are the fixed-point solutions of the beta functions. Our 5-loop fixed point solutions, provided in Appendix \S\ref{appendix:fixed-point}, are in perfect agreement with the literature \cite{Kompaniets:2021hwg}. By substituting these fixed-point solutions into the anomalous dimensions yields $\Delta_Q$ in the large $N$ limit. The leading, sub-leading and sub-sub-leading order contributions of the large $N$ expansion are as follows,
\begin{equation}
\Delta_{Q\tiny\mbox{FP}}=\left(2-\frac{\epsilon}{2}\right)Q+\frac{Q}{N}\Delta_{Q\tiny\mbox{FP}}^1
+\frac{Q}{N^2}\Delta_{Q\tiny\mbox{FP}}^2+\frac{Q}{N^3}\Delta_{Q\tiny\mbox{FP}}^3
+\frac{Q}{N^4}\Delta_{Q\tiny\mbox{FP}}^4+\frac{Q}{N^5}\Delta_{Q\tiny\mbox{FP}}^5+\mathcal{O}(\frac{1}{N^6})~,
\end{equation}
where
\begin{eqnarray}
\Delta_{Q{\tiny\mbox{FP}}}^{1}&=&\bigg[4-3 Q\bigg]\epsilon + \bigg[-\frac{8}{3}+\frac{7 Q}{4}\bigg]\epsilon^2+ \bigg[-\frac{7}{9}+\frac{11 Q}{16}\bigg]\epsilon^3\nonumber\\
&&+ \bigg[\zeta_3-\frac{8}{27}+\left(\frac{19 }{64}-\frac{3 \zeta_3 }{4}\right)Q\bigg]\epsilon^4+\bigg[\frac{\pi ^4}{120}-\frac{43}{324} -\frac{2 \zeta_3}{3}+\left(\frac{7\zeta_3 }{16}-\frac{\pi ^4 }{160}+\frac{35 }{256}\right)Q\bigg]\epsilon^5~,
\end{eqnarray}
\begin{eqnarray}
\Delta_{Q{\tiny\mbox{FP}}}^{2}&=& \bigg[176-132 Q\bigg]\epsilon+ \bigg[-\frac{1568}{3}+\frac{857 Q}{2}-45 Q^2\bigg]\epsilon^2\nonumber\\
&&+\bigg[\frac{4105}{18}+72 \zeta_3-\left(108 \zeta_3 +\frac{3743 }{24}\right)Q+\left(36 \zeta_3 +\frac{93 }{4}\right)Q^2\bigg]\epsilon^3 \nonumber\\
&&+ \bigg[74 \zeta_3+\frac{3 \pi ^4}{5}+\frac{19315}{216}+\left(21 \zeta_3 -\frac{9 \pi ^4 }{10}-\frac{7465 }{72}\right)Q+\left(\frac{3 \pi ^4 }{10}-51 \zeta_3 +\frac{167 }{8}\right)Q^2\bigg]\epsilon^4\nonumber\\
&&+ \bigg[\frac{12641}{324}-\frac{2281 \zeta_3}{6}+\frac{37 \pi ^4}{60}+36 \zeta_5+\left(\frac{577 \zeta_3 }{2}-54 \zeta_5 +\frac{7 \pi ^4 }{40}-\frac{44459 }{864}\right)Q\nonumber\\
&&~~~~~~~~~~~~~~~~~~~~~~~~~~~~~~~~~~~~~~~~~~~~~~~~~~~~~~~~~~~+\left( 18 \zeta_5 -\frac{65 \zeta_3 }{4}+\frac{339 }{32}-\frac{17 \pi ^4 }{40}\right)Q^2\bigg]\epsilon^5 ~,
\end{eqnarray}
\begin{eqnarray}
\Delta_{Q{\tiny\mbox{FP}}}^{3}&=&  \bigg[7744-5808 Q\bigg]\epsilon+ \bigg[-66872+59520 Q-9000 Q^2\bigg]\epsilon^2\nonumber\\
&&+\bigg[\frac{193141}{2}-4212 \zeta_3-\bigg(3996 \zeta_3+88575\bigg) Q+\bigg(4536 \zeta_3+20538\bigg) Q^2-1350 Q^3\bigg]\epsilon^3\nonumber\\
&&+\bigg[61920 \zeta_3-14040 \zeta_5-\frac{351 \pi ^4}{10}-\frac{107537}{216}+\bigg(14580 \zeta_5-22146 \zeta_3-\frac{63913}{6}-\frac{333 \pi ^4}{10}\bigg) Q\nonumber\\
&&~~~~~~~~~~~~~~~~~~~~~~+\bigg(\frac{2101}{2}-18558 \zeta_3-1080 \zeta_5+\frac{189 \pi ^4}{5}\bigg) Q^2+\bigg(\frac{1107}{2}+1728 \zeta_3+540 \zeta_5\bigg) Q^3\bigg]\epsilon^4\nonumber\\
&&+\bigg[516 \pi ^4-540 \zeta_3^2-166574 \zeta_3-\frac{35397 \zeta_5}{2}-\frac{130 \pi ^6}{7}-\frac{14477777}{864} \nonumber\\
&&~~~~~~~~+\bigg(\frac{9276203}{432}-2106 \zeta_3^2+\frac{245019 \zeta_3}{2}-\frac{981 \zeta_5}{2}+\frac{135 \pi ^6}{7}-\frac{3691 \pi ^4}{20}\bigg) Q\nonumber\\
&&~~~~~~~~~~~~~~~~~~~~~~+\bigg(3348 \zeta_3^2+\frac{549 \zeta_3}{2}+3663 \zeta_5-\frac{10 \pi ^6}{7}-\frac{3093 \pi ^4}{20}-\frac{190189}{24}\bigg) Q^2\nonumber\\
&&~~~~~~~~~~~~~~~~~~~~~~~~~~~~~~~~~~~~~~~+\bigg(\frac{7551}{8}-702 \zeta_3^2-2322 \zeta_3-1215 \zeta_5+\frac{5 \pi ^6}{7}+\frac{72 \pi ^4}{5}\bigg) Q^3\bigg]\epsilon^5~.
\end{eqnarray}
Expressions $\Delta_{Q{\tiny\mbox{FP}}}^{4}$ and $\Delta_{Q{\tiny\mbox{FP}}}^{5}$ are presented in Appendix \S\ref{appendix:5-loop-large-N}. These results are further validated by the following consistency checks. The 3-loop contribution up to $\epsilon^3$ order is consistent with literature findings \cite{Vasiliev:1981yc,Vasiliev:1981dg,Vasiliev:1982dc,Jack:2021ziq}\footnote{The value $a_t^{(3)}$ in paper \cite{Jack:2021ziq} that computes $\Delta_Q^{(3)}$ is incorrect. After using the correct value $a_t^{(3)}=\frac{g^6}{3}+g^5h$, the $\Delta_Q^{(3)}$ and the $1/N^3$ term in their paper are consistent with the 3-loop order part of our results.}. The corresponding terms $\frac{Q^{i+1}\epsilon^i}{N^i},\frac{Q^{i+1}\epsilon^{i+1}}{N^i}, i=1,\ldots,5$ are also consistent with those reported in the semi-classical literature \cite{Antipin:2021jiw}. The 4-loop contribution up to $1/N^2$ order has been previously provided in \cite{Huang:2024hsn}, while the 5-loop contributions up to $1/N^5$ order presented here are new. All the above results have been provided in ancillary files to the arXiv submission of this article.

\section{Conclusion}
\label{sec:conclusion}

In this paper, we employ the OPE based renormalization algorithm proposed in \cite{Huang:2024hsn} to compute the 5-loop scaling dimensions of the $\phi^Q$ operator in the six-dimensional cubic scalar theory. This algorithm represents a non-trivial implementation of the connection between Wilson coefficients of OPE and anomalous dimensions, enabling the determination of renormalization $Z$-factors by treating UV divergences globally.
A key advantage is its avoidance of cumbersome sub-divergence subtraction. Instead, it systematically extracts the essential UV divergences for renormalization directly from two-point integrals. Advancements in modern computational techniques, particularly the graphical function method \cite{Schnetz:2013hqa,Golz:2015rea,Borinsky:2021gkd,Schnetz:2024qqt,HP} adopted herein, have rendered feasible the evaluation of these two-point integrals even beyond five loops. The transformative impact of the graphical function method, combined with the efficiency and robustness of the proposed OPE based algorithm, has facilitated the successful completion of renormalization computations up to five loops on a standard desktop computer. These results are derived from over 40,000 two-point propagator-type integrals computed using the HyperlogProcedures package \cite{HP}, augmented by non-trivial summation protocols inherent to the algorithm. We have successfully obtained the 5-loop corrections to the anomalous dimension of the $\phi^Q$ operator, as well as the large $N$ expansion of the scaling dimension at the Wilson-Fisher fixed point up to $1/N^5$ order. It stands for a new record for the scaling dimensions of the $\phi^Q$ operator in cubic scalar theory. Our results exhibit perfect consistency with lower-loop perturbative calculations reported in the literature and align with corresponding terms derived via semi-classical methods. Furthermore, the non-trivial cancellation of $\epsilon$ poles in the final result provides a stringent  consistency check, validating its correctness. This work further demonstrates the efficiency and versatility of the OPE based algorithm for high-precision renormalization computations in quantum field theories.

Efforts are currently underway to apply this OPE based algorithm to tackle the challenges presented by more sophisticated problems. One promising initiative aims to break the existing 6-loop benchmark in quartic scalar theory by computing the 7-loop scaling dimensions of the $\phi^Q$ operator. Such research will not only push the precision of renormalization computations but also drive the evolution of modern techniques for multi-loop integral evaluations. Also as previously noted, the OPE based algorithm boasts strong generality and extensibility beyond scalar field theories. It would be particularly insightful to explore its application to renormalization computations in other theoretical frameworks, such as the Gross-Neveu-Yukawa theory and scalar QED, among others. In these theories, Feynman integrals are no longer purely scalar in nature, but instead involve non-trivial numerator structures. While future methodological refinements may enable the calculation of such non-trivial master integrals via a modified graphical function method tailored to handle numerator-dependent topologies, the most robust approach available at present relies on computational algorithms \cite{Smirnov:2008iw,Maierhofer:2017gsa,Wu:2023upw,Guan:2024byi} based on the integration-by-parts (IBP) method \cite{Tkachov:1981wb,Chetyrkin:1981qh,Laporta:2000dsw}. Another potential challenge lies in extending this algorithm to compute the anomalous dimensions of more intricate operators, for instance, those containing higher derivatives, Lorentz indices, or in scenarios involving operator mixing \cite{Derkachov:1997ch,Cao:2021cdt}. Although the conceptual application of the OPE based algorithm remains straightforward in these cases, as demonstrated in \cite{Huang:2024hsn}, practical computations necessitate more advanced technical implementations. Further exploration of the algorithm’s diverse applications in renormalization and targeted enhancements of its computational capabilities will undoubtedly open new avenues for high-precision quantum field theory calculations.

\section*{Acknowledgments}

We would like to thank Oliver Schnetz for stimulating conversations on the graphical function method and the usage of HyperlogProcedures package, and Hugh Osborn for useful comments on this paper. 
This work is supported by the Science Challenge Project (No. TZ2025012), and NSAF No.U2330401.

\appendix

\section{Five-loop corrections to the scaling dimensions}
\label{appendix:5-loop-result}

In this section, we collect our results of the scaling dimension $\Delta_Q$ in six-dimensional cubic scalar theory. Note that the 4-loop result has been presented in \cite{Huang:2024hsn}. The 5-loop correction of $\Delta_Q$ is given by
\begin{equation}
\Delta_Q^{{\tiny\mbox{5-loop}}}=Q\sum_{i=2}^{10} g^i h^{10-i}\delta^5_i~,
\end{equation}
where,
{\footnotesize
\begin{eqnarray}
\delta^5_2&=&-\frac{3336089}{13436928}+\frac{25882183 Q}{161243136}+\left(-\frac{685}{7776}+\frac{15341 Q}{248832}\right) \zeta_3+ \left(-\frac{1177}{1866240}+\frac{23 Q}{55296}\right)\pi ^4\nonumber\\
&&~~~~~~~~~~~~~~~~~~~~~~~~~~~~~~~+\left(\frac{311}{1296}-\frac{271 Q}{1728}\right) \zeta_5+\left(\frac{5}{40824}-\frac{5 Q}{54432}\right)\pi ^6 +\left(-\frac{1}{108}+\frac{Q}{144}\right) \zeta_3^{~2}~,
\end{eqnarray}
\begin{eqnarray}
\delta^5_3&=&-\frac{133429553}{161243136}+\frac{26603591 Q}{13436928}-\frac{228283 Q^2}{279936}+\left(\frac{231613}{62208}-\frac{28829 Q}{6912}+\frac{13091 Q^2}{15552}\right) \zeta_3+\left(-\frac{241}{46080}\right.\nonumber\\
&&\left.+\frac{121 Q}{11520}-\frac{2599 Q^2}{622080}\right)\pi ^4 +\left(-\frac{29}{16}+\frac{421 Q}{576}+\frac{719 Q^2}{1728}\right) \zeta_5+ \left(-\frac{5}{3402}+\frac{5 Q}{4536}\right)\pi ^6+\left(\frac{1}{9}-\frac{Q}{12}\right) \zeta_3^{~2}~,
\end{eqnarray}
\begin{eqnarray}
\delta^5_4&=&\frac{4152458077}{241864704}+\frac{1256311 N}{20155392}-\frac{1679472427 Q}{80621568}-\frac{2125781 N Q}{53747712}+\frac{24768415 Q^2}{4478976}-\frac{20291 Q^3}{41472}\nonumber\\
&&+\left(\frac{104597}{3456}-\frac{65 N}{2592}-\frac{2478701 Q}{62208}+\frac{3919 N Q}{248832}+\frac{169543 Q^2}{15552}-\frac{131 Q^3}{144}\right)\zeta_3 + \left(-\frac{31489}{1866240}\right.\nonumber\\
&&\left.-\frac{11 N}{622080}+\frac{307 Q}{13824}+\frac{11 N Q}{497664}-\frac{2023 Q^2}{311040}+\frac{Q^3}{1080}\right)\pi ^4+\left(-\frac{33983}{2592}+\frac{N}{324}+\frac{6473 Q}{288}-\frac{N Q}{432}\right.\nonumber\\
&&\left.-\frac{3041 Q^2}{288}+\frac{85 Q^3}{72}\right)\zeta_5+ \left(\frac{205}{20412}-\frac{25 Q}{1944}+\frac{25 Q^2}{6804}-\frac{5 Q^3}{13608}\right)\pi ^6 +\left(\frac{1195}{216}-\frac{323 Q}{72}-\frac{37 Q^2}{36}\right.\nonumber\\
&&\left.+\frac{4 Q^3}{9}\right) \zeta_3^{~2}+\left(-\frac{5467}{96}+\frac{6055 Q}{96}-\frac{147 Q^2}{16}\right) \zeta_7~,
\end{eqnarray}
\begin{eqnarray}
\delta^5_5&=&-\frac{705928679}{26873856}+\frac{25722139 N}{161243136}+\frac{519118363 Q}{13436928}-\frac{3257 N Q}{7776}-\frac{38531567 Q^2}{2239488}+\frac{129851 N Q^2}{746496}\nonumber\\
&&+\frac{10871 Q^3}{3072}-\frac{641 Q^4}{2304}+ \left(-\frac{1604333}{31104}-\frac{1145 N}{1536}+\frac{581737 Q}{6912}+\frac{23861 N Q}{20736}-\frac{1196249 Q^2}{31104}\right.\nonumber\\
&&\left.-\frac{3719 N Q^2}{10368}+\frac{7727 Q^3}{1152}-\frac{11 Q^4}{24}\right)\zeta_3+ \left(\frac{4411}{155520}+\frac{607 N}{124416}-\frac{11137 Q}{207360}-\frac{17 N Q}{2592}+\frac{1997 Q^2}{62208}\right.\nonumber\\
&&\left.+\frac{43 N Q^2}{23040}-\frac{161 Q^3}{34560}\right)\pi ^4+ \left(\frac{12157}{576}+\frac{43 N}{216}-\frac{15383 Q}{576}-\frac{49 N Q}{192}+\frac{9767 Q^2}{1728}+\frac{107 N Q^2}{1728}\right.\nonumber\\
&&\left.+\frac{5 Q^3}{9}-\frac{Q^4}{3}\right)\zeta_5+ \left(\frac{305}{54432}+\frac{25 Q}{36288}-\frac{95 Q^2}{15552}+\frac{5 Q^3}{3402}\right)\pi ^6+\left(-\frac{223}{144}-\frac{371 Q}{96}+\frac{1723 Q^2}{288}\right.\nonumber\\
&&\left.-\frac{10 Q^3}{9}\right) \zeta_3^{~2}+\left(\frac{917}{16}-\frac{581 Q}{6}+\frac{2401 Q^2}{48}-\frac{35 Q^3}{3}+\frac{7 Q^4}{6}\right) \zeta_7~,
\end{eqnarray}
\begin{eqnarray}
\delta^5_6&=&\frac{101332127}{1889568}-\frac{377140433 N}{80621568}+\frac{20375 N^2}{13436928}-\frac{151156525 Q}{1679616}+\frac{923790881 N Q}{161243136}-\frac{84245 N^2 Q}{80621568}+\frac{3532885 Q^2}{69984}\nonumber\\
&&-\frac{3589273 N Q^2}{2239488}-\frac{14759 Q^3}{1024}+\frac{9473 N Q^3}{82944}+\frac{545 Q^4}{256}-\frac{7 Q^5}{48}+\left(\frac{1035907}{7776}+\frac{4147 N}{62208}+\frac{41 N^2}{46656}-\frac{6811451 Q}{31104}\right.\nonumber\\
&&\left.+\frac{44599 N Q}{62208}-\frac{197 N^2 Q}{248832}+\frac{1258969 Q^2}{10368}-\frac{36589 N Q^2}{62208}-\frac{19235 Q^3}{576}+\frac{17 N Q^3}{144}+\frac{65 Q^4}{24}\right)\zeta_3\nonumber\\
&&+\left(\frac{50047}{466560}-\frac{719 N}{311040}-\frac{17 N^2}{1866240}-\frac{35393 Q}{311040}+\frac{7 N Q}{3072}+\frac{17 N^2 Q}{2488320}+\frac{349 Q^2}{20736}-\frac{103 N Q^2}{622080}+\frac{73 Q^3}{17280}-\frac{N Q^3}{2160}\right)\pi ^4 \nonumber\\
&&+ \left(\frac{26623}{144}+\frac{7325 N}{648}-\frac{296567 Q}{864}-\frac{25771 N Q}{1728}+\frac{186683 Q^2}{864}+\frac{851 N Q^2}{192}-\frac{1435 Q^3}{24}-\frac{65 N Q^3}{144}+\frac{35 Q^4}{6}\right)\zeta_5\nonumber\\
&&+\left(\frac{415}{163296}-\frac{185 N}{40824}+\frac{5 Q}{2268}+\frac{55 N Q}{9072}-\frac{295 Q^2}{54432}-\frac{25 N Q^2}{13608}+\frac{25 Q^3}{13608}+\frac{5 N Q^3}{27216}\right)\pi ^6\nonumber\\
&&+ \left(-\frac{15473}{432}-\frac{367 N}{216}+71 Q+\frac{53 N Q}{24}-\frac{5899 Q^2}{144}-\frac{11 N Q^2}{18}+\frac{223 Q^3}{36}+\frac{N Q^3}{36}\right)\zeta_3^{~2}\nonumber\\
&&+\left(-\frac{8827}{24}+\frac{19817 Q}{32}-\frac{17129 Q^2}{48}+\frac{413 Q^3}{4}-\frac{35 Q^4}{3}\right) \zeta_7~,
\end{eqnarray}
\begin{eqnarray}
\delta^5_7&=&-\frac{1727695009}{13436928}+\frac{1219348333 N}{80621568}-\frac{2802343 N^2}{53747712}+\frac{1827570041 Q}{6718464}-\frac{38406623 N Q}{2239488}+\frac{42845 N^2 Q}{2239488}\nonumber\\
&&-\frac{360106991 Q^2}{2239488}+\frac{22560479 N Q^2}{4478976}+\frac{35173 N^2 Q^2}{4478976}+\frac{1125739 Q^3}{41472}-\frac{20573 N Q^3}{20736}+\frac{3223 Q^4}{4608}+\frac{133 N Q^4}{2304}-\frac{7 Q^5}{16}\nonumber\\
&&+\left(-\frac{48245}{3888}+\frac{1642177 N}{31104}+\frac{3011 N^2}{62208}+\frac{128345 Q}{972}-\frac{4686905 N Q}{62208}-\frac{331 N^2 Q}{5184}-\frac{1930387 Q^2}{10368}+\frac{406627 N Q^2}{15552}\right.\nonumber\\
&&\left.+\frac{1165 N^2 Q^2}{62208}+\frac{14215 Q^3}{192}-\frac{1451 N Q^3}{576}-\frac{141 Q^4}{16}\right)\zeta_3 +\left(-\frac{177979}{311040}-\frac{71161 N}{622080}-\frac{95 N^2}{248832}+\frac{162677 Q}{155520}\right.\nonumber\\
&&\left.+\frac{103393 N Q}{622080}+\frac{17 N^2 Q}{23040}-\frac{19751 Q^2}{34560}-\frac{3713 N Q^2}{62208}-\frac{91 N^2 Q^2}{311040}+\frac{1781 Q^3}{17280}+\frac{89 N Q^3}{17280}\right)\pi ^4 +\left(-\frac{2030531}{1728}\right.\nonumber\\
&&\left.-\frac{47359 N}{576}-\frac{N^2}{144}+\frac{1301549 Q}{576}+\frac{203777 N Q}{1728}+\frac{N^2 Q}{96}-\frac{824411 Q^2}{576}-\frac{26117 N Q^2}{576}-\frac{N^2 Q^2}{288}+\frac{6655 Q^3}{18}\right.\nonumber\\
&&\left.+\frac{445 N Q^3}{72}-\frac{110 Q^4}{3}\right)\zeta_5 +\left(\frac{3725}{27216}+\frac{55 N}{3402}-\frac{8185 Q}{36288}-\frac{3085 N Q}{108864}+\frac{11695 Q^2}{108864}+\frac{1405 N Q^2}{108864}-\frac{205 Q^3}{13608}\right.\nonumber\\
&&\left.-\frac{25 N Q^3}{13608}\right)\pi ^6+ \left(\frac{9959}{72}+\frac{629 N}{72}-\frac{22685 Q}{96}-\frac{3493 N Q}{288}+\frac{34195 Q^2}{288}+\frac{1273 N Q^2}{288}-\frac{685 Q^3}{36}-\frac{ 11 N Q^3}{18}\right)\zeta_3^{~2}\nonumber\\
&&+\left(\frac{12887}{12}-\frac{147 N}{16}-\frac{18291 Q}{8}+\frac{147 N Q}{16}+\frac{9793 Q^2}{6}-\frac{3731 Q^3}{8}+\frac{140 Q^4}{3}\right)\zeta_7 ~,
\end{eqnarray}
\begin{eqnarray}
\delta^5_8&=&\frac{26586982301}{60466176}-\frac{779370073 N}{60466176}+\frac{26353453 N^2}{80621568}-\frac{3827 N^3}{20155392}-\frac{8024021965 Q}{10077696}+\frac{79115735 N Q}{5038848}+\frac{1162525 N^2 Q}{20155392}\nonumber\\
&&+\frac{2029 N^3 Q}{17915904}+\frac{514446007 Q^2}{1119744}-\frac{4617385 N Q^2}{559872}-\frac{25159 N^2 Q^2}{165888}-\frac{2234915 Q^3}{20736}+\frac{237013 N Q^3}{82944}+\frac{127 N^2 Q^3}{27648}\nonumber\\
&&+\frac{15847 Q^4}{1536}-\frac{163 N Q^4}{768}-\frac{3 Q^5}{4}+ \left(-\frac{8364127}{15552}-\frac{929849 N}{23328}-\frac{18653 N^2}{31104}-\frac{43 N^3}{31104}+\frac{7945621 Q}{7776}+\frac{2483159 N Q}{31104}\right.\nonumber\\
&&\left.+\frac{33251 N^2 Q}{62208}+\frac{245 N^3 Q}{248832}-\frac{8598169 Q^2}{15552}-\frac{163897 N Q^2}{3456}-\frac{1465 N^2 Q^2}{20736}+\frac{9743 Q^3}{108}+\frac{899 N Q^3}{144}-\frac{149 Q^4}{24}\right.\nonumber\\
&&\left.+\frac{3 N Q^4}{8}\right)\zeta_3+\left(\frac{18179}{15552}+\frac{89461 N}{311040}+\frac{2899 N^2}{622080}+\frac{N^3}{622080}-\frac{314647 Q}{155520}-\frac{145487 N Q}{311040}-\frac{4417 N^2 Q}{622080}-\frac{N^3 Q}{829440}\right.\nonumber\\
&&\left.+\frac{31291 Q^2}{31104}+\frac{7607 N Q^2}{34560}+\frac{481 N^2 Q^2}{207360}-\frac{151 Q^3}{1152}-\frac{29 N Q^3}{864}\right)\pi ^4 + \left(\frac{6140413}{2592}+\frac{784319 N}{5184}+\frac{43 N^2}{864}-\frac{4206853 Q}{864}\right.\nonumber\\
&&\left.-\frac{406439 N Q}{1728}-\frac{7 N^2 Q}{64}+\frac{2890897 Q^2}{864}+\frac{7891 N Q^2}{72}+\frac{103 N^2 Q^2}{1728}-\frac{44575 Q^3}{48}-\frac{1735 N Q^3}{96}+\frac{260 Q^4}{3}\right)\zeta_5\nonumber\\
&&+\left(-\frac{9425}{40824}-\frac{7015 N}{163296}+\frac{7565 Q}{18144}+\frac{4255 N Q}{54432}-\frac{3995 Q^2}{18144}-\frac{325 N Q^2}{7776}+\frac{965 Q^3}{27216}+\frac{365 N Q^3}{54432}\right)\pi ^6 \nonumber\\
&&+ \left(-\frac{5089}{216}-\frac{5359 N}{432}+\frac{2083 Q}{48}+\frac{4207 N Q}{144}-\frac{1003 Q^2}{48}-\frac{2935 N Q^2}{144}+\frac{155 Q^3}{72}+\frac{581 N Q^3}{144}\right)\zeta_3^{~2}\nonumber\\
&&+ \left(-\frac{200585}{96}-\frac{1631 N}{24}+\frac{415247 Q}{96}+\frac{371 N Q}{6}-\frac{148813 Q^2}{48}+\frac{11123 Q^3}{12}-\frac{280 Q^4}{3}\right)\zeta_7~,
\end{eqnarray}
\begin{eqnarray}
\delta^5_9&=&-\frac{4988268911}{10077696}+\frac{158410849 N}{4478976}-\frac{1951205 N^2}{20155392}+\frac{162107 N^3}{161243136}+\frac{2881576087 Q}{3359232}-\frac{213374875 N Q}{6718464}-\frac{720205 N^2 Q}{746496}\nonumber\\
&&-\frac{1213 N^3 Q}{2239488}-\frac{237912325 Q^2}{559872}+\frac{2904071 N Q^2}{746496}+\frac{2387851 N^2 Q^2}{4478976}+\frac{11 N^3 Q^2}{1492992}+\frac{146737 Q^3}{2592}+\frac{2015 N Q^3}{13824}-\frac{209 N^2 Q^3}{82944}\nonumber\\
&&+\frac{673 Q^4}{144}-\frac{1915 N Q^4}{4608}-\frac{5 Q^5}{6}+\left(\frac{11111}{32}+\frac{992585 N}{15552}+\frac{47023 N^2}{31104}+\frac{1397 N^3}{124416}-\frac{14576275 Q}{15552}-\frac{6227 N Q}{54}-\frac{5197 N^2 Q}{7776}\right.\nonumber\\
&&\left.-\frac{23 N^3 Q}{3456}+\frac{2176717 Q^2}{2592}+\frac{674003 N Q^2}{10368}-\frac{659 N^2 Q^2}{10368}-\frac{7 N^3 Q^2}{6912}-\frac{39937 Q^3}{144}-\frac{6193 N Q^3}{576}-\frac{35 N^2 Q^3}{384}+\frac{215 Q^4}{8}\right.\nonumber\\
&&\left.-\frac{5 N Q^4}{24}\right) \zeta_3+\left(\frac{1703}{5184}-\frac{30931 N}{62208}-\frac{203 N^2}{69120}-\frac{N^3}{51840}-\frac{12353 Q}{31104}+\frac{4771 N Q}{5760}+\frac{233  N^2Q}{31104}+\frac{ N^3Q}{69120}+\frac{1427 Q^2}{25920}\right.\nonumber\\
&&\left.-\frac{511 N Q^2 }{1280}-\frac{41 N^2 Q^2 }{10368}+\frac{7 Q^3}{240}+\frac{73N Q^3 }{1152}+\frac{N^2Q^3 }{2304}\right) \pi ^4+\left(-\frac{1144559}{432}-\frac{321965 N}{1728}-\frac{67 N^2}{96}+\frac{599057 Q}{108}\right.\nonumber\\
&&\left.+\frac{534791 N Q}{1728}+\frac{847 N^2 Q}{864}-\frac{1665455 Q^2}{432}-\frac{93157 N Q^2}{576}-\frac{37 N^2 Q^2}{72}+\frac{76435 Q^3}{72}+\frac{235 N Q^3}{9}-105 Q^4+\frac{5 N Q^4}{6}\right) \zeta_5\nonumber\\
&&+\left(-\frac{155}{6804}+\frac{2735 N}{27216}-\frac{25 N^2}{54432}+\frac{55 Q}{1512}-\frac{18325 N Q}{108864}+\frac{25 N^2 Q}{54432}-\frac{5 Q^2}{432}+\frac{2995 N Q^2}{36288}+\frac{5 Q^3}{2268}-\frac{5 N Q^3}{378} \right) \pi^6\nonumber\\
&&+\left(-\frac{3037}{9}+\frac{2369 N}{72}-\frac{19 N^2}{144}+\frac{6817 Q}{12}-\frac{16729 N Q}{288}+\frac{19 N^2 Q}{144}-\frac{6445 Q^2}{24}+\frac{2951 N Q^2}{96}+\frac{112 Q^3}{3}-5 N Q^3\right) \zeta_3^{~2}\nonumber\\
&&+\left(\frac{78659}{24}-\frac{203 N}{12}-\frac{299551 Q}{48}+\frac{259 N Q}{16}+\frac{181027 Q^2}{48}+\frac{35 N Q^2}{48}-\frac{21077 Q^3}{24}+70 Q^4\right) \zeta_7~,
\end{eqnarray}
\begin{eqnarray}
\delta^5_{10}&=&\frac{3873526685}{15116544}-\frac{1261849117 N}{60466176}-\frac{1111975 N^2}{30233088}+\frac{2132959 N^3}{241864704}-\frac{43 N^4}{2519424}-\frac{4120586761 Q}{10077696}+\frac{217226153 N Q}{6718464}\nonumber\\
&&+\frac{24155359 N^2 Q}{26873856}-\frac{1574323 N^3 Q}{161243136}+\frac{35 N^4 Q}{1990656}+\frac{12775867 Q^2}{69984}-\frac{16632703 N Q^2}{1119744}-\frac{239203 N^2 Q^2}{559872}+\frac{41 N^3 Q^2}{31104}\nonumber\\
&&-\frac{3373 Q^3}{144}+\frac{3283 N Q^3}{1152}-\frac{133 N^2 Q^3}{4608}+\frac{7 Q^4}{64}-\frac{469 N Q^4}{1536}-\frac{Q^5}{2}+\left(\frac{2181835}{11664}-\frac{33815 N}{1458}-\frac{246587 N^2}{46656}-\frac{3619 N^3}{186624}\right.\nonumber\\
&&\left.-\frac{N^4}{11664}-\frac{2534795 Q}{15552}+\frac{502549N Q }{5184}+\frac{119575N^2 Q }{20736}+\frac{37N^3 Q }{4608}+\frac{7N^4 Q }{124416}-\frac{9701 Q^2}{81}-\frac{390451 NQ^2 }{5184}-\frac{11453 N^2 Q^2}{7776}\right.\nonumber\\
&&\left.+\frac{31 N^3 Q^2}{7776}+\frac{50953 Q^3}{432}+\frac{4409 NQ^3 }{432}+\frac{143 N^2Q^3 }{576}-\frac{47 Q^4}{2}-\frac{19 NQ^4 }{48}\right) \zeta_3+\left(-\frac{10277}{23328}+\frac{17249 N}{58320}+\frac{667 N^2}{933120}\right.\nonumber\\
&&\left.-\frac{23 N^3}{233280}+\frac{N^4}{933120}+\frac{27649 Q}{31104}-\frac{153709 N Q}{311040}-\frac{1493 N^2Q }{311040}+\frac{97N^3 Q }{622080}-\frac{N^4Q }{1244160}-\frac{199 Q^2}{360}+\frac{13021 NQ^2 }{51840}\right.\nonumber\\
&&\left.+\frac{137 N^2Q^2 }{38880}-\frac{17 N^3Q^2 }{311040}+\frac{77 Q^3}{720}-\frac{137 NQ^3 }{3456}-\frac{13  N^2Q^3}{17280}\right) \pi ^4+\left(\frac{185473}{144}+\frac{88807 N}{864}+\frac{28717 N^2}{5184}+\frac{N^3}{216}\right.\nonumber\\
&&\left.-\frac{378875 Q}{144}-\frac{345655 N Q}{1728}-\frac{3679 N^2 Q}{576}-\frac{N^3 Q}{144}+\frac{387313 Q^2}{216}+\frac{99619 N Q^2}{864}+\frac{475 N^2 Q^2}{288}+\frac{N^3 Q^2}{432}-\frac{3915 Q^3}{8}\right.\nonumber\\
&&\left.-\frac{2825 N Q^3}{144}-\frac{5 N^2 Q^3}{32}+\frac{125 Q^4}{3}-\frac{N Q^4}{3}\right) \zeta_5+\left(\frac{445}{3402}-\frac{145 N}{2268}-\frac{3155 Q}{13608}+\frac{6025 N Q}{54432}+\frac{10 Q^2}{81}-\frac{3055 N Q^2}{54432}\right.\nonumber\\
&&\left.-\frac{5 N^2 Q^2}{27216}-\frac{65 Q^3}{3402}+\frac{25 N Q^3}{3024}+\frac{5 N^2 Q^3}{54432}\right) \pi ^6+\left(\frac{2342}{9}-\frac{209 N}{24}+\frac{23 N^2}{36}-\frac{4145 Q}{9}+\frac{3271 N Q}{144}-N^2 Q+240 Q^2\right.\nonumber\\
&&\left.-\frac{2287 N Q^2}{144}+\frac{31 N^2 Q^2}{72}-\frac{344 Q^3}{9}+\frac{49 N Q^3}{24}-\frac{13 N^2 Q^3}{144}\right) \zeta_3^{~2}+\left(-\frac{216181}{96}-\frac{567 N}{32}+\frac{131411 Q}{32}-\frac{105 N Q}{4}\right.\nonumber\\
&&\left.-2296 Q^2+\frac{301 N Q^2}{8}+\frac{3563 Q^3}{8}-\frac{119 N Q^3}{48}-\frac{49 Q^4}{3}+\frac{7 N Q^4}{6}\right) \zeta_7 ~.
\end{eqnarray}
}

\section{Solutions at the Wilson-Fisher fixed point}
\label{appendix:fixed-point}

The beta functions can be solved order by order in $\epsilon$, which leads to the solutions at fixed point as,
{\footnotesize
\begin{eqnarray}
g^{\ast}&=&\sqrt{\frac{6\epsilon}{N}}\bigg\{ 1+\frac{22}{N}+\frac{726}{N^2}-\frac{326180}{N^3}-\frac{349658330}{N^4}
+\bigg[-\frac{155}{6 N}-\frac{1705}{N^2}+\frac{912545}{N^3}+\frac{3590574890}{3 N^4}\bigg]\epsilon+ \bigg[\frac{1777}{144 N}\nonumber\\
&&+\frac{1}{N^2}\bigg(\frac{29093}{36}-1170 \zeta_3\bigg)+\frac{1}{N^3}\bigg(-\frac{106755739}{72}-228060 \zeta_3\bigg)+\frac{92543220 \zeta_3-1920046163}{N^4}\bigg]\epsilon^2+ \bigg[-\frac{217}{2592 N}\nonumber\\
&&+\frac{1}{N^2}\bigg(\frac{709151}{864}+3126 \zeta_3-\frac{13 \pi ^4}{2}+360 \zeta_5\bigg)+\frac{1}{N^3}\bigg(\frac{779869165}{432}+2229915 \zeta_3-1267 \pi ^4-2438640 \zeta_5\bigg)\nonumber\\
&&+\frac{1}{N^4}\bigg(\frac{2437481884309}{1296}+278186790 \zeta_3+864409 \pi ^4-1469512080 \zeta_5\bigg)\bigg]\epsilon^3+\bigg[\frac{1}{N}\bigg(\frac{6973}{41472}-\frac{155 \zeta_3}{96}\bigg)\nonumber\\
&&+\frac{1}{N^2}\bigg(-\frac{49050023}{124416}-\frac{40871 \zeta_3}{24}+\frac{1563 \pi ^4}{80}-4320 \zeta_5+\frac{5 \pi ^6}{14}+81 \zeta_3^2\bigg)+\frac{1}{N^3}\bigg(-\frac{17059272503}{15552}-\frac{77297317 \zeta_3}{48}\nonumber\\
&&+\frac{667967 \pi ^4}{48}+\frac{23014935 \zeta_5}{4}-\frac{16935 \pi ^6}{7}-366930 \zeta_3^2-\frac{2941029 \zeta_7}{2}\bigg)+\frac{1}{N^4}\bigg(-\frac{68647291122307}{62208}+\frac{40450011887 \zeta_3}{12}\nonumber\\
&&-\frac{1345559 \pi ^4}{24}+\frac{13461052935 \zeta_5}{2}-\frac{9381745 \pi ^6}{7}-37063008 \zeta_3^2-6633588717 \zeta_7\bigg)\bigg]\epsilon^4 \bigg\}~,
\end{eqnarray}
}
and
{\footnotesize
\begin{eqnarray}
h^{\ast}&=&6\sqrt{\frac{6\epsilon}{N}}\bigg\{1+\frac{162}{N}+\frac{68766}{N^2}+\frac{41224420}{N^3}+\frac{28762554870}{N^4}+ \bigg[-\frac{215}{2 N}-\frac{86335}{N^2}-\frac{75722265}{N^3}-\frac{69633402510}{N^4}\bigg]\epsilon+ \bigg[\frac{927}{16 N}\nonumber\\
&&+\frac{1}{N^2}\bigg(\frac{270911}{6}-26190 \zeta_3\bigg)+\frac{1}{N^3}\bigg(\frac{1156981601}{24}-26334180 \zeta_3\bigg)+\frac{1}{N^4}\bigg(\frac{116908872997}{2}-23638808340 \zeta_3\bigg)\bigg]\epsilon^2\nonumber\\
&&+ \bigg[-\frac{3461}{864 N}+\frac{1}{N^2}\bigg(\frac{25209239}{864}+115860 \zeta_3-\frac{291 \pi ^4}{2}-70200 \zeta_5\bigg)+\frac{1}{N^3}\bigg(\frac{18166643735}{432}+198108185 \zeta_3-175491 \pi ^4\nonumber\\
&&-169538400 \zeta_5\bigg)+\frac{1}{N^4}\bigg(\frac{5465613593485}{144}+235067598534 \zeta_3-172334673 \pi ^4-211183698960 \zeta_5\bigg)\bigg]\epsilon^3\nonumber\\
&&+ \bigg[\frac{1}{N}\bigg(\frac{5945}{13824}-\frac{215 \zeta_3}{32}\bigg)+\frac{1}{N^2}\bigg(-\frac{1204206197}{41472}-\frac{973091 \zeta_3}{8}+\frac{5793 \pi ^4}{8}+104895 \zeta_5-\frac{975 \pi ^6}{14}-15795 \zeta_3^2\bigg)+\nonumber\\
&&\frac{1}{N^3}\bigg(-\frac{35896723781}{648}-\frac{1912729075 \zeta_3}{16}+\frac{10821883 \pi ^4}{8}+\frac{2017087425 \zeta_5}{4}-\frac{1245950 \pi ^6}{7}-23961852 \zeta_3^2-\frac{637290423 \zeta_7}{2}\bigg)\nonumber\\
&&+\frac{1}{N^4}\bigg(-\frac{179864809668917}{2304}-\frac{552249422829 \zeta_3}{4}+\frac{34006612103 \pi ^4}{20}+\frac{1850885090325 \zeta_5}{2}-\frac{1654305365 \pi ^6}{7}\nonumber\\
&&-15580792368 \zeta_3^2-725681165454 \zeta_7\bigg)\bigg]\epsilon^4\bigg\}~.
\end{eqnarray}
}

\section{Large $N$ expansion at the Wilson-Fisher fixed point}
\label{appendix:5-loop-large-N}

Here we present the results of the $1/N^4$ and $1/N^5$ orders in the large $N$ expansion of $\gamma_{\varphi^Q}$ at Wilson-Fisher fixed point. 
We have,
{\footnotesize
\begin{eqnarray}
\Delta_{Q{\tiny\mbox{FP}}}^{4}&=&\bigg[-2481664+1861248 Q\bigg]\epsilon+\bigg[-\frac{21069760}{3}+8828264 Q-3043440 Q^2\bigg]\epsilon^2+\bigg[\frac{259232542}{9}-861840 \zeta_3+\bigg(-30411798\nonumber\\
&&-1662768 \zeta_3\bigg) Q+\bigg(8984340+1478304 \zeta_3\bigg) Q^2-495720 Q^3
\bigg]\epsilon^3+\bigg[-\frac{25851511}{4}+44545394 \zeta_3-7182 \pi ^4-40335840 \zeta_5\nonumber\\
&&+\bigg(\frac{124128583}{24}-31054098 \zeta_3-\frac{69282 \pi ^4}{5}+38233080 \zeta_5\bigg) Q+\bigg(-4410989-2172420 \zeta_3+\frac{61596 \pi ^4}{5}-4110480 \zeta_5\bigg) Q^2\nonumber\\
&&+\bigg(1024488+388854 \zeta_3+186840 \zeta_5\bigg) Q^3-\frac{213597 Q^4}{4}\bigg]\epsilon^4+\bigg[-\frac{29834168837}{1296}-\frac{498959309 \zeta_3}{6}+\frac{22272697 \pi ^4}{60}\nonumber\\
&&+110252862 \zeta_5-9004500 \zeta_3^2-\frac{373480 \pi ^6}{7}-47146617 \zeta_7+\bigg(\frac{23354482969}{864}+48023731 \zeta_3-\frac{5175683 \pi ^4}{20}-89443629 \zeta_5\nonumber\\
&&+\frac{354010 \pi ^6}{7}+10983168 \zeta_3^2+42115059 \zeta_7\bigg) Q+\bigg(-\frac{27985893}{4}+8241600 \zeta_3-\frac{36207 \pi ^4}{2}-1692738 \zeta_5-\frac{38060 \pi ^6}{7}\nonumber\\
&&-2645568 \zeta_3^2+326592 \zeta_7\bigg) Q^2+\bigg(\frac{1270377}{4}-3338073 \zeta_3+\frac{64809 \pi ^4}{20}+1459530 \zeta_5+\frac{1730 \pi ^6}{7}+11124 \zeta_3^2-19278 \zeta_7\bigg) Q^3\nonumber\\
&&+\bigg(\frac{252477}{16}+92178 \zeta_3+36288 \zeta_5+9072 \zeta_7\bigg) Q^4 \bigg]\epsilon^5~,
\end{eqnarray}
\begin{eqnarray}
\Delta_{Q{\tiny\mbox{FP}}}^{5}&=&\bigg[-2852566016
+2139424512 Q\bigg]\epsilon+\bigg[\frac{9747487616}{3}-424181248 Q-1627608960 Q^2
\bigg]\epsilon^2+\bigg[\frac{77119431952}{9}\nonumber\\
&&+1600458624 \zeta_3+\bigg(-\frac{36719656688}{3}-2589857280 \zeta_3\bigg) Q+\bigg(5365027104+905601024 \zeta_3\bigg) Q^2-223974720 Q^3
\bigg]\epsilon^3\nonumber\\
&&+\bigg[-\frac{18452489398}{3}+16928531152 \zeta_3+\frac{66685776 \pi ^4}{5}-25646146560 \zeta_5+\bigg(\frac{62653214582}{9}-12498672264 \zeta_3\nonumber\\
&&-21582144 \pi ^4+24847024320 \zeta_5\bigg) Q+\bigg(-4161747404-280446528 \zeta_3+\frac{37733376 \pi ^4}{5}-3130401600 \zeta_5\bigg) Q^2\nonumber\\
&&+\bigg(640256832+94633056 \zeta_3+135475200 \zeta_5\bigg) Q^3-28653588 Q^4
\bigg]\epsilon^4+\bigg[-\frac{12059984664415}{648}-\frac{98689756858 \zeta_3}{3}\nonumber\\
&&+\frac{2116066394 \pi ^4}{15}+141568450614 \zeta_5-\frac{237464320 \pi ^6}{7}-4542718032 \zeta_3^2-92577398445 \zeta_7+\bigg(\frac{2294461732357}{108}\nonumber\\
&&+21292069758 \zeta_3-\frac{520778011 \pi ^4}{5}-123620853240 \zeta_5+\frac{230065040 \pi ^6}{7}+7314253776 \zeta_3^2+79133304537 \zeta_7\bigg) Q \nonumber\\
&&+\bigg(-\frac{27277501615}{6}-237707128 \zeta_3-\frac{11685272 \pi ^4}{5}+8399505492 \zeta_5-\frac{28985200 \pi ^6}{7}-2829860928 \zeta_3^2\nonumber\\
&&-1796643828 \zeta_7\bigg) Q^2+\bigg(-215265924-1832415606 \zeta_3+\frac{3943044 \pi ^4}{5}+1094776020 \zeta_5+179200 \pi ^6+151863552 \zeta_3^2\nonumber\\
&&-230521788 \zeta_7\bigg) Q^3+\bigg(56025189+32423328 \zeta_3+10158048 \zeta_5+10360224 \zeta_7\bigg) Q^4-2457216 Q^5\bigg]\epsilon^5~,
\end{eqnarray}
}

\bibliographystyle{JHEP}
\bibliography{Hbib}

\providecommand{\href}[2]{#2}\begingroup\raggedright\begin{thebibliography}{10}

\bibitem{Wilson:1971dc}
K.~G. Wilson and M.~E. Fisher, {\it {Critical exponents in 3.99 dimensions}},
  {\em Phys. Rev. Lett.} {\bf 28} (1972) 240--243.

\bibitem{PELISSETTO2002549}
A.~Pelissetto and E.~Vicari, {\it Critical phenomena and renormalization-group
  theory},  {\em Phys. Rept.} {\bf 368} (2002), no.~6 549--727.

\bibitem{Moshe:2003xn}
M.~Moshe and J.~Zinn-Justin, {\it {Quantum field theory in the large N limit: A
  Review}},  {\em Phys. Rept.} {\bf 385} (2003) 69--228,
  [\href{http://arxiv.org/abs/hep-th/0306133}{{\tt hep-th/0306133}}].

\bibitem{Fei:2014yja}
L.~Fei, S.~Giombi, and I.~R. Klebanov, {\it {Critical $O(N)$ models in
  $6-\epsilon$ dimensions}},  {\em Phys. Rev. D} {\bf 90} (2014), no.~2 025018,
  [\href{http://arxiv.org/abs/1404.1094}{{\tt arXiv:1404.1094}}].

\bibitem{Fei:2014xta}
L.~Fei, S.~Giombi, I.~R. Klebanov, and G.~Tarnopolsky, {\it {Three loop
  analysis of the critical O(N) models in 6-\ensuremath{\varepsilon}
  dimensions}},  {\em Phys. Rev. D} {\bf 91} (2015), no.~4 045011,
  [\href{http://arxiv.org/abs/1411.1099}{{\tt arXiv:1411.1099}}].

\bibitem{Wilson:1971vs}
K.~G. Wilson, {\it {Feynman graph expansion for critical exponents}},  {\em
  Phys. Rev. Lett.} {\bf 28} (1972) 548--551.

\bibitem{BREZIN1973227}
E.~Brezin, J.~{Le Guillou}, J.~Zinn-Justin, and B.~Nickel, {\it Higher order
  contributions to critical exponents},  {\em Physics Letters A} {\bf 44}
  (1973), no.~3 227--228.

\bibitem{Brezin:1973igb}
E.~Brezin, J.~C. Le~Guillou, and J.~Zinn-Justin, {\it {Wilson's theory of
  critical phenomena and callan-symanzik equations in 4-epsilon dimensions}},
  {\em Phys. Rev. D} {\bf 8} (1973) 434--440. [Addendum: Phys.Rev.D 9,
  1121--1124 (1974), Erratum: Phys.Rev.D 10, 2046--2046 (1974)].

\bibitem{Kazakov:1979ik}
D.~I. Kazakov, O.~V. Tarasov, and A.~A. Vladimirov, {\it {Calculation of
  Critical Exponents by Quantum Field Theory Methods}},  {\em Sov. Phys. JETP}
  {\bf 50} (1979) 521.

\bibitem{Chetyrkin:1981jq}
K.~G. Chetyrkin, A.~L. Kataev, and F.~V. Tkachov, {\it {Five Loop Calculations
  in the $g \phi^4$ Model and the Critical Index $\eta$}},  {\em Phys. Lett. B}
  {\bf 99} (1981) 147. [Erratum: Phys.Lett.B 101, 457 (1981)].

\bibitem{Gorishnii:1983gp}
S.~G. Gorishnii, S.~A. Larin, F.~V. Tkachov, and K.~G. Chetyrkin, {\it {Five
  Loop Renormalization Group Calculations in the $g \phi^4$ in Four-dimensions
  Theory}},  {\em Phys. Lett. B} {\bf 132} (1983) 351.

\bibitem{Kazakov:1983dyk}
D.~I. Kazakov, {\it {THE METHOD OF UNIQUENESS, A NEW POWERFUL TECHNIQUE FOR
  MULTILOOP CALCULATIONS}},  {\em Phys. Lett. B} {\bf 133} (1983) 406--410.

\bibitem{Gorishnii:1983jr}
S.~G. Gorishnii, S.~A. Larin, and F.~V. Tkachov, {\it {$\epsilon$ Expansion for
  Critical Exponents: The O ($\epsilon^5$) Approximation}},  {\em Phys. Lett.
  A} {\bf 101} (1984) 120.

\bibitem{Kleinert:1991rg}
H.~Kleinert, J.~Neu, V.~Schulte-Frohlinde, K.~G. Chetyrkin, and S.~A. Larin,
  {\it {Five loop renormalization group functions of O(n) symmetric phi**4
  theory and epsilon expansions of critical exponents up to epsilon**5}},  {\em
  Phys. Lett. B} {\bf 272} (1991) 39--44,
  [\href{http://arxiv.org/abs/hep-th/9503230}{{\tt hep-th/9503230}}]. [Erratum:
  Phys.Lett.B 319, 545 (1993)].

\bibitem{Adzhemyan:2013jra}
L.~T. Adzhemyan and M.~V. Kompaniets, {\it {Five-loop numerical evaluation of
  critical exponents of the $\phi^4$ theory}},  {\em J. Phys. Conf. Ser.} {\bf
  523} (2014) 012049, [\href{http://arxiv.org/abs/1309.5621}{{\tt
  arXiv:1309.5621}}].

\bibitem{Vladimirov:1979zm}
A.~A. Vladimirov, {\it {Method for Computing Renormalization Group Functions in
  Dimensional Renormalization Scheme}},  {\em Theor. Math. Phys.} {\bf 43}
  (1980) 417.

\bibitem{Chetyrkin:1980pr}
K.~G. Chetyrkin, A.~L. Kataev, and F.~V. Tkachov, {\it {New Approach to
  Evaluation of Multiloop Feynman Integrals: The Gegenbauer Polynomial x Space
  Technique}},  {\em Nucl. Phys. B} {\bf 174} (1980) 345--377.

\bibitem{Caswell:1981ek}
W.~E. Caswell and A.~D. Kennedy, {\it {A Simple Approach to Renormalization
  Theory}},  {\em Phys. Rev. D} {\bf 25} (1982) 392.

\bibitem{Chetyrkin:1982nn}
K.~G. Chetyrkin and F.~V. Tkachov, {\it {Infrared R Operation and Ultraviolet
  Counterterms in the MS Scheme}},  {\em Phys. Lett. B} {\bf 114} (1982)
  340--344.

\bibitem{CHETYRKIN1984419}
K.~Chetyrkin and V.~Smirnov, {\it R*-operation corrected},  {\em Physics
  Letters B} {\bf 144} (1984), no.~5 419--424.

\bibitem{Larin:2002sc}
S.~Larin and P.~van Nieuwenhuizen, {\it {The Infrared R* operation}},
  \href{http://arxiv.org/abs/hep-th/0212315}{{\tt hep-th/0212315}}.

\bibitem{Broadhurst:1995km}
D.~J. Broadhurst and D.~Kreimer, {\it {Knots and numbers in Phi**4 theory to 7
  loops and beyond}},  {\em Int. J. Mod. Phys. C} {\bf 6} (1995) 519--524,
  [\href{http://arxiv.org/abs/hep-ph/9504352}{{\tt hep-ph/9504352}}].

\bibitem{Derkachov:1997pf}
S.~E. Derkachov, J.~A. Gracey, and A.~N. Manashov, {\it {Four loop anomalous
  dimensions of gradient operators in phi**4 theory}},  {\em Eur. Phys. J. C}
  {\bf 2} (1998) 569--579, [\href{http://arxiv.org/abs/hep-ph/9705268}{{\tt
  hep-ph/9705268}}].

\bibitem{Batkovich:2016jus}
D.~V. Batkovich, K.~G. Chetyrkin, and M.~V. Kompaniets, {\it {Six loop
  analytical calculation of the field anomalous dimension and the critical
  exponent $\eta$ in $O(n)$-symmetric $\varphi^4$ model}},  {\em Nucl. Phys. B}
  {\bf 906} (2016) 147--167, [\href{http://arxiv.org/abs/1601.01960}{{\tt
  arXiv:1601.01960}}].

\bibitem{Kompaniets:2016hct}
M.~Kompaniets and E.~Panzer, {\it {Renormalization group functions of $\phi^4$
  theory in the MS-scheme to six loops}},  {\em PoS} {\bf LL2016} (2016) 038,
  [\href{http://arxiv.org/abs/1606.09210}{{\tt arXiv:1606.09210}}].

\bibitem{Kompaniets:2017yct}
M.~V. Kompaniets and E.~Panzer, {\it {Minimally subtracted six loop
  renormalization of $O(n)$-symmetric $\phi^4$ theory and critical exponents}},
   {\em Phys. Rev. D} {\bf 96} (2017), no.~3 036016,
  [\href{http://arxiv.org/abs/1705.06483}{{\tt arXiv:1705.06483}}].

\bibitem{Schnetz:2013hqa}
O.~Schnetz, {\it {Graphical functions and single-valued multiple
  polylogarithms}},  {\em Commun. Num. Theor. Phys.} {\bf 08} (2014) 589--675,
  [\href{http://arxiv.org/abs/1302.6445}{{\tt arXiv:1302.6445}}].

\bibitem{Golz:2015rea}
M.~Golz, E.~Panzer, and O.~Schnetz, {\it {Graphical functions in parametric
  space}},  {\em Lett. Math. Phys.} {\bf 107} (2017), no.~6 1177--1192,
  [\href{http://arxiv.org/abs/1509.07296}{{\tt arXiv:1509.07296}}].

\bibitem{Borinsky:2021gkd}
M.~Borinsky and O.~Schnetz, {\it {Graphical functions in even dimensions}},
  {\em Commun. Num. Theor. Phys.} {\bf 16} (2022), no.~3 515--614,
  [\href{http://arxiv.org/abs/2105.05015}{{\tt arXiv:2105.05015}}].

\bibitem{Schnetz:2024qqt}
O.~Schnetz and S.~Theil, {\it {Notes on graphical functions with numerator
  structure}},  {\em PoS} {\bf LL2024} (2024) 026,
  [\href{http://arxiv.org/abs/2407.17133}{{\tt arXiv:2407.17133}}].

\bibitem{HP}
O.~Schnetz, {\em {\verb!HyperlogProcedures! {\sl ver.0.8}}}.
\newblock Maple package available on the homepage of the author,
  https://www.math.fau.de/person/oliver-schnetz.

\bibitem{Schnetz:2016fhy}
O.~Schnetz, {\it {Numbers and Functions in Quantum Field Theory}},  {\em Phys.
  Rev. D} {\bf 97} (2018), no.~8 085018,
  [\href{http://arxiv.org/abs/1606.08598}{{\tt arXiv:1606.08598}}].

\bibitem{Schnetz:2025opm}
O.~Schnetz, {\it {Graphical functions with spin}},  {\em JHEP} {\bf 06} (2025)
  053, [\href{http://arxiv.org/abs/2504.05850}{{\tt arXiv:2504.05850}}].

\bibitem{Macfarlane:1974vp}
A.~J. Macfarlane and G.~Woo, {\it {$\phi^3$ Theory in Six Dimensions and the
  Renormalization Group}},  {\em Nucl. Phys. B} {\bf 77} (1974) 91--108.
  [Erratum: Nucl.Phys.B 86, 548--548 (1975)].

\bibitem{Ma:1975vn}
E.~Ma, {\it {Asymptotic Freedom and a Quark Model in Six-Dimensions}},  {\em
  Prog. Theor. Phys.} {\bf 54} (1975) 1828.

\bibitem{deAlcantaraBonfim:1980pe}
O.~F. de~Alcantara~Bonfim, J.~E. Kirkham, and A.~J. McKane, {\it {Critical
  Exponents to Order $\epsilon^3$ for $\phi^3$ Models of Critical Phenomena in
  Six $\epsilon$-dimensions}},  {\em J. Phys. A} {\bf 13} (1980) L247.
  [Erratum: J.Phys.A 13, 3785 (1980)].

\bibitem{deAlcantaraBonfim:1981sy}
O.~F. de~Alcantara~Bonfim, J.~E. Kirkham, and A.~J. McKane, {\it {Critical
  Exponents for the Percolation Problem and the Yang-lee Edge Singularity}},
  {\em J. Phys. A} {\bf 14} (1981) 2391.

\bibitem{Gracey:2015tta}
J.~A. Gracey, {\it {Four loop renormalization of $\phi^3$ theory in six
  dimensions}},  {\em Phys. Rev. D} {\bf 92} (2015), no.~2 025012,
  [\href{http://arxiv.org/abs/1506.03357}{{\tt arXiv:1506.03357}}].

\bibitem{Kompaniets:2021hwg}
M.~Kompaniets and A.~Pikelner, {\it {Critical exponents from five-loop scalar
  theory renormalization near six-dimensions}},  {\em Phys. Lett. B} {\bf 817}
  (2021) 136331, [\href{http://arxiv.org/abs/2101.10018}{{\tt
  arXiv:2101.10018}}].

\bibitem{Borinsky:2021jdb}
M.~Borinsky, J.~A. Gracey, M.~V. Kompaniets, and O.~Schnetz, {\it {Five-loop
  renormalization of \ensuremath{\phi}3 theory with applications to the
  Lee-Yang edge singularity and percolation theory}},  {\em Phys. Rev. D} {\bf
  103} (2021), no.~11 116024, [\href{http://arxiv.org/abs/2103.16224}{{\tt
  arXiv:2103.16224}}].

\bibitem{Schnetz:2025wtu}
O.~Schnetz, {\it {{\ensuremath{\phi}}3 theory at six loops}},  {\em Phys. Rev.
  D} {\bf 112} (2025), no.~1 016028,
  [\href{http://arxiv.org/abs/2505.15485}{{\tt arXiv:2505.15485}}].

\bibitem{DePrato:2003yd}
M.~De~Prato, A.~Pelissetto, and E.~Vicari, {\it {Third harmonic exponent in
  three-dimensional N vector models}},  {\em Phys. Rev. B} {\bf 68} (2003)
  092403, [\href{http://arxiv.org/abs/cond-mat/0302145}{{\tt
  cond-mat/0302145}}].

\bibitem{Calabrese:2004ca}
P.~Calabrese and P.~Parruccini, {\it {Harmonic crossover exponents in O(n)
  models with the pseudo-epsilon expansion approach}},  {\em Phys. Rev. B} {\bf
  71} (2005) 064416, [\href{http://arxiv.org/abs/cond-mat/0411027}{{\tt
  cond-mat/0411027}}].

\bibitem{Arias-Tamargo:2019xld}
G.~Arias-Tamargo, D.~Rodriguez-Gomez, and J.~G. Russo, {\it {The large charge
  limit of scalar field theories and the Wilson-Fisher fixed point at
  $\epsilon=0$}},  {\em JHEP} {\bf 10} (2019) 201,
  [\href{http://arxiv.org/abs/1908.11347}{{\tt arXiv:1908.11347}}].

\bibitem{Badel:2019oxl}
G.~Badel, G.~Cuomo, A.~Monin, and R.~Rattazzi, {\it {The Epsilon Expansion
  Meets Semiclassics}},  {\em JHEP} {\bf 11} (2019) 110,
  [\href{http://arxiv.org/abs/1909.01269}{{\tt arXiv:1909.01269}}].

\bibitem{Antipin:2020abu}
O.~Antipin, J.~Bersini, F.~Sannino, Z.-W. Wang, and C.~Zhang, {\it {Charging
  the $O(N)$ model}},  {\em Phys. Rev. D} {\bf 102} (2020), no.~4 045011,
  [\href{http://arxiv.org/abs/2003.13121}{{\tt arXiv:2003.13121}}].

\bibitem{Giombi:2020enj}
S.~Giombi and J.~Hyman, {\it {On the large charge sector in the critical O(N)
  model at large N}},  {\em JHEP} {\bf 09} (2021) 184,
  [\href{http://arxiv.org/abs/2011.11622}{{\tt arXiv:2011.11622}}].

\bibitem{Arias-Tamargo:2020fow}
G.~Arias-Tamargo, D.~Rodriguez-Gomez, and J.~G. Russo, {\it {On the UV
  completion of the $O(N)$ model in $6-\epsilon$ dimensions: a stable
  large-charge sector}},  {\em JHEP} {\bf 09} (2020) 064,
  [\href{http://arxiv.org/abs/2003.13772}{{\tt arXiv:2003.13772}}].

\bibitem{Antipin:2021jiw}
O.~Antipin, J.~Bersini, F.~Sannino, Z.-W. Wang, and C.~Zhang, {\it {More on the
  cubic versus quartic interaction equivalence in the $O(N)$ model}},  {\em
  Phys. Rev. D} {\bf 104} (2021) 085002,
  [\href{http://arxiv.org/abs/2107.02528}{{\tt arXiv:2107.02528}}].

\bibitem{Jack:2021ypd}
I.~Jack and D.~R.~T. Jones, {\it {Anomalous dimensions at large charge in d=4
  O(N) theory}},  {\em Phys. Rev. D} {\bf 103} (2021), no.~8 085013,
  [\href{http://arxiv.org/abs/2101.09820}{{\tt arXiv:2101.09820}}].

\bibitem{Jin:2022nqq}
Q.~Jin and Y.~Li, {\it {Five-loop anomalous dimensions of
  \ensuremath{\phi}$^{Q}$ operators in a scalar theory with O(N) symmetry}},
  {\em JHEP} {\bf 10} (2022) 084, [\href{http://arxiv.org/abs/2205.02535}{{\tt
  arXiv:2205.02535}}].

\bibitem{Bednyakov:2022guj}
A.~Bednyakov and A.~Pikelner, {\it {Six-loop anomalous dimension of the
  \ensuremath{\phi}Q operator in the O(N) symmetric model}},  {\em Phys. Rev.
  D} {\bf 106} (2022), no.~7 076015,
  [\href{http://arxiv.org/abs/2208.04612}{{\tt arXiv:2208.04612}}].

\bibitem{Jack:2021ziq}
I.~Jack and D.~R.~T. Jones, {\it {Scaling dimensions at large charge for cubic
  \ensuremath{\phi}3 theory in six dimensions}},  {\em Phys. Rev. D} {\bf 105}
  (2022), no.~4 045021, [\href{http://arxiv.org/abs/2112.01196}{{\tt
  arXiv:2112.01196}}].

\bibitem{Huang:2024hsn}
R.~Huang, Q.~Jin, and Y.~Li, {\it {From operator product expansion to anomalous
  dimensions}},  {\em JHEP} {\bf 06} (2025) 135,
  [\href{http://arxiv.org/abs/2410.03283}{{\tt arXiv:2410.03283}}].

\bibitem{Huang:2025ree}
R.~Huang, Q.~Jin, and Y.~Li, {\it {On the seven-loop renormalization of
  Gross-Neveu model}},  {\em JHEP} {\bf 06} (2025) 134,
  [\href{http://arxiv.org/abs/2504.00713}{{\tt arXiv:2504.00713}}].

\bibitem{Collins_1984}
J.~C. Collins, {\em Renormalization: An Introduction to Renormalization, the
  Renormalization Group and the Operator-Product Expansion}.
\newblock Cambridge Monographs on Mathematical Physics. Cambridge University
  Press, 1984.

\bibitem{Eden:2012fe}
B.~Eden, P.~Heslop, G.~P. Korchemsky, V.~A. Smirnov, and E.~Sokatchev, {\it
  {Five-loop Konishi in N=4 SYM}},  {\em Nucl. Phys. B} {\bf 862} (2012)
  123--166, [\href{http://arxiv.org/abs/1202.5733}{{\tt arXiv:1202.5733}}].

\bibitem{Marino:2024uco}
M.~Marino and R.~Miravitllas, {\it {Trans-series from condensates}},
  \href{http://arxiv.org/abs/2402.19356}{{\tt arXiv:2402.19356}}.

\bibitem{Liu:2025bqq}
Y.~Liu and M.~Mari{\~n}o, {\it {Trans-series from condensates in the non-linear
  sigma model}},  \href{http://arxiv.org/abs/2507.02605}{{\tt
  arXiv:2507.02605}}.

\bibitem{tHooft:1973mfk}
G.~'t~Hooft, {\it {Dimensional regularization and the renormalization group}},
  {\em Nucl. Phys. B} {\bf 61} (1973) 455--468.

\bibitem{Huang:2016bmv}
R.~Huang, Q.~Jin, and B.~Feng, {\it {Form Factor and Boundary Contribution of
  Amplitude}},  {\em JHEP} {\bf 06} (2016) 072,
  [\href{http://arxiv.org/abs/1601.06612}{{\tt arXiv:1601.06612}}].

\bibitem{Huang:2022qnx}
R.~Huang, Q.~Jin, and Y.~Li, {\it {Wilson lines and boundary operators of BCFW
  shifts}},  {\em JHEP} {\bf 12} (2022) 023,
  [\href{http://arxiv.org/abs/2210.07025}{{\tt arXiv:2210.07025}}].

\bibitem{Vasiliev:1981yc}
A.~N. Vasiliev, Y.~M. Pismak, and Y.~R. Khonkonen, {\it {Simple Method of
  Calculating the Critical Indices in the 1/$N$ Expansion}},  {\em Theor. Math.
  Phys.} {\bf 46} (1981) 104--113.

\bibitem{Vasiliev:1981dg}
A.~N. Vasiliev, Y.~M. Pismak, and Y.~R. Khonkonen, {\it {1/$N$ Expansion:
  Calculation of the Exponents $\eta$ and Nu in the Order 1/$N^2$ for Arbitrary
  Number of Dimensions}},  {\em Theor. Math. Phys.} {\bf 47} (1981) 465--475.

\bibitem{Vasiliev:1982dc}
A.~N. Vasiliev, Y.~M. Pismak, and Y.~R. Khonkonen, {\it {1/N expansion:
  calculation of the exponent $\eta$ in the order 1/$N^3$ by the conformal
  bootstrap method}},  {\em Theor. Math. Phys.} {\bf 50} (1982) 127--134.

\bibitem{Smirnov:2008iw}
A.~V. Smirnov, {\it {Algorithm FIRE -- Feynman Integral REduction}},  {\em
  JHEP} {\bf 10} (2008) 107, [\href{http://arxiv.org/abs/0807.3243}{{\tt
  arXiv:0807.3243}}].

\bibitem{Maierhofer:2017gsa}
P.~Maierh\"ofer, J.~Usovitsch, and P.~Uwer, {\it {Kira\textemdash{}A Feynman
  integral reduction program}},  {\em Comput. Phys. Commun.} {\bf 230} (2018)
  99--112, [\href{http://arxiv.org/abs/1705.05610}{{\tt arXiv:1705.05610}}].

\bibitem{Wu:2023upw}
Z.~Wu, J.~Boehm, R.~Ma, H.~Xu, and Y.~Zhang, {\it {NeatIBP 1.0, a package
  generating small-size integration-by-parts relations for Feynman integrals}},
   {\em Comput. Phys. Commun.} {\bf 295} (2024) 108999,
  [\href{http://arxiv.org/abs/2305.08783}{{\tt arXiv:2305.08783}}].

\bibitem{Guan:2024byi}
X.~Guan, X.~Liu, Y.-Q. Ma, and W.-H. Wu, {\it {Blade: A package for
  block-triangular form improved Feynman integrals decomposition}},  {\em
  Comput. Phys. Commun.} {\bf 310} (2025) 109538,
  [\href{http://arxiv.org/abs/2405.14621}{{\tt arXiv:2405.14621}}].

\bibitem{Tkachov:1981wb}
F.~V. Tkachov, {\it {A theorem on analytical calculability of 4-loop
  renormalization group functions}},  {\em Phys. Lett. B} {\bf 100} (1981)
  65--68.

\bibitem{Chetyrkin:1981qh}
K.~G. Chetyrkin and F.~V. Tkachov, {\it {Integration by parts: The algorithm to
  calculate $\beta$-functions in 4 loops}},  {\em Nucl. Phys. B} {\bf 192}
  (1981) 159--204.

\bibitem{Laporta:2000dsw}
S.~Laporta, {\it {High-precision calculation of multiloop Feynman integrals by
  difference equations}},  {\em Int. J. Mod. Phys. A} {\bf 15} (2000)
  5087--5159, [\href{http://arxiv.org/abs/hep-ph/0102033}{{\tt
  hep-ph/0102033}}].

\bibitem{Derkachov:1997ch}
S.~E. Derkachov and A.~N. Manashov, {\it {The Simple scheme for the calculation
  of the anomalous dimensions of composite operators in the 1/N expansion}},
  {\em Nucl. Phys. B} {\bf 522} (1998) 301--320,
  [\href{http://arxiv.org/abs/hep-th/9710015}{{\tt hep-th/9710015}}].

\bibitem{Cao:2021cdt}
W.~Cao, F.~Herzog, T.~Melia, and J.~R. Nepveu, {\it {Renormalization and
  non-renormalization of scalar EFTs at higher orders}},  {\em JHEP} {\bf 09}
  (2021) 014, [\href{http://arxiv.org/abs/2105.12742}{{\tt arXiv:2105.12742}}].

\end{thebibliography}\endgroup

\end{document}